\documentclass[a4paper]{article}
\usepackage[top=2cm,bottom=2cm,left=2.5cm,right=2.5cm]{geometry}
\usepackage[utf8]{inputenc}
\usepackage{amsmath,amssymb}
\usepackage[colorlinks=true]{hyperref}
\usepackage{graphicx}
\usepackage{sidecap}
\sidecaptionvpos{figure}{c}
\usepackage{subcaption}
\usepackage[dvipsnames]{xcolor}

\usepackage{pgfplots}
\usepgfplotslibrary{external}
\numberwithin{equation}{section}

\newcommand{\ubar}{\overline{u}}

\newcommand{\sgn}{\mathrm{sgn}}

\newcommand{\cc}{{\mathrm{c.c.}}}

\newcommand{\rmd}{\mathrm{d}}

\title{Dispersive Riemann problem for the Benjamin-Bona-Mahony equation}
\author{\textsc{T. Congy, G.~A. El, M.~A. Hoefer \& M. Shearer}}

\begin{document}

\maketitle

\begin{abstract}
Long time dynamics of the smoothed step initial value problem or
dispersive Riemann problem for the Benjamin-Bona-Mahony (BBM) equation
$u_t + uu_x = u_{xxt}$ are studied using asymptotic methods and
numerical simulations. The catalog of solutions of the dispersive
Riemann problem for the BBM equation is much richer than for the
related, integrable, Korteweg-de Vries equation
$u_t + uu_x + u_{xxx} =0.$
% In contrast to its integrable counterpart, the Korteweg-de Vries
% equation $u_t + uu_x + u_{xxx} =0$, the classification of the
% dispersive Riemann problem for the BBM equation is significantly
% richer.
The transition width of the initial smoothed step is found to
significantly impact the dynamics.  Narrow width gives rise to
rarefaction and dispersive shock wave (DSW) solutions that are
accompanied by the generation of two-phase linear wavetrains, solitary
wave shedding, and expansion shocks.  Both narrow and broad initial
widths give rise to two-phase nonlinear wavetrains or DSW implosion
and a new kind of dispersive Lax shock for symmetric data.  The
dispersive Lax shock is described by an approximate self-similar
solution of the BBM equation whose limit as $t \to \infty$ is a
stationary, discontinuous weak solution.  By introducing a slight
asymmetry in the data for the dispersive Lax shock, the generation of
an incoherent solitary wavetrain is observed.  Further asymmetry leads
to the DSW implosion regime that is effectively described by a pair of
coupled nonlinear Schr\"{o}dinger equations.  The complex interplay
between nonlocality, nonlinearity and dispersion in the BBM equation
underlies the rich variety of nonclassical dispersive hydrodynamic
solutions to the dispersive Riemann problem.
\end{abstract}

\tableofcontents\vspace{1cm}

\section{Introduction}  

The unidirectional regularized shallow water equation, often called
the Benjamin-Bona-Mahony (BBM) equation \cite{benjamin_model_1972},
was originally introduced by Peregrine
\cite{peregrine_calculations_1966} to numerically model the
development of shallow-water undular bores.  The canonical
representation of the BBM equation as an asymptotic model for shallow
water waves has the form
\begin{equation}
  \label{BBM_original}
  u_t+u_x+\varepsilon (uu_x - \alpha u_{xxt})=0,
\end{equation}
where $\varepsilon \ll 1$ and $\alpha={\cal O}(1)$ are constants.
Equation \eqref{BBM_original} shows that, to leading order in
$\varepsilon$, $u_t=-u_x$ so that \eqref{BBM_original} is
asymptotically equivalent to the Korteweg-de Vries (KdV) equation
\begin{equation}
  \label{kdv_asymp}
  u_t+u_x+\varepsilon (uu_x  + \alpha u_{xxx})=0.
\end{equation}
Despite the asymptotic equivalence of the KdV and BBM equations, their
mathematical properties are drastically different. In particular, the
BBM equation \eqref{BBM_original} yields more satisfactory short-wave
behavior, owing to its regularization of the unbounded growth in the
frequency, phase and group velocity values present in the linear
dispersion relation for the KdV equation.  This leads to a number of
advantages of the BBM equation in the context of well-posedness and
computational convenience.  On the other hand, the BBM equation lacks
integrability, which is a prominent feature of the KdV equation with
many remarkable consequences including an infinite number of conserved
quantities and the existence of solitons, localized solutions
exhibiting elastic interactions.

In addition to the particular application of the KdV and the BBM
equations as asymptotic models for weakly nonlinear shallow water
waves and other dispersive media, these equations exemplify two
qualitatively different ways to regularize a scalar nonlinear
conservation law, acutely displayed by their normalized versions:
\begin{align}
  \label{kdv}
  \text{KdV:}&  \qquad u_t + uu_x =- u_{xxx}, \\
  \label{bbm}
  \text{BBM:}& \qquad  u_t + uu_x = u_{xxt}.
\end{align}
Equations \eqref{kdv} and \eqref{bbm} can be obtained from
\eqref{kdv_asymp} and \eqref{BBM_original}, respectively, by
introducing the change of variables
\begin{equation}
  \label{change1}
  \tilde x = x/\sqrt{\varepsilon\alpha}, \quad \tilde t =
  t/\sqrt{\varepsilon\alpha}, \quad \tilde u = \varepsilon u +1,
\end{equation}
and dropping the tildes.

In this paper, we study the dispersive regularization of an initial
step-like transition for the BBM equation \eqref{bbm}.  This kind of
dispersive Riemann problem was introduced and studied for the KdV
equation \eqref{kdv} by Gurevich and Pitaevskii
\cite{gurevich_nonstationary_1974} using Whitham modulation theory
\cite{whitham_linear_1999}. The most prominent feature of the KdV
dispersive Riemann problem is the occurrence of a {\em dispersive
  shock wave} (DSW), a smooth, expanding nonlinear wavetrain that
replaces the discontinuous, traveling shock solution of the hyperbolic
conservation law $u_t + (\tfrac12 u^2)_x=0$---the Hopf
equation---subject to the initial data
\begin{equation}
  \label{step2}
  u(x,0) =
  \begin{cases}
    u_- &\text{if } x < 0\\[6pt]
    u_+ &\text{if } x \ge 0
  \end{cases},
\end{equation}
when $u_->u_+$ (the Lax entropy condition) and the shock speed $c$
satisfies the Rankine-Hugoniot jump condition
$c = \tfrac12 (u_-+u_+)$.  If $u_-<u_+$, the initial discontinuity
\eqref{step2} for the Hopf equation evolves into an expanding,
continuous {\em rarefaction wave} (RW), which retains its general
structure under KdV dispersive regularization subject to smoothing of
its weak discontinuities and small-amplitude oscillations at the left
corner of the classical RW.  DSWs and RWs are the only possible KdV
dispersive regularizations of the Riemann problem.  Because
qualitatively similar DSWs and RWs have been found to arise in a
variety of ``KdV-like'' dispersive regularizations of hyperbolic
conservation laws, such DSWs and RWs are referred to as classical or
convex.

It turns out that the BBM dispersive Riemann problem exhibits a number
of features that are markedly different from the KdV dispersive
Riemann problem, i.e., are nonclassical. To elucidate them, we first
note that the consideration of discontinuous Riemann data
\eqref{step2} for a dispersive equation leads to anomalous features,
such as the generation of waves with unbounded phase and group
velocities in the KdV equation \cite{bilman_numerical_2020}.  One way
to avoid this unphysical behavior, is to introduce smoothed Riemann
data, e.g.,
\begin{equation}
  \label{step}
  u(x,0) = \frac{u_+-u_-}{2} \tanh \left( \frac{x}{\xi} \right) +
  \frac{u_++u_-}{2},
\end{equation}
where the parameter $\xi$ represents the characteristic width of the
initial transition. If $\xi \to 0$, \eqref{step} converges to the
discontinuous data \eqref{step2}. For applications, the consideration
of a nonlinear dispersive equation with smoothed step data of the type
\eqref{step} constitutes a physically meaningful dispersive Riemann
problem.  Henceforth, we will refer to the data \eqref{step} as a
smoothed step.

Drawing upon the theory of Riemann problems for hyperbolic
conservation laws, one may hypothesize that the leading order,
long-time asymptotic solution of the dispersive Riemann problem is
independent of the value of $\xi$. Indeed, the evolution of a smooth,
compressive step with $u_->u_+$ in the KdV equation is asymptotically
described by a self-similar solution of the Whitham modulation
equations, which does not involve the initial step's width or shape
and only depends on the boundary parameters $u_\pm$ where
$x \to \pm \infty$ \cite{ablowitz_dispersive_2013}. However, solutions
of BBM Riemann problems turn out to be quite sensitive to the value of
$\xi$, leading to qualitative differences in the wave patterns
generated by the evolution of the smoothed step \eqref{step} with the
same values of $u_{\pm}$ but different values of $\xi$. This can be
understood by noting that the BBM equation~\eqref{bbm} can be put in
the integro-differential form \cite{benjamin_model_1972}
\begin{equation}
  \label{bbm2}
  u_t(x,t) =  \frac14 \int \limits_{-\infty}^{+\infty} {\rm
    sgn}(x-y) e^{-|x-y|} u(y,t)^2 dy ,
\end{equation}
explicitly reflecting its nonevolutionary and nonlocal character.  The
factor $e^{-|x-y|}$ in \eqref{bbm2} introduces the intrinsic BBM
nonlocality length scale $\ell = 1$.  Consequently, the regimes of the
initial value problem~\eqref{bbm}, \eqref{step} characterized by $\xi
\gg 1$ and $\xi \ll 1$ are expected to be qualitatively different. The
nonevolutionary character of the BBM equation also results in
nonconvexity of the linear dispersion relation, with zero dispersion
points for certain values of the wavenumber and background. Generally,
we find that the complex interplay between nonlocality, nonlinearity
and dispersion in the BBM equation gives rise to remarkably rich
nonclassical dynamics that are revealed in the study of the Riemann
problem.

It is worth commenting briefly on the relevance of the BBM equation
\eqref{BBM_original} and its reduced version \eqref{bbm} for physical
applications.  While \eqref{BBM_original} is a generic asymptotic
model of weakly nonlinear, dispersive waves, it is not an
asymptotically resolved equation.  The small parameter $\epsilon$
cannot be scaled out of the equation while maintaining its asymptotic
validity.  For example, the transformation \eqref{change1} breaks the
asymptotic long wave assumption.  However, there are geophysical
scenarios in which the reduced BBM equation \eqref{bbm} is a
reasonable physical model.  An example scenario is the circular, free
interface between two viscous, Stokes fluids whose long wave evolution
is well-approximated by the conduit equation $u_t + uu_x - u u_{txx} +
u_t u_{xx} = 0$ \cite{olson_solitary_1986,lowman_dispersive_2013-1}.
Other examples include a class of models for magma transport in the
upper mantle $u_t + (u^n)_x - (u^n(u^{-m}u_t)_x)_x = 0$ where $n \in
[2,5]$, $m \in [0,1]$ are parameters
\cite{scott_magma_1984,whitehead_magma_1990} and channelized water
flow beneath glaciers $u_t + (u^\alpha (1 - (\sgn (u_t) |u_t
u^{-1}|^{1/n})_x)^\beta)_x = 0$ where $n \ge 1$, $\alpha > 1$, $\beta
> 0$ are parameters \cite{stubblefield_solitary_2020}.  The BBM
equation \eqref{bbm} captures the quadratic hydrodynamic flux $\left (
  \frac{1}{2} u^2 \right )_x$ and nonlocal, nonevolutionary dispersion
$-u_{xxt}$ inherent to these physical models in certain parameter
regimes.  Therefore, the BBM equation in its reduced, normalized form
\eqref{bbm} represents a significant, physically inspired dispersive
regularization of the Hopf equation that is distinct from the KdV
dispersive regularization \eqref{kdv}.

In this paper, we report on wave structures and phenomena associated
with the Riemann problem for the BBM equation.  Solutions of
\eqref{bbm}, \eqref{step} include familiar wave patterns such as
classical RWs and DSWs that also appear in the dispersive Riemann
problem for the KdV equation.  However, the BBM dispersive Riemann
problem additionally exhibits a variety of nonclassical waves that do
not appear as solutions to KdV-type equations.  These include dynamics
that depend strongly upon the smooth step transition width $\xi$ and
those that do not.  When $\xi \ll 1$, two-phase linear wavepackets
accompany the RWs and DSWs.  When $u_+ = -u_-$, the BBM equation
\eqref{bbm} admits a stationary, discontinuous weak solution that is
stable according to the Lax entropy condition
\cite{lax_hyperbolic_1957, lax_hyperbolic_1973} if $0<u_-$.  For smooth initial data
\eqref{step} in which $u_+ = -u_- < 0$, the numerical solution evolves
toward the Lax shock solution but is accompanied by increasingly short
waves.  These waves are described by an approximate self-similar
solution to the BBM equation that we term a dispersive Lax shock.
However, the solution is observed to be structurally unstable to
slightly asymmetric initial conditions ($u_+ \ne -u_-$), numerically
evolving into an incoherent collection of small amplitude, short waves
and solitary waves. Remarkably, for $u_+ = -u_- >0$, for which the
discontinuous solution violates the Lax entropy condition and is an
expansive shock wave solution of the Hopf equation, the solution of
the corresponding initial value problem \eqref{bbm}, \eqref{step} with
$0 < \xi \ll 1$ exhibits a smooth solution that approximates the
discontinuity.  In this case, the smooth expansion shock decays
algebraically in time, giving way to a RW as shown by an asymptotic
analysis in \cite{el_expansion_2016}.  In that paper, it was also
observed that when the initial data fail to be symmetric, specifically
($u_+\ne -u_-,$), then the expansion shock may be accompanied by a
train of one or more solitary waves.  Finally, for arbitrary $\xi >
0$, we identify a regime of the BBM Riemann problem that gives rise to
DSW implosion in which a nonlinear two-phase interaction develops from
the small amplitude edge of the DSW. DSW implosion was previously
predicted and observed in the conduit and magma equations
\cite{lowman_dispersive_2013}.

As we have noted, the BBM equation can be viewed as a prototypical
model for wave phenomena that arises in other physical systems
described by nonevolutionary, nonlinear dispersive wave
equations. While DSW implosion has been observed in the conduit and
magma equations, other intriguing BBM wave patterns such as expansion
shocks, solitary wave shedding, and dispersive Lax shocks await their
realization in physical systems.  These nonclassical wave patterns
require new mathematical approaches for their interpretation.  We
utilize detailed numerical simulation, asymptotic methods, and
properties of the BBM equation in order to provide a relatively
complete classification of solutions of the BBM dispersive Riemann
problem.

The paper is organized as follows.  Owing to the classification's
richness and complexity, we begin with a high-level description of our
findings in Section \ref{bbm_riemann}.  We review the long-time
asymptotic description of linear waves in Section \ref{linear_wave} as
they play a prominent role in BBM dispersive Riemann problem dynamics.
This is followed by Section \ref{sec:shocks-rarefactions} in which
rarefaction waves, dispersive Lax shocks, and expansion shocks are
studied. This leads naturally to the description and analysis of the
shedding of solitary waves in Section \ref{shedding}.  Section
\ref{DSW} is devoted to DSW solutions, DSW implosion that includes a
new two-phase description of the dynamics, and the generation of
incoherent solitary wavetrains.  We conclude in Section
\ref{sec:conclusions-outlook} with a discussion and future outlook.

\section{The BBM Riemann problem: summary of solutions and analysis}
\label{bbm_riemann}
With the scaling of variables,
\begin{equation}
\label{change_var}
u\to \alpha v,\quad x \to \beta X, \quad t \to \gamma T,
\end{equation}
the BBM equation becomes:
\begin{equation}
v_T + \frac{\alpha \gamma}{\beta} vv_X - \frac{1}{\beta^2} v_{XXT} = 0.
\end{equation}
Thus~\eqref{bbm} is invariant under the change of
variables~\eqref{change_var} if $\alpha\gamma = \beta = \pm 1$. In
particular the BBM equation remains unchanged after $(u,x)\to
(-u,-x).$ We therefore restrict the study of the dispersive Riemann
problem~\eqref{bbm}, \eqref{step} to initial data satisfying
$u_-+u_+\ge 0$.

\begin{figure}
\centering
\includegraphics{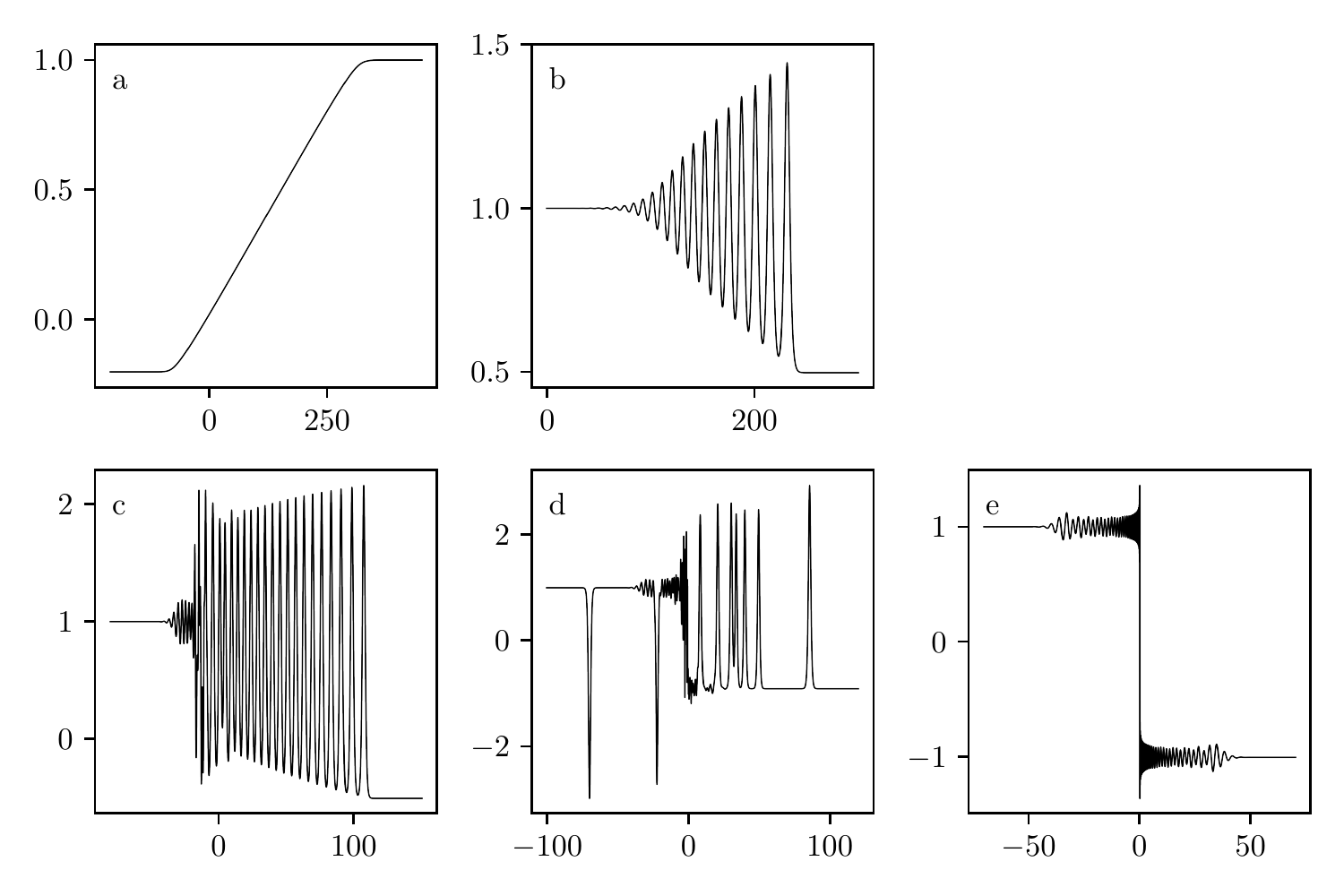}
\hspace*{-20mm}
\begin{minipage}{0.47\linewidth}
\includegraphics{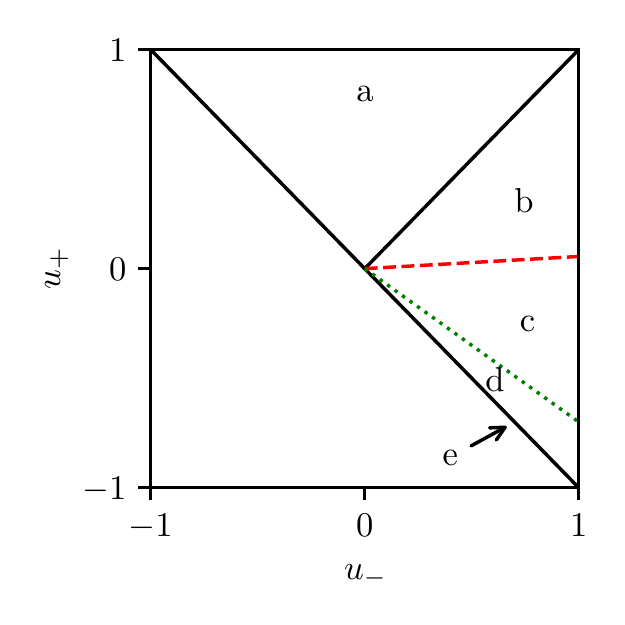}
\end{minipage}
\begin{minipage}{0.47\linewidth}
\def\arraystretch{1.5}
\begin{tabular}{l|l|l}
\hline
a & $|u_-| < u_+$ & convex RW \\
\hline
b & $\mu \,u_- < u_+ < u_-$ & convex DSW \\
\hline
c & $\sigma_3(\xi) \,u_- < u_+ < \mu \,u_-$ & DSW
implosion \\
\hline
d & $-u_- < u_+ < \sigma_3(\xi) \,u_-$ & incoherent waves \\
\hline
e & $u_+ = - u_- < 0$ & dispersive Lax shock \\
\hline
\end{tabular}
\end{minipage}
\caption{Classification of the BBM dispersive Riemann
  problem~\eqref{bbm}, \eqref{step} in the $u_++u_-\geq 0$ half plane
  (triangle) when $\xi \gg 1$.  Corresponding representative numerical
  simulations for $\xi = 10$ in each case are shown.  The table is a
  description of the partitioning and wave patterns. $\mu = e^{-3/2}/4
  \approx 0.056$, $\sigma_3(\xi)$ is generally unknown but empirically
  $\sigma_3(10) \approx -0.70$.}
\label{fig:xi10}
\end{figure}

\begin{figure}
\centering
\includegraphics{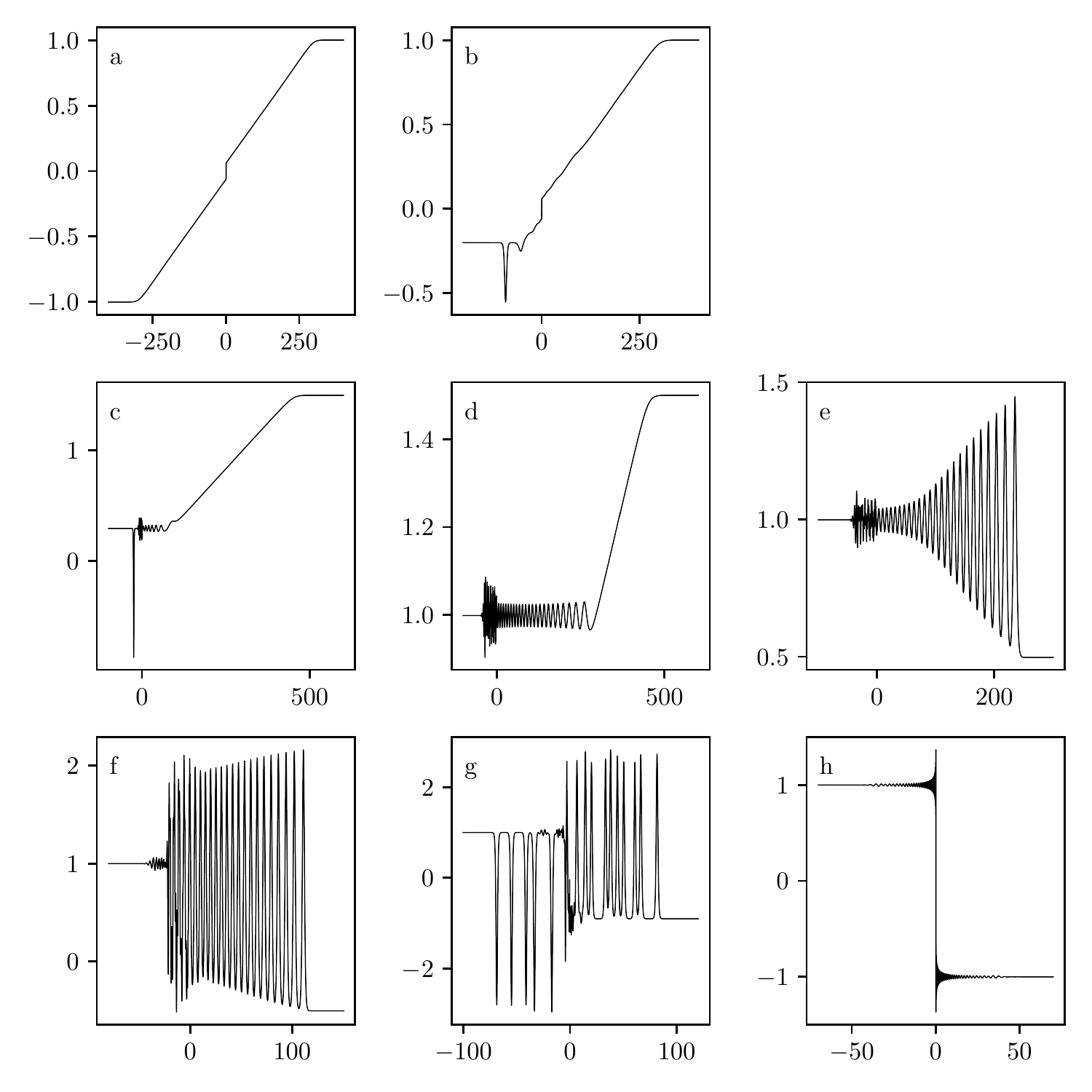}
\hspace*{-20mm}
\begin{minipage}{0.42\linewidth}
\includegraphics{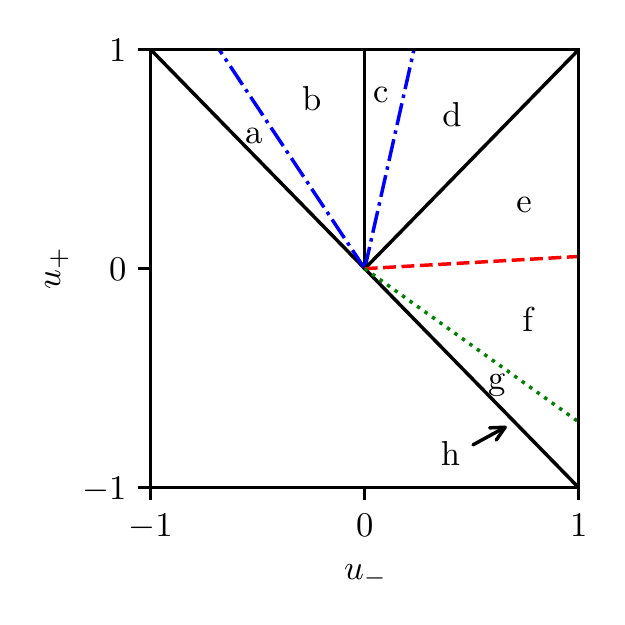}
\end{minipage}
\begin{minipage}{0.52\linewidth}
\def\arraystretch{1.5}
\begin{tabular}{l|l|l}
\hline
a & $0 < \! -u_- \! \le \! u_+ \! < \! -\sigma_1(\xi) \, u_-$ & expansion shock \\
\hline
b & $0 < -\sigma_1(\xi) \, u_- < u_+$ & exp shock
$\!\!+\!\!$ solis \\
\hline
c & $0 < \sigma_2(\xi) \, u_- < u_+$ & RW $\!\!+\!\!$ solis
$\!\!+\!\!$ lin waves \\
\hline
d & $u_- < u_+ < \sigma_2(\xi) \, u_-$ & RW $\!\!+\!\!$ lin
waves \\
\hline
e & $\mu \, u_- < u_+ < u_-$ & DSW $\!\!+\!\!$ lin waves
\\
\hline
f & $\sigma_3(\xi) \, u_- < u_+ < \mu \, u_-$ & DSW
implosion \\
\hline
g & $-u_- < u_+ < \sigma_3(\xi) \, u_-$ & incoherent waves \\
\hline
h & $u_+ = -u_- < 0$ & dispersive Lax shock \\
\hline
\end{tabular}
\end{minipage}
\caption{Classification of the BBM dispersive Riemann
  problem~\eqref{bbm}, \eqref{step} when $\xi \ll 1$.  The triangle is
  a partitioning of the boundary data in the $u_+ + u_- \ge 0$ half
  plane.  Corresponding representative numerical simulations for $\xi
  = 0.1$ in each case are shown.  The table is a description of the
  partitioning and wave patterns.  $\mu = e^{-3/2}/4 \approx 0.056$,
  $\sigma_2(\xi) = 1 + \sqrt{6\beta} \xi^{1/2} + \sqrt{27\beta/8}
  \xi^{5/2} + \cdots$, $\beta \approx 9.5$, $\sigma_3(\xi)$ is
  generally unknown but empirically $\sigma_3(0.1) \approx -0.70$.
  $\sigma_1(\xi)$ is generally unknown but empirically $\sigma_1(0.1)
  \approx 0.68$.}
\label{fig:xi01}
\end{figure}

In this section and throughout the paper, we refer to Figures
\ref{fig:xi10} and \ref{fig:xi01}, which
concisely capture the
BBM dispersive Riemann problem classification.  The figures include a
partitioning of the $u_+ \ge - u_-$ half plane for the smoothed step
initial data \eqref{step} and corresponding representative numerical
simulations of each qualitatively distinct wave pattern that emerges
during the course of BBM \eqref{bbm} evolution.  Figure \ref{fig:xi10}
corresponds to the broad, slowly varying smoothed initial step in
which the transition width parameter $\xi \gg 1$ ($\xi = 10$ in all
numerical simulations).  Figure \ref{fig:xi01} represents solutions
for which the smoothing is narrow and sharp, corresponding to $\xi \ll
1$ ($\xi = 0.1$ in all numerical simulations).  Some wave patterns
persist for both large and small transition width, but many do not.
We do not investigate the subtleties of the crossover $\xi \sim 1$ in
which the transition width coincides with the nonlocality length,
choosing instead to utilize scale separation for $\xi \gg 1$ and $\xi
\ll 1$ where asymptotic methods are available.

\bigskip
\noindent{\it Large transition width $\xi \gg 1$, Figure \ref{fig:xi10}}

\medskip When $u_+ > \max(\mu u_-, -u_-)$ where
$\mu = e^{-3/2}/4$, convex RWs (region $a$) and DSWs (region $b$)
are generated, qualitatively similar to those generated by KdV
\eqref{kdv} evolution.  In Sec.~\ref{sec:BBM_fitting}, we use the DSW
fitting approach to make quantitative predictions for the DSW's
properties (edge velocities, harmonic edge wavenumber, soliton edge
amplitude, and DSW structure near the harmonic edge).  We also
identify the loss of convexity at the harmonic edge (where a zero
linear dispersion point is attained) as the progenitor for
nonclassical wave patterns when $u_+$ is below the line
$\mu u_-$.

For initial data sufficiently close to the line $u_+ = \mu u_-$
(region $c$), DSW implosion occurs. A weakly nonlinear, two-phase
modulation theory using coupled nonlinear Schr\"{o}dinger equations is
developed in Sec.~\ref{implosion}, the two wave envelopes associated
with modulations of short and long waves, respectively.  We find that
there is a spatial redistribution of energy from the long waves in a
partial DSW on the right to a short wave wavepacket on the left.  The
partial DSW is typically accompanied by some number of depression
envelope solitary waves.  Very similar DSW implosion dynamics are also
observed in the small transition width regime $\xi \ll 1$ (see region
$f$ in Fig.~\ref{fig:xi01}).

There exists a stationary, discontinuous compressive shock solution of
the BBM equation when $u_+ = -u_- < 0$ (region $e$).  In
Sec.~\ref{sec:stat-lax-shocks}, we show that smoothed step initial
data for this case exhibits the usual nonlinear self-steepening but is
also accompanied by the continued production of shorter and shorter
waves during evolution.  In a neighborhood of $x = 0$, an oscillatory
overshoot appears that quickly saturates to a constant magnitude.  An
approximate self-similar solution in the form $u(x,t) = g(xt)$ is
obtained that describes these oscillations as $t \to \infty$.  This
feature is reminiscent of Gibbs' phenomenon in the theory of Fourier
series and linear dispersive partial differential equations with
discontinuous initial data \cite{biondini_gibbs_2017}.  We refer to
this wave pattern as a dispersive Lax shock.  However, if the
initial data is perturbed asymmetrically so that $-u_- < u_+ <
\sigma_3(\xi) u_-$ (region $d$), then the evolution exhibits
a large number of solitary waves and highly oscillatory wavepackets
that do not exhibit a discernible, coherent pattern.  We refer to this
as the incoherent solitary wave regime and describe it in
Subsection~\ref{sec:incoh-solit-wave}.  The dispersive Lax shock and
the incoherent solitary wave patterns also occur for small transition
width $\xi \ll 1$ (see regions $h$ and $g$ in Fig.~\ref{fig:xi01}).

\bigskip
\noindent{\it Small transition width $\xi \ll 1$, Figure \ref{fig:xi01}}

\medskip Generally, the features observed for large transition width
in Fig.~\ref{fig:xi10} persist into the small transition width regime
shown in Fig.~\ref{fig:xi01} but with modification.  Rarefaction waves
are accompanied by a decaying expansion shock in regions $a$ and $b$,
depression solitary waves propagating to the left in regions $b$ and $c$,
and small amplitude linear waves propagating to the left in regions
$c$ and $d$.

From the ideal expansion shock solution for symmetric data ($u_- =
-u_+ < 0$) (reviewed in Sec.~\ref{expansion}) into region $a$ for
asymmetric data, a short-time analysis of the solution reveals an
asymmetry that favors the generation of a depression wave for $x < 0$
(Sec.~\ref{early_time}).  When the threshold $u_+ = -\sigma_1(\xi)
u_-$ is exceeded, the asymmetry produces a solitary wave that is shed
from the expansion shock. Greater asymmetry generates the shedding of
more depression solitary waves until $u_+ \ge 0$ and an expansion
shock is no longer generated.  Nevertheless, solitary wave shedding persists
until $u_+ = \sigma_2(\xi) u_-$.  Solitary Wave shedding thresholds
$\sigma_1(\xi)$ and $\sigma_2(\xi)$ are analyzed in
Sec.~\ref{threshold}.

When $u_- \ge 0$ (regions $c$--$e$), the small transition width gives
rise to the generation of small amplitude wavepackets that accompany
the RW or DSW.  In the absence of solitary wave shedding, we can
conveniently partition the initial smoothed step in the Fourier domain
to prescribe initial data for the linearized BBM equation.  In
Sec.~\ref{linear_wave}, long-time asymptotic analysis of the solution
integral yields a precise prediction for three distinct wavepacket
regimes corresponding to one-phase, two-phase, and Airy modulations.
The analysis of DSW implosion (regions $c$ and $f$ in
Figs.~\ref{fig:xi10} and \ref{fig:xi01}, respectively) can be viewed
as a weakly nonlinear generalization of the linear wavepacket
analysis.

The dynamics of DSW implosion (region $f$), incoherent waves (region
$g$), and the dispersive Lax shock (region $h$) are qualitatively
similar to the respective regions in the large transition width case
identified in Fig.~\ref{fig:xi10}.

A general conclusion from our analysis, in both the large and small
transition width regimes, is that the loss of linear dispersion
convexity underlies the generation of nonclassical wave patterns.

\section{Linear wavetrains and Airy modulation near the zero
  dispersion point}
\label{linear_wave}

In this section, we investigate the asymptotic structure of the
approximately linear wavetrains generated in BBM dispersive Riemann
problems with smoothed step initial data \eqref{step} in the parameter
domain $(u_-,u_+, \xi)$ defined by: $0<u_-$, $\mu < u_+/u_- <
\sigma_2(\xi)$, where $\mu = {\rm e}^{-3/2}/4$ and $\sigma_2 (\xi)$ is
defined later in equation \eqref{umax}; see regions $d$ and $e$ in
Fig.~\ref{fig:xi01}.  The significance of $\sigma_{1,2}$ will be
clarified later. Depending on the relative values of $u_-$ and $u_+$,
the initial step generates either a RW ($u_-<u_+$ ) or a DSW
($u_->u_+$). In both cases, an approximately linear wavetrain develops
on the background $u=u_-$ behind the RW or DSW and occupies an
expanding region $x \in [x_0^-(t), x_0^+(t)]$, where the right
boundary $x_0^+$ coincides with the trailing edge of the RW or DSW and
$x_0^-$ will be determined.  In this section, we use linear theory to
analyze the corresponding wavetrains, hence we refer to them as
``linear wavetrains'', with an understanding that this is an
approximation.

\subsection{Linear wavetrains in the BBM Riemann problem}

We start by linearizing the BBM equation \eqref{bbm} about a constant
background $\ubar >0$, $u(x,t) = \ubar + \varphi(x,t)$, $|\varphi| \ll
\ubar$, and obtain
\begin{equation}
\label{bbm_lin}
\varphi_t + \ubar \varphi_x - \varphi_{xxt} = 0.
\end{equation}
For $\varphi \propto \exp[i(k x - \omega t)]$, we obtain the
BBM linear dispersion relation:
\begin{equation}
\label{dispersion}
\omega =\omega_0(k,\ubar)=\ubar\frac{k}{1+ k^2} .
\end{equation}
Then the {\it group velocity} is given by the first derivative of $\omega_0$
\begin{equation}
\label{vg}
\partial_k \omega_0 = \ubar\frac{1-k^2}{(1+ k^2)^2},
\end{equation}
and the {\it dispersion sign} is defined by the sign of the second
derivative
\begin{equation}
\label{disp_curv}
\sgn \left ( \partial_{kk} \omega_0 \right ) = \sgn \left ( \ubar\frac{2 k
(k^2-3)}{(1+ k^2)^3} \right ).
\end{equation}
The dispersion relation is nonconvex and the dispersion sign is zero
for $k=0$, $k=\sqrt{3}$ or $\ubar =0$. We stress that it is the
non-convexity of the linear dispersion relation \eqref{dispersion}
that gives rise to crucial differences between the BBM and KdV
dispersive Riemann problems.  The KdV linear dispersion relation is
strictly convex for $k > 0$.

The group velocity \eqref{vg} has a minimum when $k=\sqrt{3}$,
\begin{equation}
\partial_k \omega_0 \ge s_{\min}(\ubar) =  -\frac{\ubar}{8} ,
\end{equation}
and a maximum at $k=0$,
\begin{equation}
\partial_k \omega_0 \le s_{\max} (\ubar) = \ubar.
\end{equation}
Consequently, a BBM linear wavetrain that develops from a localized
initial disturbance {\it considered in isolation} on the constant
background $\overline{u}$ is confined to the region $x \in [s_{\min}t,
s_{\max} t]$ in the asymptotic regime $t \gg 1$, although some rapidly
decaying oscillations are possible outside this region.  We now place
this in the context of the dispersive Riemann problem where the linear
wavetrain is generated on the background $\ubar = u_-$ as part of the
nonlinear-dispersive regularization of smoothed step initial data.  We
consider two cases $u_-<u_+$ and $u_->u_+$ separately, distinguishing
between a linear wave trailing a RW or a DSW (see
Fig.~\ref{fig:xi01}(d,e)).

(i) First, consider the case $u_-<u_+$ that leads to a RW. As we shall
see later in Sec.~\ref{RW}, the trailing edge of the RW is located at
$x=u_- t = s_{\max}(u_-) t$. Provided $u_+/u_-< \sigma_2(\xi) $, no
solitary wave propagates in the region $x<s_{\max} t$
(cf. Sec.~\ref{RW}). Thus the two kinds of structures generated in
this class of BBM dispersive Riemann problems---the linear wavetrain
for $x \in [s_{\min}(u_-) t, s_{\max}(u_-)t]$, and the RW for $x \in
[u_-t,u_+t]$---do not overlap and can be described separately, except
in the vicinity of the point $x=u_-t$ where the two solutions could be
matched.  A linear wavetrain can also be generated for $u_+/u_- \ge
\sigma_2$ but we do not consider this case because one or more
solitary waves may accompany the wavetrain, complicating its analysis.

(ii) For the case $u_->u_+$, which leads to a DSW, we show in
Sec.~\ref{DSW} that the trailing edge of the DSW is located at $x =
s_- t < s_{\max} (u_-)t$ so that the trailing linear wavetrain is
necessarily attached to the DSW.  For $u_+/u_- >\mu$, we have
$s_{\min}(u_-) < s_-$ and our main concern here will be the
description of the linear wavetrain located outside the vicinity of
the DSW trailing edge point $x = s_- t$, where the two modulated
wavetrains could be matched.

Summarizing, we assume that for dispersive Riemann problems with $\mu
< u_+/u_- < \sigma_2(\xi)$, the generation and propagation of a
small-amplitude wavetrain is governed by the linearized BBM
equation~\eqref{bbm_lin} with $\overline{u} = u_-$, and its dynamics
are fully decoupled from the nonlinear DSW or RW dynamics. In other
words, we suppose that the initial profile's Fourier spectrum can be
partitioned into two essentially distinct components.  The short-wave
(large wavenumber) component of the spectrum is assumed to evolve
according to the linear wave equation \eqref{bbm_lin}.  The long-wave
(small wavenumber) component of the spectrum is assumed to evolve
according to the nonlinear BBM equation \eqref{bbm}.  Consequently,
the linear wave evolution is superposed with the nonlinear evolution.
We justify our partition of the initial Fourier spectrum by comparison
of our linear analysis with numerical simulations.  Since the Fourier
transform of the initial condition~\eqref{step} is given by the
formula,
\begin{equation}
{\cal F}[u_0(x)] = \int_\mathbb{R} u_0(x) e^{-ikx} dx
= \frac{u_+-u_-}{2i} \frac{ \pi \xi}{\sinh(\pi \xi
k/2)} + \pi(u_++u_-) \delta(k),
\end{equation}
the partitioning assumption implies that the Fourier transform of the linear
wave $\varphi(x,t)$ at $t=0$ is approximated by
\begin{equation}
\label{phi0}
\hat \varphi_0 (k) = {\cal F} [\varphi(x,0)] \sim \displaystyle
\frac{u_+-u_-}{2i} \frac{ \pi \xi}{\sinh(\pi \xi k/2)}, \quad \forall
k \in \mathbb{R}.
\end{equation}
The initial condition~\eqref{phi0} simplifies for a step ($\xi \to 0$)
to $\hat \varphi_0 (k) = (u_+-u_-)/i k$, $k>k_0$.  Note that $|\hat
\varphi_0 (k)|$ is a rapidly decaying function of $\xi $, and so no
linear wavetrain is expected to be generated for $\xi \gg 1$. This
agrees with Fig.~\ref{fig:xi10}, which shows the evolution for the
dispersive Riemann problem with $\xi =10$.

The solution of~\eqref{bbm_lin}, is given by
\begin{equation}
\label{spm3}
\varphi(x,t) =  \frac{1}{2\pi} \int_\mathbb{R} \hat \varphi_0(k) e^{-i
\chi(k) t} dk,\quad \chi(k) = \omega_0(k, u_-) - k \frac{x}{t},
\end{equation}
For notational convenience, we suppress the dependence of $\chi$ on
$\frac{x}{t}$.

\subsection{Stationary phase analysis}
Consider the asymptotic regime $t \to \infty$ with $x/t$ fixed.
We can evaluate the integral~\eqref{spm3} using the stationary phase
method, cf. for instance \cite{whitham_linear_1999}.  Fast
oscillations average out in the integral, and the main
contribution comes from the neighborhood of the stationary points
$k_n$ defined as solutions of $\chi'(k)=0$, namely
\begin{equation}
\label{stat}
\partial_k \omega_0 = u_- \frac{1-k_n^2}{(1+ k_n^2)^2} = x/t.
\end{equation}
A first approximation of $\varphi(x,t)$ is obtained by substituting
the Taylor expansions:
\begin{equation}
\label{Phi1}
\hat \varphi_0(k) = \hat \varphi_0(k_n)+
{\cal O}\left[k-k_n \right],
\quad\chi(k) = \chi(k_n) + \frac{\partial_k^2\chi(k_n)}{2} (k-k_n)^2 +
{\cal O}\left[(k-k_n)^3 \right].
\end{equation}
Equation \eqref{stat} implies that $k_n$ depends on $s \equiv
x/t$. The computation of~\eqref{spm3} thus reduces to the integration
of a Gaussian, and the corresponding solution $\varphi$ only depends
on the number of stationary points and their respective positions in
$k$-space:
\begin{equation}
\label{spm1}
\varphi(x,t) \sim \frac{1}{2\pi} \sum_{n} \hat \varphi_0(k_n)
e^{-i \chi(k_n) t}
\int_\mathbb{R} e^{-i \, \partial_{kk}\omega_0(k_n,u_-)
(k-k_n)^2 t/2} dk.
\end{equation}
Real solutions of~\eqref{stat} come in pairs $(-k_n,k_n)$ and, since
$\hat \varphi_0(-k) = \hat \varphi^*_0(k)$ and $\omega_0(-k,u_-) =
-\omega_0(k,u_-)$, the sum in~\eqref{spm1} simplifies to
\cite{whitham_linear_1999}:
\begin{equation}
\label{spm0}
\varphi(x,t) \sim  \frac{1}{\pi} \sum_{k_n>0} \Re \left[
\hat \varphi_0(k_n)
e^{-i \chi(k_n) t}
\int_\mathbb{R} e^{-i \, \partial_{kk}\omega_0(k_n,u_-)
k^2 t/2} dk\right].
\end{equation}

The stationary phase equation~\eqref{stat} divides space-time into two
regions.

(a) In the region $s_{\rm max} > s=x/t >0$, denoted hereafter as Region I,
\eqref{stat} has only one positive solution
\begin{equation}
\label{k1}
0 < k_1 = \sqrt{\frac{-u_--2s+\sqrt{u_-(u_-+8s)}}{2s}} \leq \sqrt 3,
\quad s \in (0,s_{\rm max}),
\end{equation}
so the series \eqref{spm0} has just one term.  This is comparable to
what we would obtain with a convex linear dispersion relation, and we
refer to this region as the {\it convex linear regime}. In fact, for
$k \ll 1$, we obtain the linear dispersion relation $\omega_0(k,\ubar)
\sim \ubar(k-k^3)$, which is the linear dispersion relation for the
KdV equation.

\begin{SCfigure}[][h]
\includegraphics{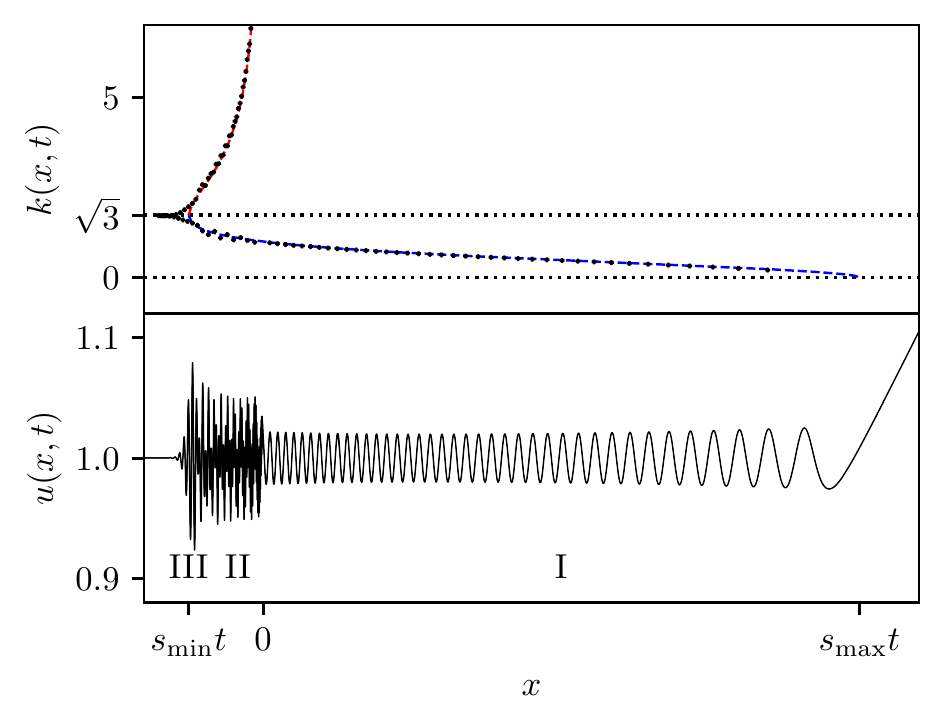}
\caption{The lower panel displays the variation of the dispersive
  Riemann problem's numerical solution with $\xi=0.1$, $u_-=1$ and
  $u_+=1.5$ at $t=500$. The solution is well-approximated by
  a one-phase linear wave in region I and a two-phase linear wave in
  region II. The variation of the waves' wavenumber extracted from the
  numerical solution is represented by black disks in the upper panel;
  the procedure to extract wavenumbers in region II is detailed in
  Appendix~\ref{wave_param}. The blue (red) dashed line corresponds to
  the variation of the analytical solutions $k_1(x,t)$ ($k_2(x,t)$).}
\label{spm_k}
\end{SCfigure}

(b) In the region $s_{\rm min} < s=x/t <0$, denoted hereafter as
Region II, \eqref{stat} has $k_1(s)$ and an additional positive
solution
\begin{equation}
\label{k2}
k_2(s) = \sqrt{\frac{-u_--2s-\sqrt{u_-(u_-+8s)}}{2s}} \geq \sqrt 3,
\quad s \in \left ( s_{\rm min}, 0\right ),
\end{equation}
so that the series \eqref{spm0} has two terms.  The coexistence of two
waves with wavenumbers $k_1$ and $k_2$ at the same position $x$ within
Region II is a direct consequence of the nonconvexity of the
dispersion relation \eqref{dispersion}. The modulation of the linear
wave in this region dramatically differs from a convex-dispersion,
KdV-type linear modulation.  In Sec.~\ref{implosion} below, we show
that the coexistence of two dominant wavenumbers also persists in the
nonlinear regime, leading to the phenomenon of DSW implosion. The
upper panel of Fig.~\ref{spm_k} displays the comparison between the
wavenumber of the waves obtained numerically from the full partial
differential equation (PDE) simulation, shown in the lower panel, and
the graph of $x$ given by \eqref{stat} as a function of $k=k_n$ for a
fixed value of $t,$ with $k_n$, $n = 1,2$ given by the formulas
\eqref{k1}, \eqref{k2}. The two regions I and II are treated
separately using the stationary phase approximation \eqref{spm0}. A
third region, denoted III in Fig.~\ref{spm_k} and defined hereafter as
$x \lesssim s_{\rm min}t$, is resolved with an expansion near
$x=s_{\min} t.$

\medskip (Region I) \ In the convex linear regime of Region I, the
stationary phase approximation yields
\begin{equation}
\varphi(x,t) \sim\sqrt{\frac{2}{\pi}} \Re \left[ \frac{\hat
\varphi_0(k_1)} {\sqrt{|\partial_{kk}\omega_0( k_1,u_-)| t}} e^{i[k_1
x-\omega_0(k_1,u_-)t]- i \, \sgn[ \partial_{kk}\omega_0(
k_1,u_-)]\pi/4}\right].
\end{equation}
Since $\arg[\hat{\varphi}_0(k_1)]=-\pi/2$ and
$\partial_{kk}\omega_0( k_1,u_-)<0$, we have
\begin{equation}
\label{sol1}
\varphi(x,t) \sim \varphi_{\rm I}(x,t)= |\hat \varphi_0(k_1)|
\sqrt{\frac{2}{\pi |\partial_{kk} \omega_0(k_1,u_-)|t}} \cos \left[
k_1 x - \omega_0(k_1,u_-)t - \frac{\pi}{4} \right].
\end{equation}
This solution is not valid close to the so-called caustic trajectory
$x = u_-t=s_{\rm max}t,$ where the two dominant wavenumbers $k_1$ and
$-k_1$ coalesce at $k_1=0$, (cf. for
instance~\cite{child_semiclassical_2014}) and $\partial_{kk}
\omega_0(k_1,u_-) = 0$. At this edge of Region I, one needs to
consider the next order in the expansion of $\chi(k)$ in the Taylor
expansion~\eqref{Phi1}.  Besides, as pointed out earlier, a RW or a
DSW develops close to the point $x/t=u_-$ or $x/t=s_-<u_-$
(respectively), and the linear wave has to be asymptotically matched
with the hydrodynamic state. Such a higher order derivation, which can
be achieved by following the matched asymptotics procedure developed
for the KdV equation
\cite{grava_numerical_2008,leach_large-time_2008}, is not important
for us here because we have a well-defined initial value problem
\eqref{bbm_lin}, \eqref{phi0}. The comparison between the approximate
solution~\eqref{sol1} and the numerical solution of the dispersive
Riemann problem in region I is displayed in Fig.~\ref{spm}.

\medskip (Region II) \ In Region II, for which $x \in [s_{\min} t ,
0]$, the modulated wave corresponding to the large wavenumber branch
$k_2$ coexists with the wave from the lower wavenumber branch $k_1$,
leading to a modulated beating pattern. The stationary phase
approximation then yields the superposition
\begin{equation}
  \begin{split}
    \label{sol3}
    \varphi(x,t)
    &\sim\sqrt{\frac{2}{\pi}} \Re \left[ \frac{\hat \varphi_0(k_1)}
      {\sqrt{-\partial_{kk}\omega_0(
          k_1,u_-) t}} e^{i[k_1 x-\omega_0(k_1,u_-)t]+ i \pi/4} + \frac{\hat
        \varphi_0(k_2)}
      {\sqrt{\partial_{kk}\omega_0(
          k_2,u_-) t}} e^{i[k_2 x-\omega_0(k_2,u_-)t]- i \pi/4}\right],\\
    &= \sqrt{\frac{2}{\pi}} \Bigg( \frac{|\hat \varphi_0(k_1)|}{\sqrt{
          |\partial_{kk} \omega_0(k_1,u_-)| t}} \cos \left[ k_1 x -
        \omega_0(k_1,u_-)t - \frac{\pi}{4} \right] \\
      &\qquad\qquad + \frac{|\hat
        \varphi_0(k_2)|}{\sqrt{ |\partial_{kk} \omega_0(k_2,u_-)| t}} \cos
      \left[ k_2 x - \omega_0(k_2,u_-)t - \frac{3\pi}{4} \right] \Bigg).
  \end{split}
\end{equation}

\medskip (Region III) \ The approximation \eqref{sol3} fails to
describe the oscillations of the linear wave for $x/t$ close to
$s_{\min}$, corresponding to the zero dispersion point where the two
branches coalesce: $k_1 = k_2 = \sqrt 3$, $\partial_{kk}
\omega_0(\sqrt 3,u_-) = 0$.  In this region, the
expansion~\eqref{Phi1} is insufficient.  We shall denote the union of
this special region $s \sim s_{\rm min}$ and $s < s_{\rm min}$ as $s
\lesssim s_{\rm min}$ and call it Region III.  In order to describe
the oscillations in this region, we expand $\chi(k)$ close to the
inflection point $k=\sqrt 3$
\begin{equation}
\chi(k) = \omega_0(\sqrt 3,u_-) - \sqrt 3 \, x/t
- (x/t -  s_{\rm min})  (k -\sqrt 3)
+\frac{\partial_{kkk} \omega_0(\sqrt 3,u_-)}{6} t (k-\sqrt 3)^3 + {\cal
O}[(k -\sqrt 3)^4 ] .
\end{equation}
Substituting this expansion into eq.~\eqref{spm3}, we obtain
\begin{equation}
\label{airy_int}
\varphi(x,t) \sim \frac{1}{\pi} \Re \left[
\hat \varphi_0(\sqrt 3) e^{i [\sqrt 3 x -\omega_0(\sqrt 3,u_-) t]}
\int_\mathbb{R} e^{i (x-s_{\rm min} t)(k-\sqrt 3) -i \,
\partial_{kkk}\omega_0( \sqrt 3,u_-)
(k-\sqrt 3)^3 t/6} dk \right] .
\end{equation}
With the change of variable $k-\sqrt 3 \to -\alpha \kappa$ with
$\alpha = (2/\partial_{kkk}\omega_0( \sqrt 3,u_-)t)^{1/3}$, the
approximation becomes
\begin{equation}
\label{airy}
\varphi(x,t)
= 2 \alpha |\hat \varphi_0(\sqrt
3) |{\rm Ai}[-\alpha(x-s_{\min} t)] \cos
\left[ \sqrt 3 \left(x - \frac{u_-}{4} t \right) - \frac{\pi}{2}
\right], \quad \alpha = \left( \frac{32}{3 u_- t} \right)^{1/3} \ll 1,
\end{equation}
where ${\rm Ai}(y)$ is the Airy function defined by
\cite{abramowitz_handbook_2013}
\begin{equation}
\label{ai_def}
{\rm Ai}(y) = \frac{1}{2\pi} \int_\mathbb{R} e^{i( \kappa y +
\kappa^3/3)} d\kappa \in \mathbb{R}.
\end{equation}
The solution \eqref{airy} represents a harmonic cosine wave modulated
by the Airy function.  Note that the Airy approximation~\eqref{airy}
is also valid for $x/t<s_{\min}$, where the Airy modulation describes
an exponential decay of the wave's amplitude:
${\rm Ai}(y) \sim y^{-1/4}\exp(-\tfrac23 y^{3/2})$ as $y \to +\infty$,
which in our case translates to $(x-s_{\min} t )\to - \infty$.

\subsection{Uniform Airy approximation}

Using the technique originally developed
in~\cite{chester_extension_1957}, we derive a uniform approximation
that is valid in the union of regions II and III whose limiting
behavior is given by~\eqref{airy} when $x \lesssim s_{\rm min} t$
and~\eqref{sol3} when $x \gg s_{\rm min} t$.  We summarize here the
derivation detailed in~\cite{child_semiclassical_2014}. First, we
suppose that the solution in region II accords with the ansatz
\begin{equation}
\label{uni_ansatz}
\varphi(x,t) = \frac{1}{\pi} \Re \left[ \int_\mathbb{R} (\phi_0+
\kappa \phi_1) e^{i(A- \kappa \zeta +
\kappa^3/3)} d\kappa \right],
\end{equation}
where $\phi_0\in \mathbb{C}$, $\phi_1 \in \mathbb{C}$,
$A \in \mathbb{R}$ and $\zeta \in \mathbb{R}$ are functions of $(x,t)$
that are to be determined. By selecting the exponential's phase
$\psi = A- \kappa \zeta + \kappa^3/3$, we will ensure that it has the
same stationary points as $\chi(k)$ for $x$ close to $s_{\rm min}
t$. Note that, although only the leading order term
$\hat \varphi_0(k_n)$ in the Taylor expansion of $\hat \varphi_0(k)$
near $k = k_n$ in the approximations leading to~\eqref{spm0}
and~\eqref{airy_int} was used, incorporating the first order term
$\hat \varphi_0'(k_n)(k-k_n)$ in the analysis is essential for
constructing a uniform approximation.  We first determine the
functions $\phi_0$, $\phi_1$, $A$ and $\zeta$ by evaluating the
integral~\eqref{uni_ansatz} using the stationary phase method. For
simplicity, we consider the computation for $\zeta>0$; the computation
is similar in the case $\zeta<0$,
cf. \cite{child_semiclassical_2014}. The phase $\psi$ has two
stationary points:
\begin{equation}
\kappa = \pm \sqrt \zeta.
\end{equation}
Close to the stationary points $\pm \sqrt \zeta$, the phase
is  given by
\begin{equation}
\psi = A \mp \frac{2}{3} \zeta^{3/2} \pm \sqrt{\zeta} (\kappa \mp
\sqrt \zeta)^2 + {\cal O}[(\kappa \mp \sqrt \zeta)^3]
\end{equation}
Substituting the Taylor expansion of $\psi$ into~\eqref{uni_ansatz},
we obtain (cf.~previous stationary phase computations)
\begin{equation}
\label{uni2}
\varphi(x,t)
\sim \frac{1}{\sqrt \pi} \Re \left[ \frac{\phi_0+
\sqrt \zeta \phi_1}
{\zeta^{1/4}} e^{i(A - 2 \zeta^{3/2}/3)+ i \pi/4} + \frac{\phi_0-
\sqrt \zeta \phi_1}
{\zeta^{1/4}} e^{i(A + 2 \zeta^{3/2}/3)- i \pi/4}\right].
\end{equation}
Since the solution~\eqref{uni_ansatz} should be asymptotically valid
in all of region II, we impose that \eqref{uni2} identify
with~\eqref{sol3} for $x \gg s_{\rm min} t$. The identification
imposes
\begin{equation}
\begin{split}
&A = \frac12\Big[ (k_1+k_2)x -
(\omega_0(k_1,u_-)+\omega_0(k_2,u_-))t\Big], \quad
\zeta^{3/2} = -\frac34\Big[ (k_1-k_2)x -
(\omega_0(k_1,u_-)-\omega_0(k_2,u_-))t \Big] ,\\
&\phi_0 = \frac{ \zeta^{1/4} \hat \varphi_0(k_1)}{\sqrt{-2
\partial_{kk} \omega_0(k_1,u_-) t}} + \frac{ \zeta^{1/4} \hat
\varphi_0(k_2)}{\sqrt{ 2\partial_{kk} \omega_0(k_2,u_-) t}},\quad
\phi_1 = \frac{ \zeta^{-1/4} \hat \varphi_0(k_1)}{\sqrt{
-2\partial_{kk} \omega_0(k_1,u_-) t}} -\frac{ \zeta^{-1/4} \hat
\varphi_0(k_2)}{\sqrt{ 2\partial_{kk} \omega_0(k_2,u_-) t}}.
\end{split}
\end{equation}
Then, direct integration of~\eqref{uni_ansatz} yields:
\begin{equation}
\label{sol4}
\varphi(x,t) = \varphi_{\rm uni}(x,t) = 2 \Re \left[ e^{i A} (\phi_0
{\rm Ai}(-\zeta) - i \phi_1 {\rm Ai}'(-\zeta) ) \right],
\end{equation}
where ${\rm Ai}(y)$ is given by~\eqref{ai_def} and ${\rm Ai}'(y)
\equiv d{\rm Ai}/dy$.  By construction, \eqref{sol4} is asymptotic
to~\eqref{sol3} when $x \gg s_{\rm min} t$, i.e.~$\zeta \gg
1$. Additionally, $\phi_0 \sim 2\alpha \hat \varphi_0(\sqrt 3)$,
$\phi_1 \sim -\alpha^2 \hat \varphi_0(\sqrt 3)/\sqrt 3$, and $\zeta =
-\alpha(x-s_{\rm min} t)$, such that the uniform
approximation~\eqref{sol4} asymptotically matches with the Airy
approximation~\eqref{airy} when $x \sim s_{\rm min} t$ (recall that
$\alpha \ll 1$ in the long time regime).  Note that the uniform
approximation~\eqref{sol4} is also valid in the region $x< s_{\rm min}
t$, or equivalently $\zeta < 0$, where the analytical
expressions~\eqref{k1} and~\eqref{k2} are complex.

\begin{figure}[h]
  \centering
  \includegraphics{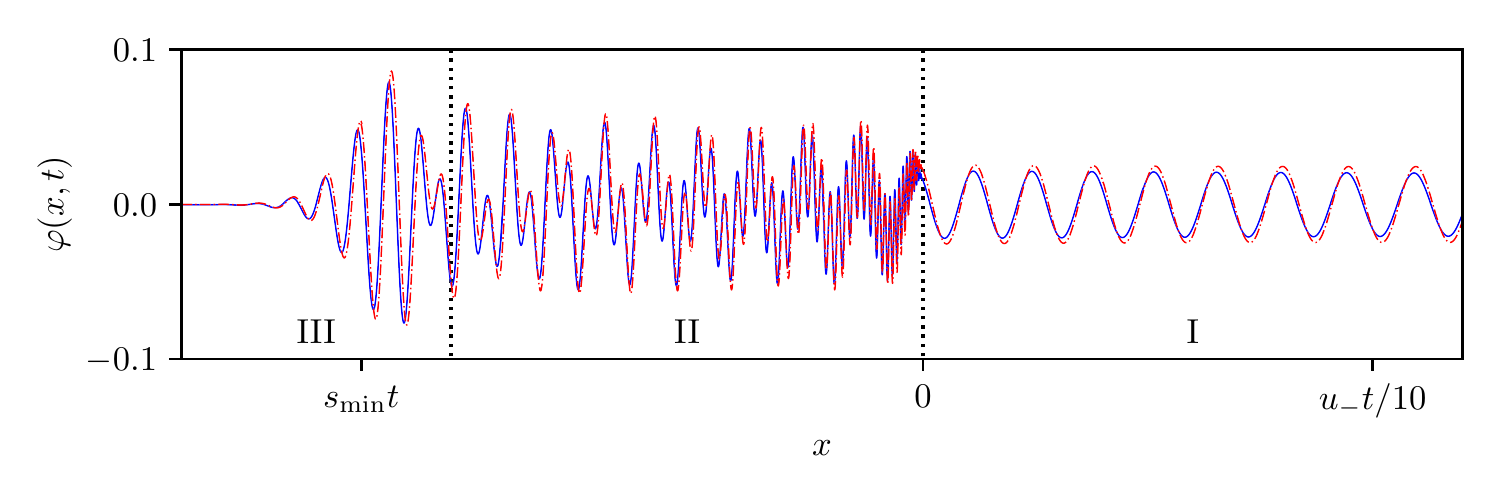}
  \caption{Comparison between the asymptotic stationary phase
    solution~\eqref{sol_lin} (red dash-dotted line) with the solution
    of the dispersive Riemann problem obtained numerically at $t=500$
    for $(\xi,u_-,u_+)=(0.1,1,1.5)$ (blue solid line).}
  \label{spm}
\end{figure}
\begin{figure}[h]
  \centering
  \includegraphics{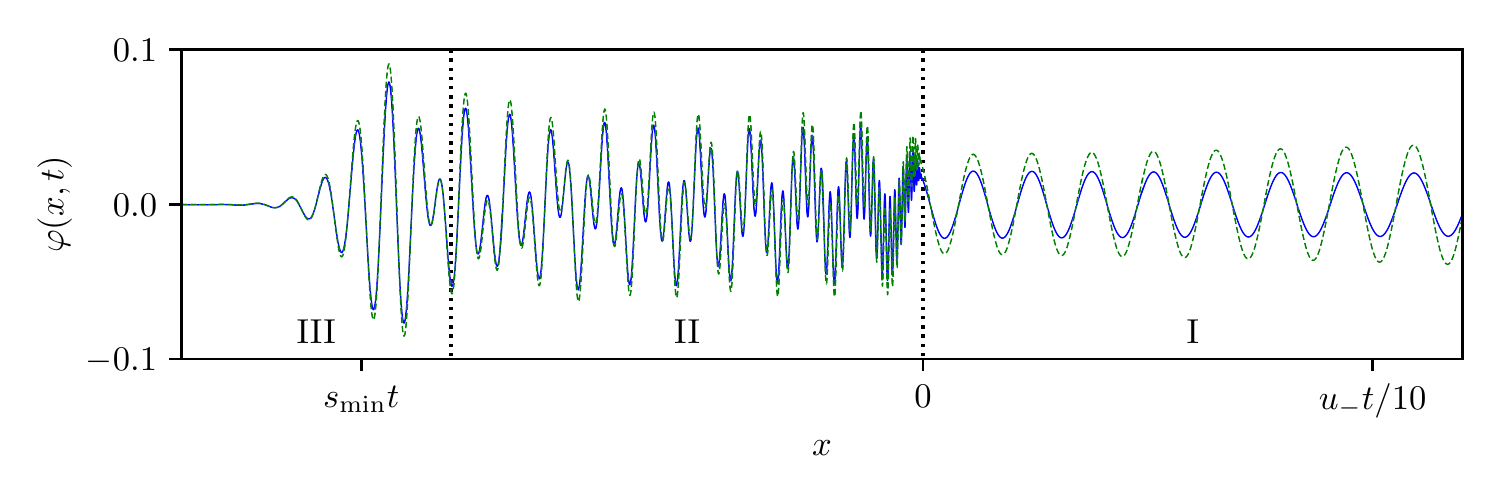}
  \caption{Comparison between the solutions $\varphi= u(x,t)-u_-$ of
    the dispersive Riemann problems $(\xi,u_-,u_+)=(0.1,1,1.5)$ (blue
    solid line) and $(\xi,u_-,u_+)=(0.1,1,0.5)$ (green dashed line) at
    $t=500$. Note that, according to~\eqref{phi0}, the linear wave
    depends on the sign of $u_+-u_-$ and the green line represents the
    field $-\varphi(x,t)$ with $u_->u_+$ in order to be compared to
    the solution $\varphi(x,t)$ with $u_-<u_+$.}
  \label{spm2}
\end{figure}

To summarize, the linear wave generated by the dispersive Riemann problem is
approximated by the piecewise-defined function
\begin{equation}
\label{sol_lin}
\varphi(x,t) \sim
\begin{cases}
\varphi_{\rm uni}(x,t), & x \leq 0,\\
\varphi_{\rm I}(x,t), & 0 < x,
\end{cases}, \quad t \gg 1, \quad \frac{x}{t} = \mathcal{O}(1),
\end{equation}
where $\varphi_{\rm I}$ and $\varphi_{\rm uni}$ are defined in
\eqref{sol1} and~\eqref{sol4}, respectively. This stationary phase
approximation compares well with the numerical solution, as displayed
in Fig.~\ref{spm}.

An important consequence of the approximate initial data \eqref{phi0}
for $\varphi$ is that the linear wave solution $\varphi(x,t)$ is valid
for both polarities of the initial jump $u_-<u_+$ and $u_->u_+$, i.e.,
for either the generation of waves accompanying a RW or the generation
of waves accompanying a DSW, respectively. Moreover, the linear waves
accompanying a RW, $\varphi_{\rm RW}(x,t)$, with jump $u_+ - u_- > 0$
are related to the linear waves accompanying a DSW with the same jump
magnitude $u_- - u_+ > 0$ by a minus sign: $\varphi_{\rm DSW}(x,t) = -
\varphi_{\rm RW}(x,t)$. Figure \ref{spm2} displays the numerical
solutions for two different initial conditions $u_-<u_+$ and $u_->u_+$
with a common difference magnitude $|u_+-u_-|$.  Although the two
numerical solutions are comparable, the discrepancy between them
increases with increasing $x$; the discrepancy is larger in
Region~I. In fact, the small amplitude assumption at the core of
linear theory is not well-fulfilled since, initially, $|\varphi(x,0)|
= {\cal O}(|u_--u_+|)$.  It is, therefore, remarkable that linear
theory still predicts the evolution of the fast oscillations with
excellent accuracy, even if the small amplitude assumption is
initially violated. The discrepancy in Fig.~\ref{spm2} is due to
different matching conditions for the linear wave with the leftmost,
trailing edge of the RW or DSW. This asymptotic matching has been
investigated for the KdV equation
in~\cite{leach_large-time_2008,grava_numerical_2008}.

Finally, we note that the asymptotic behavior \eqref{airy} near the
zero dispersion point is general and can be obtained for other wave
equations exhibiting a non-convex linear dispersion relation. For
example, it was obtained (with appropriate modifications due to a
different form of the dispersion relation) in
\cite{whitfield_wave-packet_2015} for linear wavepackets in the
Gardner-Ostrovsky equation.

\section{Shocks and rarefactions}
\label{sec:shocks-rarefactions}

In this section, we consider shock wave and rarefaction wave solutions
of the underlying conservation law $u_t+uu_x=0$ and their role in the
structure of solutions of the dispersive Riemann problem for the BBM
equation.  Stationary shocks play a significant part in this
description (regions $e$ and $h$ in Figs.~\ref{fig:xi10} and
\ref{fig:xi01}, respectively), since they are also weak solutions of
the BBM equation. Of particular interest are stationary shocks that
are either {\em compressive} (satisfying the Lax entropy condition) or
{\em expansive} (violating the Lax condition). Stationary compressive
shocks are a major feature of our analysis of dispersive Riemann
problem solutions for data $(u_-,u_+)$ in regions d, e of
Fig.~\ref{fig:xi10} and g, h of Fig.~\ref{fig:xi01} while stationary
expansion shocks are important in regions a and b of
Fig.~\ref{fig:xi01}.  Without loss of generality, we suppose that
$|u_+|=1$; the solution with $|u_+| =b \neq 1$ is given by the change
of variable $(x,t,u) \to (x,t/b, b u)$.

\subsection{Stationary Lax shocks}
\label{sec:stat-lax-shocks}

We first consider the solution when $u_-=-u_+=1$. Both the Hopf $u_t +
uu_x = 0$ and the BBM equation \eqref{bbm} admit the stationary,
discontinuous, weak solution
\begin{equation}
  \label{eq:9}
  u(x,t) =
  \begin{cases}
    1 & x \le 0 \\
    -1 & x > 0
  \end{cases},
\end{equation}
or stationary Lax shock.  The shock wave is compressive in the sense
that the Hopf characteristics (with speed $u_{\pm} = \mp 1$) on each
side of the shock located at $x=0$ propagate into the shock \cite{lax_hyperbolic_1973}.  Although
this BBM solution has been acknowledged elsewhere,
e.g.~\cite{el_expansion_2016}, it has remained mostly a curiosity.

\begin{figure}
  \centering
  \includegraphics{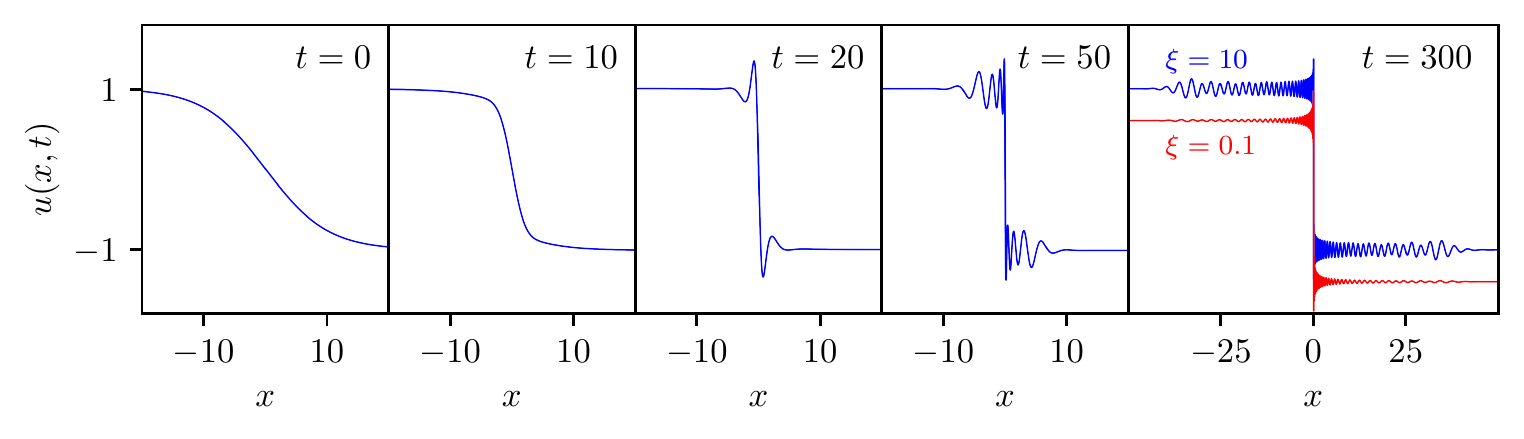}
  \caption{Numerical solution of the dispersive Riemann problem with
    $u_-=-u_+=1$ and $\xi=10$ at times shown. The last plot displays
    the numerical solution starting with a smoothed, steep step where
    $\xi=0.1$; a different color and an offset have been added to
    distinguish the variations of the two solutions.}
  \label{spm_jump1}
\end{figure}

We investigate the dynamics when \eqref{eq:9} is smoothed at $t = 0$
according to \eqref{step} with $u_\pm = \mp 1$.  The numerical
solution $u(x,t)$ is displayed in Fig.~\ref{spm_jump1} at different
times.
Evolution initially leads to the usual nonlinear self steepening ($t =
10$) due to broad initial data ($\xi = 10$).  This long wave, Hopf
equation-like evolution, is soon accompanied by oscillations ($t =
20$) when the solution is steep enough and the BBM dispersive term
$u_{xxt}$ is important.  Dispersive regularization of shock
formation typically results in a DSW (cf.~Sec.~\ref{DSW}), with an
expanding collection of oscillations.  The long-time dynamics of a
generic DSW consists of finite amplitude waves that are bounded in
wavenumber \cite{el_dispersive_2016}.  But here, oscillations of
increasing, apparently unbounded wavenumber develop as time increases,
concentrating near the origin.  The last panel of Fig.~\ref{spm_jump1}
($t=300$) shows that the large $t$ behavior for both broad and narrow
initial smoothed step profiles approach approximately the same
structure, suggesting the existence of some self-similar asymptotic
configuration.  We remark that a previous numerical study of BBM has
depicted similar behavior for broad, periodic initial data
\cite{francius_wave_2001}, associating it with discontinuity
formation.

To further investigate these dynamics, we introduce the short space
$\rho = x/\epsilon$ and long time $T = \epsilon t$ scaling where $0 <
\epsilon \ll 1$ is a small parameter representing the characteristic
length scale of the oscillations in a neighborhood of the origin.
With this scaling, the BBM equation \eqref{bbm} becomes
\begin{equation}
  \label{eq:3}
  \epsilon u_T + \frac{1}{\epsilon} u u_\rho = \frac{1}{\epsilon}
  u_{\rho\rho T} .
\end{equation}
Expanding $u$ in $\epsilon$ as $u = U_0(\rho,T) + \epsilon U_1(\rho,T)
+ \cdots$, we obtain the leading order equation
\begin{equation}
  \label{eq:4}
  U_0 \partial_\rho U_0 = \partial_{\rho\rho T} U_0 .
\end{equation}
This equation can be integrated once in $\rho$, yielding a nonlinear
Klein-Gordon equation
\begin{equation}
  \label{eq:5}
  \frac{1}{2} U_0^2 - \alpha = \partial_{\rho T} U_0 .
\end{equation}
In general, $\alpha$ is an arbitrary function of $T$.  Motivated by
the numerical observations in Fig.~\ref{spm_jump1}, we seek a
self-similar solution that must be independent of the small but
otherwise arbitrary parameter $\epsilon$, thus can only depend on
$\rho T = xt$:
\begin{equation}
  \label{eq:6}
  U_0(\rho,T) = g(\eta), \quad \eta = \rho T = x t .
\end{equation}
In order for $g$ to satisfy a well-defined ordinary differential
equation (ODE)
\begin{equation}
  \label{eq:7}
  \left ( \eta g' \right )' + \alpha - \frac{1}{2} g^2 = 0 , \quad '
  \equiv \frac{d}{d\eta} ,
\end{equation}
we require $\alpha \in \mathbb{R}$ to be constant.  The boundary
condition $u(x,t) \to -1$ as $x \to \infty$ implies $g(\eta) \to -1$
as $\eta \to \infty$. Linearizing equation \eqref{eq:7} about the
boundary condition $g(\eta) = -1 + h(\eta)$ where $|h| \ll 1$, we
obtain the equation
\begin{equation}
  \label{eq:11}
  \left ( \eta h' \right )' + \alpha - \frac{1}{2} + h = 0 .
\end{equation}
If the integration constant is set to $\alpha = 1/2$, then $h(\eta) =
0$ is a stable fixed point of the homogeneous equation \eqref{eq:11}
with the general, Bessel-type solution $h(\eta) = C_1
J_0(2\sqrt{\eta}) + C_2 Y_0(2\sqrt{\eta})$, for some $C_1,C_2 \in
\mathbb{R}$.  Consequently, the large $\eta$ asymptotics of the Bessel
solution demonstrate that the sought solution to eq.~\eqref{eq:7} with
$\alpha = 1/2$ exhibits algebraic, oscillatory decay to the requisite
boundary condition
\begin{equation}
  \label{eq:13}
  g(\eta) \sim -1 + \frac{C}{\eta^{1/4}} \cos \left ( 2\sqrt{\eta} -
    \phi \right ), \quad
  \eta \to \infty ,
\end{equation}
for some constants $C$ and $\phi$.

\begin{figure}
  \centering
  \includegraphics[scale=0.333333]{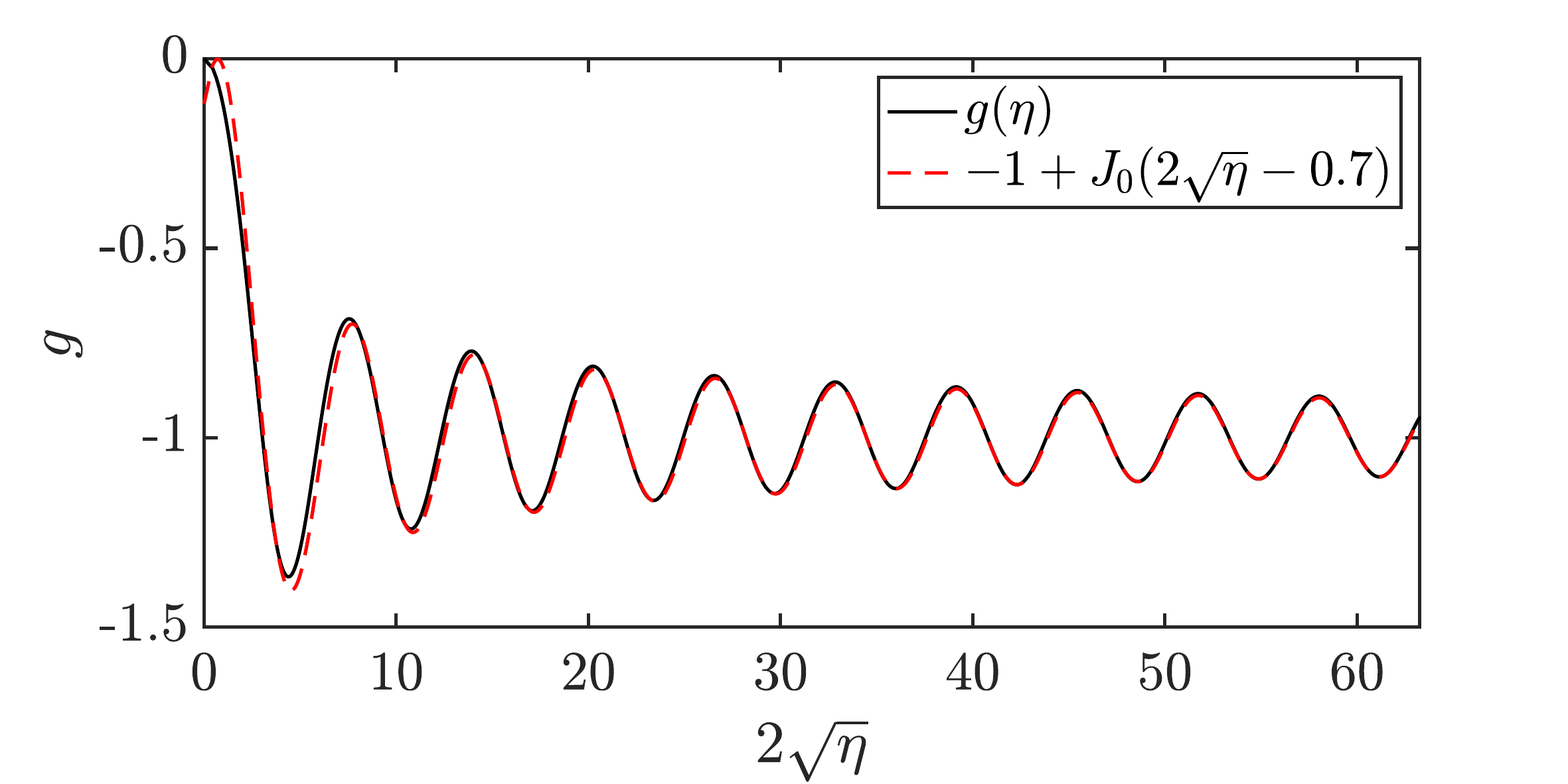}
  \caption{Self-similar dispersive Lax shock profile (solid, black) and
    a Bessel function approximation (dashed, red).}
  \label{fig:self_similar_bessel}
\end{figure}

Since the symmetric, smoothed step initial data considered here is an
odd function of $x$, we seek a solution to \eqref{eq:7} that is an odd
function of its argument so that $g(0) = 0$.  Evaluating
eq.~\eqref{eq:7} at $\eta = 0$ also determines $g'(0) = -1/2$.  The
initial conditions
\begin{equation}
  \label{eq:8}
  g(0) = 0, \quad g'(0) = - \frac{1}{2} ,
\end{equation}
uniquely determine the solution of eq.~\eqref{eq:7} with $\alpha =
1/2$ that decays to $-1$ according to \eqref{eq:13}.  We compute the
numerical solution to this initial value problem and display the
result in Fig.~\ref{fig:self_similar_bessel}.  Motivated by the
linearized equation \eqref{eq:11}, we also plot in
\ref{fig:self_similar_bessel} the function $-1 +
J_0(2\sqrt{\eta}-0.7)$, which provides a good empirical fit to
$g(\eta)$ when $2\sqrt{\eta} \gtrsim 8$.  By the asymptotics of the
Bessel function $J_0$, we can infer from
Fig.~\ref{fig:self_similar_bessel} that $C \approx \pi^{-1/2}$ and
$\phi \approx \pi/4+0.7$ in eq.~\eqref{eq:13}.

\begin{figure}
\centering
\begin{subfigure}[t]{.5\textwidth}
\centering
\includegraphics[scale=0.33333]{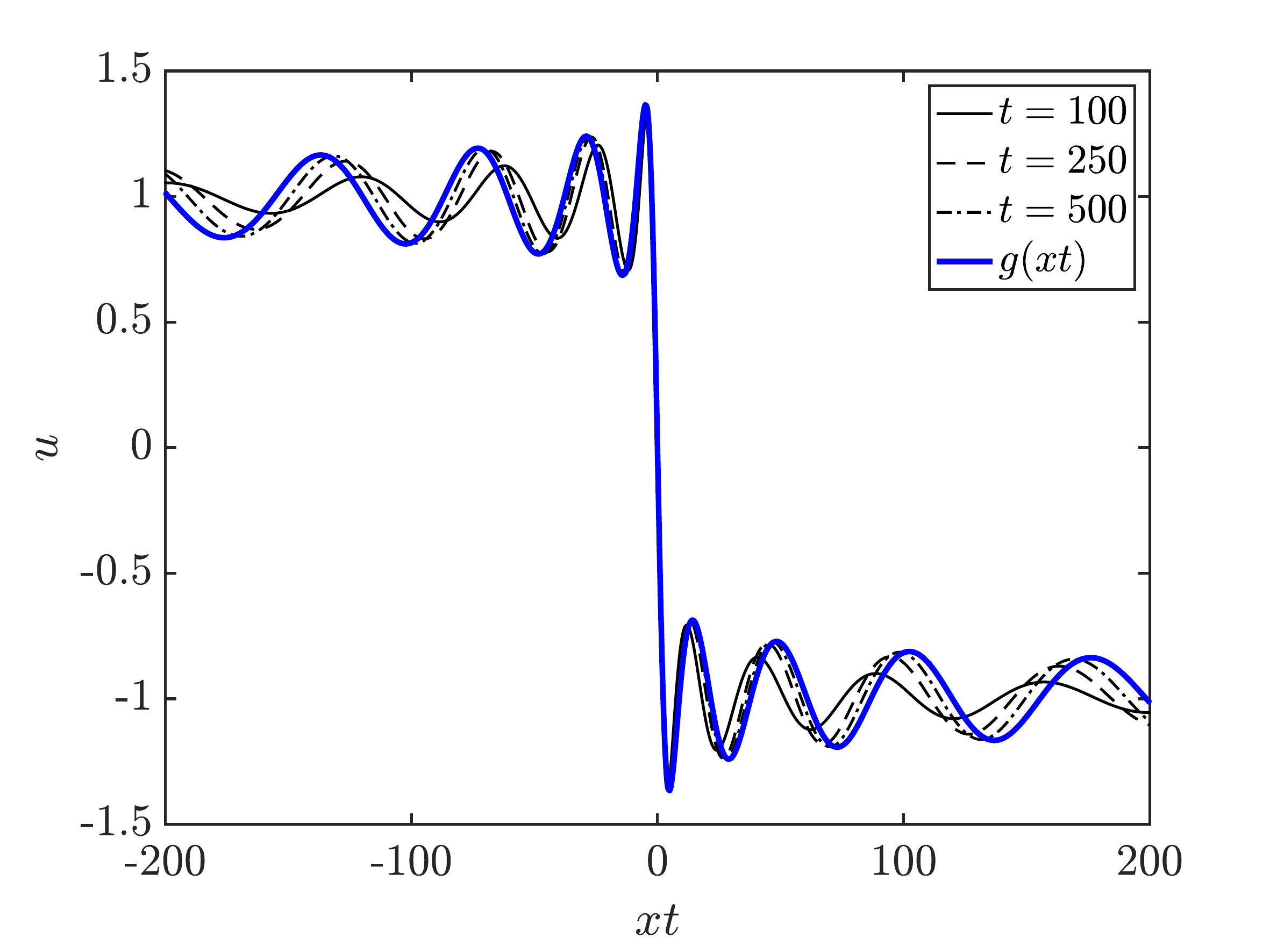}
\caption{$\xi = 0.1$}
\label{fig:laxa}
\end{subfigure}%
\begin{subfigure}[t]{.5\textwidth}
\centering
\includegraphics[scale=0.33333]{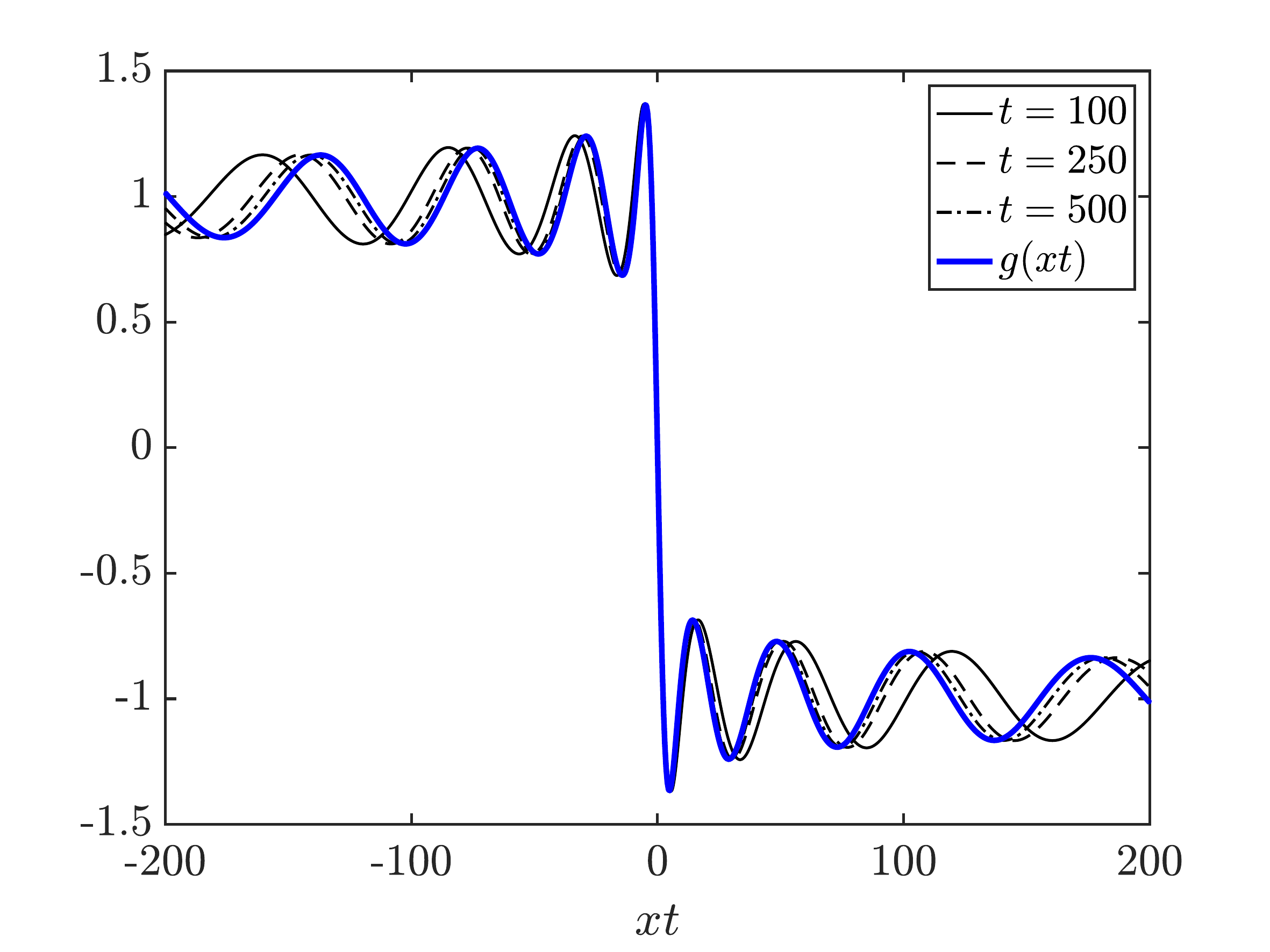}
\caption{$\xi = 10$}
\label{fig:laxb}
\end{subfigure}
\caption{Numerical solution for $u(x,t)$ at different times (solid,
  dashed, dash-dotted) tending toward the self-similar solution
  $g(xt)$ (thick, blue) for narrow (a) and broad (b) symmetric
  smoothed step initial data $u_\pm = \mp 1$.}
\label{lax}
\end{figure}

In Fig.~\ref{lax}, the numerical solution of the dispersive Riemann
problem for both narrow and broad initial data approaches the
self-similar solution $g(xt)$ (with an odd extension) for large $t$.
The narrow case in \ref{fig:laxa} leads to the development of
oscillations with wavelengths that are shorter than the self-similar
profile whereas the broad case in \ref{fig:laxb} exhibits longer
wavelength oscillations than $g(\eta)$.  In both cases, there is no
indication of discontinuity formation in finite time.

Since the odd-extended self-similar solution $g(\eta) \to \pm 1$ as
$\eta \to \mp \infty$, we observe that it converges to the stationary
Lax shock \eqref{eq:9} as $t \to \infty$ for each fixed $x$
\begin{equation}
  \label{eq:10}
  \lim_{t\to \infty} g(xt) =
  \begin{cases}
    1, & x < 0 \\
    0, & x = 0 \\
    -1, & x > 0
  \end{cases} ,
\end{equation}
although the convergence is not uniform in $x$.  The self-similar,
oscillatory profile $g(xt)$ describes how the stationary Lax shock
develops as $t \to \infty$. Based on these observations, we call the
profile $g(xt)$ a \textit{dispersive Lax shock}.  This asymptotic
solution represents a completely new type of dispersive shock
structure.

The simulations in Figs.~\ref{spm_jump1} and \ref{lax} exhibit
behavior reminiscent of Gibbs phenomenon in the theory of Fourier
series.  Perhaps a more apt comparison is to initial value problems
for linear dispersive PDE in which the data is discontinuous
\cite{biondini_gibbs_2017}.  For a broad class of constant
coefficient, linear dispersive PDE, the convergence of the solution as
$t \to 0^+$ for initial data with a discontinuity is not uniform.
Moreover, it exhibits Gibbs phenomenon whereby the solution has an
overshoot that, in proportionality to the jump, converges to the
Wilbraham-Gibbs constant
\begin{equation}
  \label{eq:12}
  \mathfrak{g} \equiv \frac{1}{\pi} \int_0^\pi \frac{\sin{z}}{z}dz -
  \frac{1}{2} \approx 0.08949 .
\end{equation}
In contrast, the dispersive Lax shock $g(xt)$ converges to the
discontinuity \eqref{eq:10} as $t \to \infty$.  While the oscillations
for $\eta$ sufficiently far from $0$ are linear and Bessel-like, the
compression of these oscillations is an inherently nonlinear process.
Narrower oscillations appear as the initial smoothed step profile
steepens and approaches a jump discontinuity.  In contrast to the
aforementioned linear initial value problems, we observe an
oscillatory overshoot that can be numerically estimated from the
self-similar profile $g(xt)$ to be, in proportion to the jump height,
\begin{equation}
  \label{eq:14}
  \max_{\eta \in \mathbb{R}} \frac{g(\eta)-1}{2} \approx 0.18342,
\end{equation}
a bit more than twice the Wilbraham-Gibbs constant $\mathfrak{g}$.

Numerical simulations are limited by the discretization and method
utilized.  The structure and self-similar scaling of $g(xt)$ will, for
large enough $t$, exceed the resolution of any fixed discretization
scheme.  Our analysis of this dispersive Riemann problem provides a
means to interpret solutions in the vicinity of dispersive Lax shocks.

We show in Sec.~\ref{implosion} that the solution of the dispersive
Riemann problem for slightly asymmetric boundary conditions $0< u_- +
u_+ \ll u_-$ is drastically different from the symmetric case
investigated here. In particular, we will show that solitary waves are
shedded in an incoherent fashion on top of a compressive shock
structure.

\subsection{Rarefaction waves}
\label{RW}

\begin{figure}
  \begin{subfigure}{0.49\textwidth}
    \centering
    \includegraphics{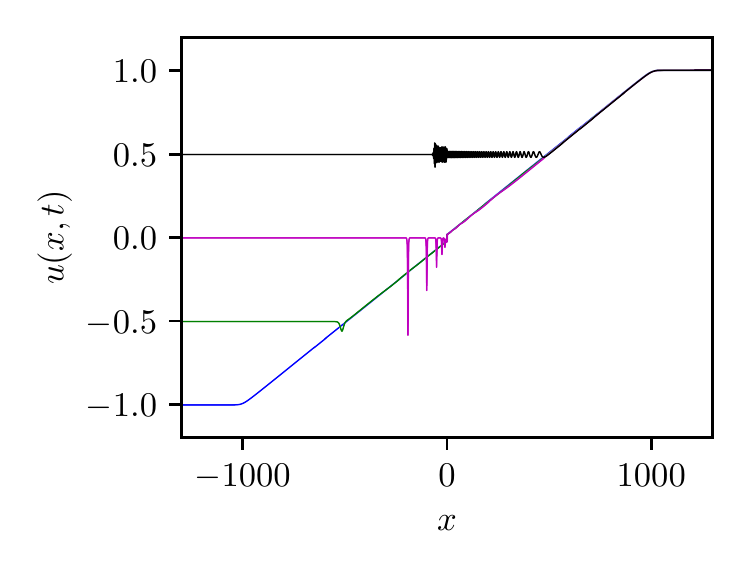}
    \caption{Narrow smoothed step initial data $\xi=0.1$. Numerical
      solutions at an earlier time are displayed in
      Figs.~\ref{minusexp2} and~\ref{RWsol}.  }
    \label{exp}
  \end{subfigure}
  \begin{subfigure}{0.49\textwidth}
    \centering
    \includegraphics{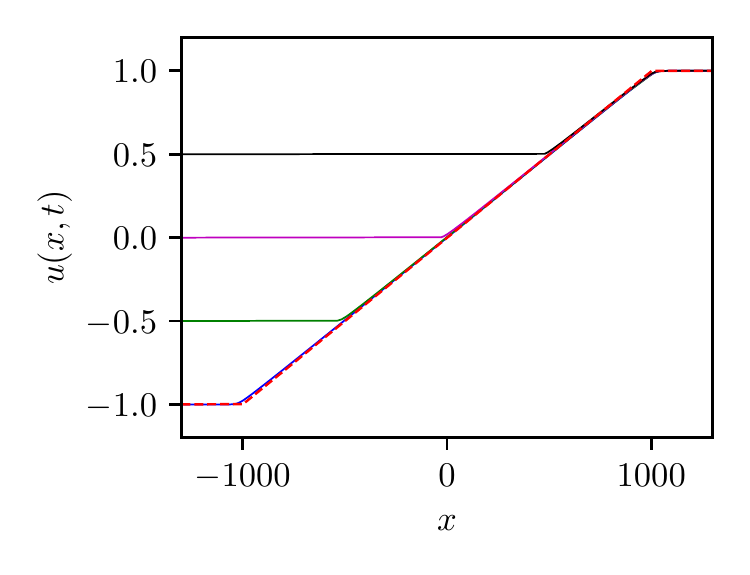}
    \caption{Broad smoothed step initial data $\xi=10$. The dashed red
      line corresponds to the analytical RW
      solution~\eqref{RW_solution} with $u_-=1$.}
    \label{RWb}
  \end{subfigure}\hfill
  \caption{Solutions of the dispersive Riemann problem
    \eqref{bbm},\eqref{step} with $u_-<1$ at $t=1000$.  Solid lines
    correspond to numerical solutions with $u_- \in
    \{-1,-0.5,0,0.5\}$.}
  \label{noRW}
\end{figure}

From now on, we consider the case $u_+=1$.  In Fig.~\ref{noRW}, we
show numerical solutions of \eqref{bbm}, \eqref{step} with
$u_-<1$. Dispersive effects are negligible for smoothed step initial
data ($\xi \gg 1$), and the asymptotic solution of the dispersive Riemann problem
is a smoothed modification of the RW solution
\begin{equation}
\label{RW_solution}
u(x,t) = u_{\rm RW}(x,t) =
\begin{cases}
u_-, &  x< u_- t,\\
x/t,  & u_- t \leq x < t,\\
1, & t \leq x,
\end{cases}
\end{equation}
of the dispersionless equation $u_t+uu_x=0.$ Fig.~\ref{RWb} displays
good agreement between the form~\eqref{RW_solution} and the smooth RW
numerical solution of equation \eqref{bbm}.

Dispersive effects are more significant for narrow smoothed step
initial data ($\xi \ll 1$), as demonstrated dramatically in
Fig.~\ref{exp} and later on in Figs.~\ref{exp2} and~\ref{RWsol}.  In
the case of symmetric initial conditions $u_-=-1$, the RW develops
with an expansion shock that is described in Sec.~\ref{expansion}
below. The evolution of asymmetric smoothed step initial data (having
$-1<u_- <1$) displays richer structure (cf.~Fig.~\ref{exp}). The
interplay between long and short waves in the linear regime (clearly
visible in Fig.~\ref{exp} with $u_-=1/2)$ is analysed in
Sec.~\ref{linear_wave}, and the emergence of a train of solitary waves
is investigated in Sec.~\ref{shedding}.  We show in
Sec.~\ref{early_time} that the development of expansion shock and
solitary wavetrain structures for $\xi \to 0$ is linked to the
atypical early time evolution of the dispersive Riemann problem.

\subsection{Expansion shocks}
\label{expansion}

Consider the initial condition~\eqref{step} with $u_-=-1$, i.e.
\begin{equation}
\label{step_ex}
u_0(x) = \tanh \left( \frac{x}{\xi} \right),
\end{equation}
with $\xi \ll 1$.  In~\cite{el_expansion_2016}, an asymptotic solution
is derived in detail, using $\xi$ as a small parameter, with long-time
scaling $T= \xi t$. Here, we outline the description in
\cite{el_expansion_2016}.

The inner structure of the solution, near $x=0,$ is described using
the same short space scaling $\chi = x/\xi$ as that utilized for the
description of the dispersive Lax shock in
Sec.~\ref{sec:stat-lax-shocks}.  However, the small parameter $\xi$ in
this case is fixed by the choice of initial data \eqref{step}.  These scalings introduced into the
BBM equation \eqref{bbm} lead to
\begin{equation}
\xi u_T + \frac{1}{\xi} uu_\chi = \frac{1}{\xi} u_{\chi\chi T} .
\end{equation}
Expanding $u$ in $\xi$: $u=U_0(\chi,T) + \xi U_1(\chi,T) + {\cal
O}(\xi^2)$, we obtain the leading order equation
\begin{equation}
  \label{eq:15}
  U_0 \partial_\chi U_0 = \partial_{\chi\chi T} U_0.
\end{equation}
In addition to the self-similar solution to this equation that we
obtained for the dispersive Lax shock, here we obtain a separated
solution describing the expansion shock.  Equation \eqref{eq:15} with
initial condition $U_0(\chi,0)=\tanh \left( \chi \right)$
corresponding to \eqref{step_ex}, admits the separated solution
\begin{equation}
  \label{inner2}
  U_0(\chi,t) = \frac{1}{1+T/2} \tanh{\chi}.
\end{equation}
Thus the inner solution of the dispersive Riemann problem~\eqref{bbm},
\eqref{step_ex} for $\xi \ll 1$ is given by
\begin{equation}
\label{inner}
u(x,t) = \frac{1}{1+ \xi t/2} \tanh \left( \frac{x}{\xi}
\right) + {\cal O}(\xi),\quad |x| = \mathcal{O}(\xi), \quad t =
\mathcal{O}\left ( 
\frac{1}{\xi} \right ).
\end{equation}

The outer structure of the solution is described using the
long space scaling $X = \xi x,$ leading to
\begin{equation}
\xi u_T + \xi uu_X = \xi^3 u_{XXT}.
\end{equation}
Looking for $u$ in the form \ $\tilde U_0(X,T) + \xi \tilde U_1(X,T) +
{\cal O}(\xi^2)$, we obtain the Hopf equation at leading order
\begin{equation}
\partial_T \tilde U_0 + \tilde U_0  \partial_X \tilde U_0  = 0.
\end{equation}
The general solution of this equation is $\tilde U_0 = f(X - \tilde
U_0 T)$. The matching of $\tilde U_0$ with the inner
solution~\eqref{inner2}, $\lim_{|X| \to 0}\tilde U_0(X,T) =
\lim_{|\chi| \to \infty}U_0(\chi,T)$, yields
\begin{equation}
\tilde U_0 (X,T) = \frac{{\rm sgn}(X)+X/2}{1+T/2}.
\end{equation}
This formula is continuously matched with the far-field conditions:
$\lim_{x\to \pm \infty} u = \pm 1$, leading to the outer solution,
\begin{equation}
\label{outer}
u(x,t) =
\begin{cases}
\dfrac{{\rm sgn}(x)+ \xi x/2}{1+\xi t/2}, &|x|<t,\\[6pt]
{\rm sgn}(x) , &|x|> t,
\end{cases} + {\cal O}(\xi),
\quad |x| = \mathcal{O} \left ( \frac{1}{\xi} \right ),\quad t
= \mathcal{O}\left (  \frac{1}{\xi}  \right ).
\end{equation}

\begin{figure}[h]
\begin{subfigure}{0.49\textwidth}
\centering
\includegraphics{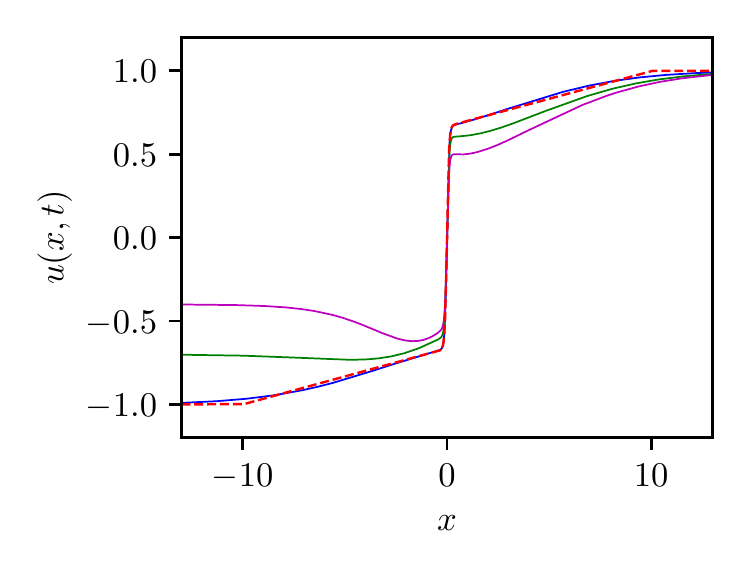}
\caption{$t=1/\xi=10$}
\end{subfigure}\hfill
\begin{subfigure}{0.49\textwidth}
\centering
\includegraphics{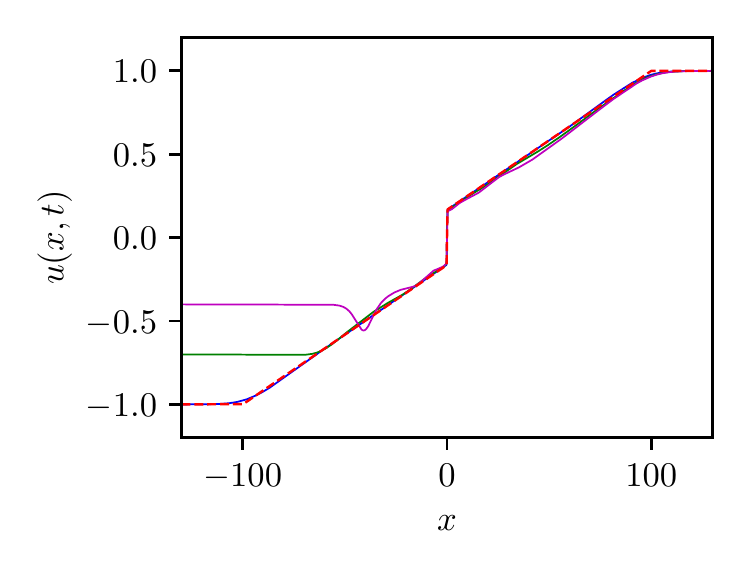}
\caption{$t=10/\xi=100$}
\end{subfigure}
\caption{Solutions of the dispersive Riemann problem \eqref{bbm}, \eqref{step}
with $u_-<0$ and $\xi=0.1$. The solid lines correspond to numerical
solutions with $u_- \in \{-1,-0.7,-0.4\}$. The red dashed line
corresponds to the expansion shock solution~$u_{\rm exp}$
\eqref{uniform_exp}; the truncated solution~\eqref{trunc} is given by
$u_{\rm exp}$ for $x\geq u_- t$ and $u_-$ for $x<u_-t $.  }
\label{exp2}
\end{figure}

Combining the descriptions~\eqref{inner} and~\eqref{outer}, we obtain
the uniformly valid asymptotic solution of~\eqref{bbm},
\eqref{step_ex} for $\xi \ll 1$
\begin{equation}
\label{uniform_exp}
u(x,t) = u_{\rm exp}(x,t) = \frac{1}{1+\xi t/2} \big[ \tanh (x/\xi)
- {\rm sgn}(x) \big] +
\begin{cases}
\dfrac{{\rm sgn}(x) + \xi x/2}{1+\xi t/2},  &|x| < t.\\[10pt]
{\rm sgn}(x), &|x| > t.
\end{cases}
\end{equation}
As remarked in~\cite{el_expansion_2016}, the accuracy of the
asymptotic solution~\eqref{uniform_exp} is excellent,
cf.~Fig.~\ref{exp2}. Note that the expansion shock structure converges
to the rarefaction solution~\eqref{RW_solution} as $t \to \infty$.

\begin{SCfigure}[][h]
\includegraphics{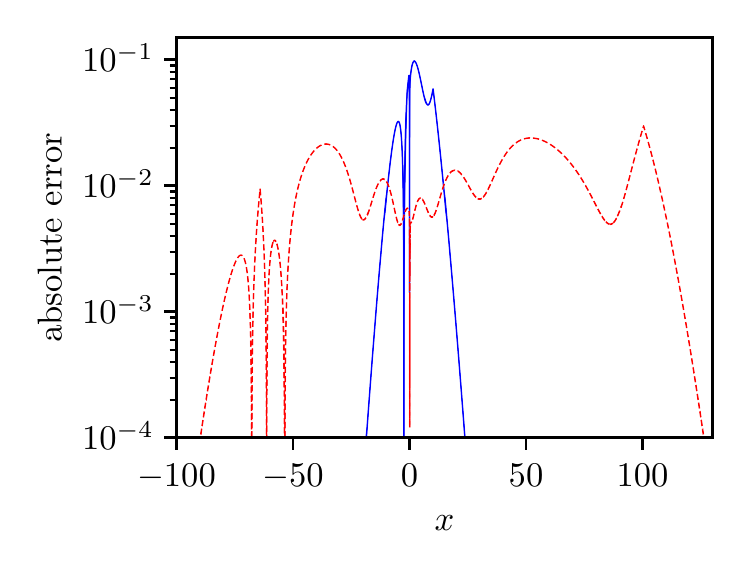}
\caption{Pointwise error between the truncated expansion
shock~\eqref{trunc} and numerical solution $\xi=0.1,\; u_-=-0.7$ at
$t=10$ (blue solid line) and $t=100$ (red dashed line).}
\label{minusexp2}
\end{SCfigure}

We also observe in Fig.~\ref{exp2} that the expansion shock
persists for slightly asymmetric boundary conditions with $-1<u_-<0$,
and $u_+=1,$ see for instance the solution for $u_-=-0.7$.  The outer
solution for such conditions is a slight modification of
\eqref{outer}
\begin{equation}
\label{outer2}
u(x,t) =
\begin{cases}
u_-, &x< u_- t+ 2\frac{u_-+1}{\xi},\\[6pt]
\dfrac{{\rm sgn}(x)+ \xi x/2}{1+\xi t/2}, &
u_- t+ 2\frac{u_-+1}{\xi} \leq  x <t,\\[6pt]
1 , & t \leq x ,
\end{cases}
\quad |x| = \mathcal{O}\left ( \frac{1}{\xi} \right ),\quad t =
\mathcal{O} \left (\frac{1}{\xi} \right ),
\end{equation}
which is continuous at $x= u_- t$. Thus the truncated solution reads:
\begin{equation}
\label{trunc}
u(x,t) = u_{\rm aexp}(x,t) = \frac{1}{1+\xi t/2} \big[ \tanh (x/\xi)
- {\rm sgn}(x) \big] +
\begin{cases}
u_-, &x< u_- t+ 2\frac{u_-+1}{\xi},\\[6pt]
\dfrac{{\rm sgn}(x)+ \xi x/2}{1+\xi t/2}, & u_- t + 2\frac{u_-+1}{\xi}
\leq  x <t,\\[6pt]
1 , & t \leq  x.
\end{cases}
\end{equation}
Fig.~\ref{exp2} shows that the asymmetric solution is in very good
agreement with the numerical solution for values of $u_-$ sufficiently
close to $-1$.  The absolute error between the analytical and the
numerical solutions is represented in Fig.~\ref{minusexp2} for
$u_-=-0.7$. The largest error at $t=1/\xi=10$ is approximately $\xi$
and decreases with time as expected.  In the next section, we analyze
the generation of solitary waves on top of the expansion shock for
larger values of $u_-$.


\section{Solitary wave shedding}
\label{shedding}

In this section, we study the generation of solitary waves observed in
numerical solutions.  Throughout this section, we keep $u_+=1$ and
consider $-1< u_- <1,$ with $\xi \ll 1$ (regions a, b, c, and d in
Fig.~\ref{fig:xi01}).  Numerical results, described in
Sec.~\ref{numsols}, exhibit a train of solitary waves for a range of
values of $u_-$. However, for $x>0,$ the solutions have either a
rarefaction wave accompanied by a region I linear wave (described in
Sec.~\ref{linear_wave}), or a portion of the expansion shock solution
(described in Sec.~\ref{expansion}). There are no solitary waves
generated in the region $x > 0$.  We can therefore restrict attention to $x <
0$.

In Sec.~\ref{early_time}, we explain the appearance of solitary waves
by analyzing the early time behavior of the solutions. In
Sec.~\ref{threshold}, we use the energy equality satisfied by smooth
solutions, to establish the values of $u_-$ for which a solitary wave
is generated from those for which the solution has no solitary wave.

\subsection{Numerical Solutions} \label{numsols}

\begin{figure}[h]
  \begin{subfigure}{0.49\textwidth}
    \centering
    \includegraphics{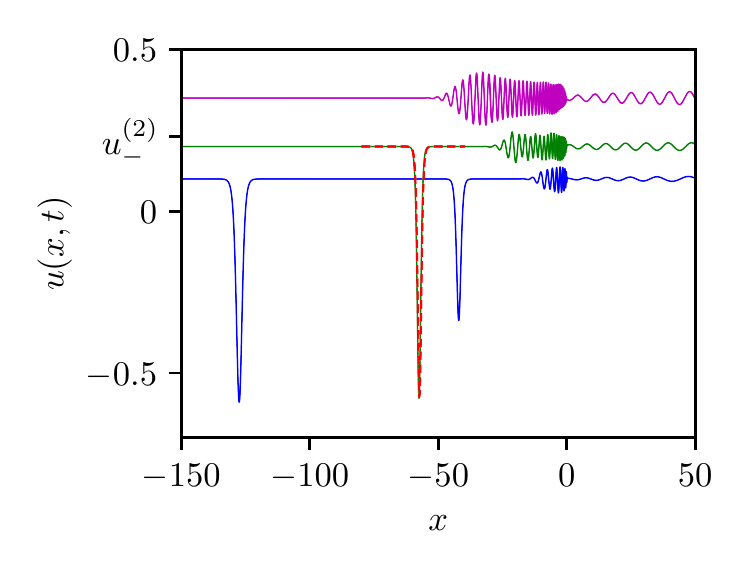}
    \caption{Positive left boundary conditions: $u_- \in
      \{0.1,0.2,0.35\}$.  The dashed red line represents the solitary
      wave solution~\eqref{sol_wave} with the parameter: $\ubar =
      u_-=0.2$ and $c = -5.9 \times 10^{-2}$.}
\label{RWsol2}
\end{subfigure}\hfill
\begin{subfigure}{0.49\textwidth}
  \centering
  \includegraphics{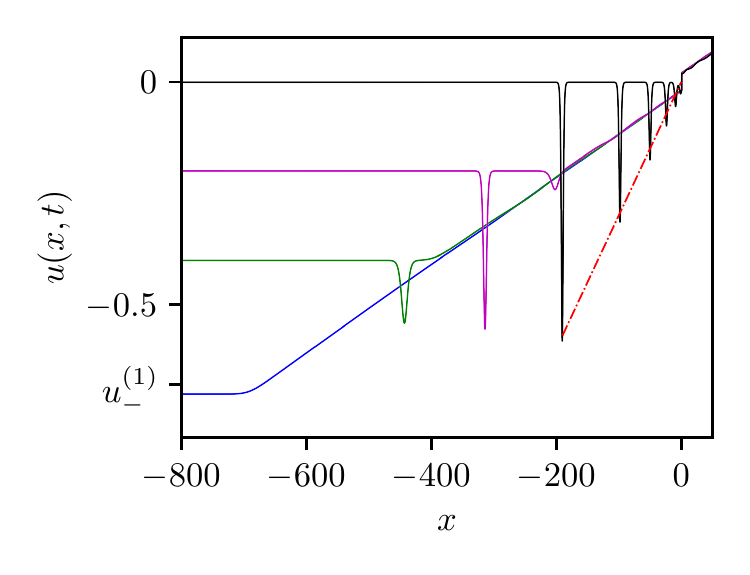}
  \caption{Negative left boundary conditions: $u_- \in
    \{-0.7,-0.4,-0.2,0\}$. The dash-dotted red line represents the
    amplitude variation given by the relation $c(a,u_-=0) = x/t$.}
  \label{RWsol3}
\end{subfigure}
\caption{Numerical solutions of the dispersive Riemann problem, with
  $\xi=0.1$ at $t=1000$.}
\label{RWsol}
\end{figure}

We start with a qualitative description of the numerical solutions
shown in Fig.~\ref{RWsol} for various  values of $u_-<1$ and
fixed $\xi =0.1$. As expected from Sec.~\ref{linear_wave}, a linear
wave develops to the left of the RW when $u_->0$.  Moreover, solitary
waves of negative polarization are additionally generated for
sufficiently small values $u_- <u_-^{(2)} \approx 0.23$ of~$u_-$ as
depicted in Fig.~\ref{RWsol2}. This threshold first appears at the
line $L_2$, $u_+ = \sigma_2(\xi) u_-$, of Fig.~\ref{fig:xi01}.  The
solitary wave solution of the BBM equation~\eqref{bbm} is given by
\cite{olver_euler_1979}
\begin{equation}
  \label{sol_wave}
  u(x,t) = \ubar + u_{\rm sol}(x,t;c,\ubar) = \ubar+3(c-\ubar){\rm
    sech}^2 \left( \sqrt{1-\frac{\ubar}{c}} \, \frac{x-ct} {2} \right),
\end{equation}
where $c$ is the wave speed and $\ubar$ is the background value.  The
wave amplitude is $a=3(c-\ubar)$ yielding the speed-amplitude relation
\begin{equation}
  \label{c}
  c = c(a,\ubar) = \ubar + a/3.
\end{equation}
Note that the solitary wave solution exists only for $c \in \mathbb{R}
\backslash [\min(\ubar,0),\max(\ubar,0)]$.  The example $\ubar =
u_-=0.2$ in Fig.~\ref{RWsol2} shows that the emitted solitary wave is
locally well described by the analytical profile~\eqref{sol_wave}.

\begin{SCfigure}[][h]
\centering
\includegraphics{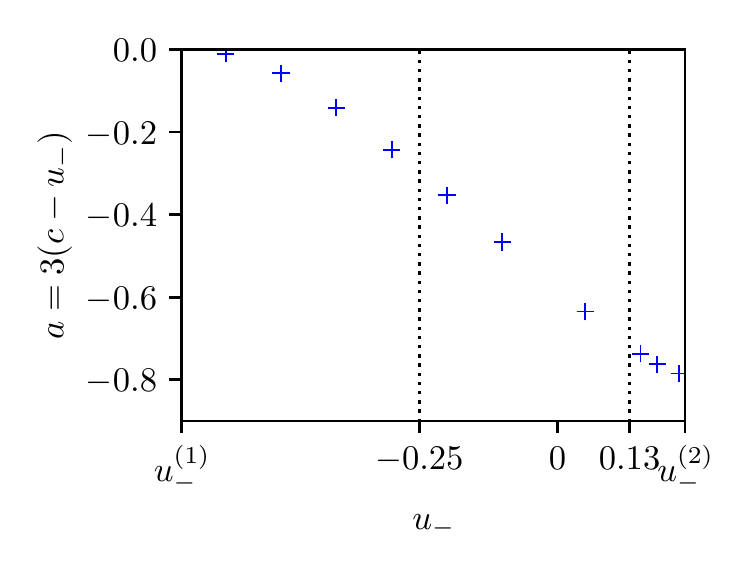}
\caption{Amplitude of the leftmost solitary wave for
$u_- \in [u_-^{(1)},u_-^{(2)}]$ with $u_-^{(1)}=-0.68$ and $u_-^{(2)}=0.23$.  The
dashed vertical lines delimit the region $u_- \in [-0.25,0.13]$ where
multiple solitary waves are generated.}
\label{soliton_amp}
\end{SCfigure}

If $\ubar=u_- = 0$, the linear dispersion relation~\eqref{dispersion}
is $\omega_0(k,\ubar)=0,$ and only a train of solitary waves is
generated for $x< 0$ as in Fig.~\ref{RWsol3}. Similarly, when $u_-<0$,
although a linear wave can propagate on a strictly negative
background, its group velocity satisfies $\partial_k \omega_0> u_-$ so
that no linear waves remain in the region $x<u_- t$ for $t \gg 1$. The
solitary wavetrain then propagates from the tail of the truncated
expansion shock solution~\eqref{trunc}, as depicted in
Fig.~\ref{RWsol3}. No solitary wave emission is observed when $u_- <
u_-^{(1)} \approx -0.68$.  The variation of the amplitude of the
leftmost solitary wave amplitude with $u_-\in [u_-^{(1)},u_-^{(2)}]$ is depicted in Fig.~\ref{soliton_amp}.

We empirically observe that when $u_- \in (u^{\rm (a)},0]$ and
multiple solitary waves are emitted, the amplitude $a(x,t)$ of the
solitary wavetrain asymptotically satisfies $c(a,u_-)=x/t$ where $c$
is the speed-amplitude relation~\eqref{c}.  A detailed study of the
solitary wavetrain using, for example, Whitham modulation
theory~\cite{whitham_linear_1999}, is left as a subject for future
work.

\subsection{Early time evolution}
\label{early_time}

The unusual generation of linear waves and solitary waves can be
qualitatively explained by investigating the early time dynamics of
the dispersive Riemann problem.  We consider in this section the small-time
expansion of the solution
\begin{equation}
\label{early}
u(x,t) = u_0(x) + t u_1(x) + {\cal O}(t^2),
\end{equation}
where $u_0(x)$ is given by~\eqref{step} with
$u_+=1$. Substituting~\eqref{early} into~\eqref{bbm}, we obtain at the
zeroth order in $t$
\begin{equation}
  \label{u1}
  u_1'' - u_1 = u_0u_0' ,
\end{equation}
where $' \equiv d/dx$.  The small parameter condition $t u_1(x) \ll
u_0(x)$ yields the boundary conditions: $\lim_{x\to \pm \infty}u_1(x)
= 0$.  Let $G(x)$ be the Green's function satisfying,
\begin{equation}
  G''(x)-G(x) = \delta(x),
\end{equation}
given by
\begin{equation}
  G(x) =  -\exp(-|x|)/2.
\end{equation}
The solution of~\eqref{u1} is then
\begin{equation}
  \label{phi12}
  u_1(x) = \int \limits_{-\infty}^{+\infty} G(x-y) \frac{d}{dy} \left(
    \frac{u_0^2(y)}{2} \right) dy = \frac14 \int
  \limits_{-\infty}^{+\infty} {\rm sgn}(x-y) e^{-|x-y|} u_0^2(y) dy.
\end{equation}
Note that $u_1(x)$ (and higher order terms) can be directly obtained
from the integro-differential form of the BBM equation~\eqref{bbm2} by
iteration, which is part of the contraction mapping argument used to
prove local existence of solutions to the BBM initial value problem
\cite{benjamin_model_1972}. The correction $u_1(x)$ is bounded by
\begin{equation}
  \label{u1max}
  |u_1(x)| \leq \frac14 \int \limits_{-\infty}^{+\infty} e^{-|y-x|} \left|
    \frac{d}{dy} \left(u_0^2(y) \right)\right| dy < \frac{1-u_-} {2 \xi},
\end{equation}
an estimate that decreases with increasing $\xi,$ so that $u_1$ is
negligible for large $\xi.$

\begin{figure}[h]
  \centering
  \begin{subfigure}[t]{.49\textwidth}
    \centering
    \includegraphics{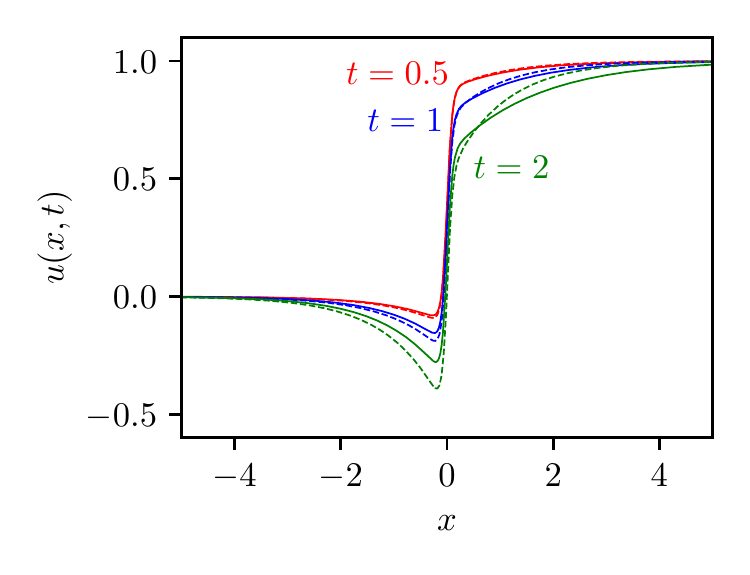}
    \caption{ $\xi=0.1$: The dashed lines correspond to the
      approximate solution $u_0(x)+ t u_1(x),$ with $u_1$ given by
      \eqref{solu1c}.}
    \label{fig:early01}
  \end{subfigure}\hfill
  \begin{subfigure}[t]{.49\textwidth}
    \centering
    \includegraphics{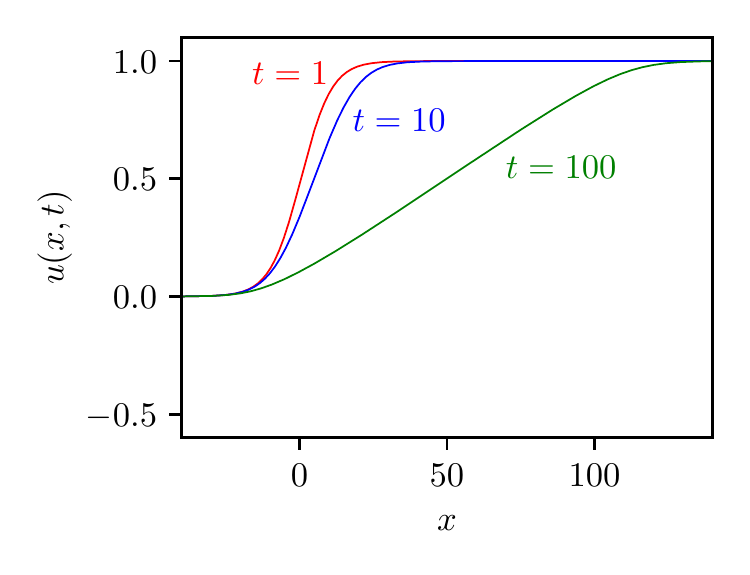}
    \caption{$\xi=10$: No dip develops for a slowly varying initial
      condition.}
    \label{fig:early10}
  \end{subfigure}
  \caption{Numerical solution (solid curves) of the smoothed step
    dispersive Riemann problem~\eqref{bbm}, \eqref{step} at different
    times.}
  \label{u1_delta}
\end{figure}

On the other hand, in the ideal case $\xi \to 0,$ the initial
condition is the step given by \eqref{step2}, and~\eqref{phi12}
simplifies to
\begin{equation}
  \label{solu1c}
  u_1(x) = \frac{u_-^2-1}{4} \, e^{-|x|} < 0,
\end{equation}
leading to a decrease in $u(x,t)$ for sufficiently small $t > 0$.
This introduces a dip in the graph of $u(x,t)$ for $x<0.$ The small
amplitude assumption $t u_1(x) \ll 1-u_-$ (recall that $1-u_-$
represents the initial smoothed step height) then reduces to $t \ll
1/(1+u_-)$.  Figure \ref{fig:early01} shows good agreement between the
approximate solution~\eqref{early} with \eqref{solu1c} and the
corresponding numerical solution of the dispersive Riemann problem
when $\xi=0.1.$ While the initial smoothed step is rarefying in the
region $x>0$, a dip develops immediately in the region $x<0$. As a
result, the initial smoothed step transition persists, contrasting
with numerical solutions for $\xi=10$ (Fig.~\ref{fig:early10}) that
more closely resemble the classical RW, with no dip developing.

\begin{figure}[h]
  \centering
  \includegraphics{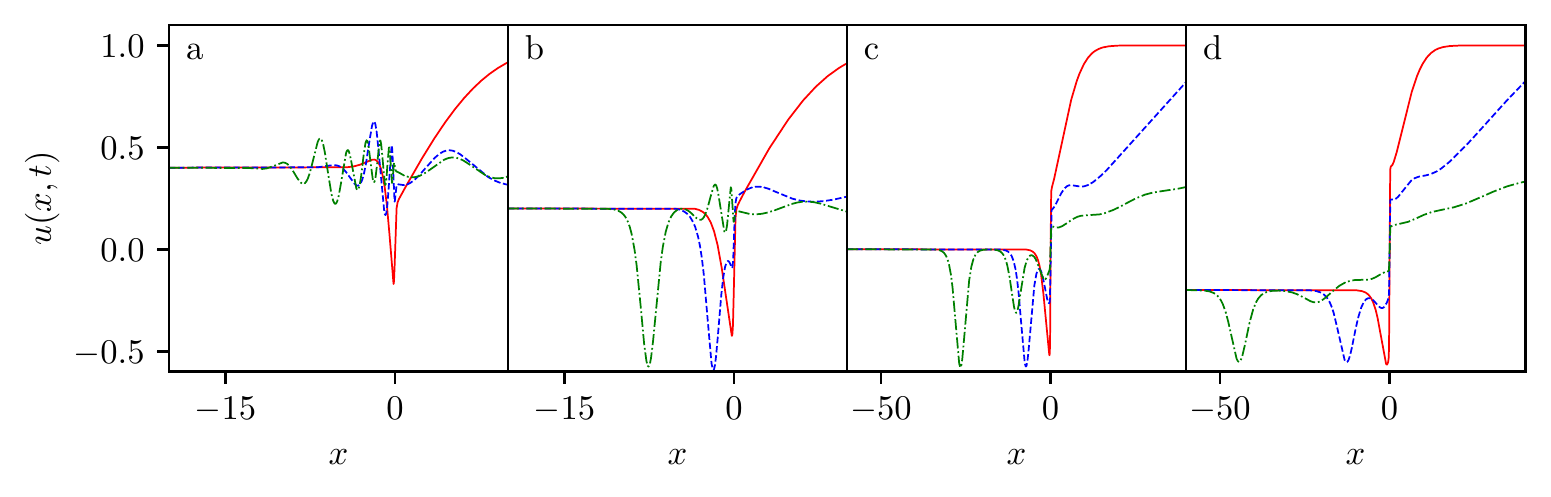}
  \caption{Initial evolution of the smoothed step dispersive Riemann
    problem~\eqref{bbm}, \eqref{step} for $\xi=0.1$. The solution is
    plotted for three different times: $t=10$ (red solid line), $t=50$
    (blue dashed line) and $t=150$ (green dash-dotted line). a)
    $u_-=0.4$, b) $u_-=0.2$, c) $u_-=0$, d) $u_-=-0.2$.}
  \label{decay}
\end{figure}

\begin{SCfigure}[][h]
  \centering
  \includegraphics{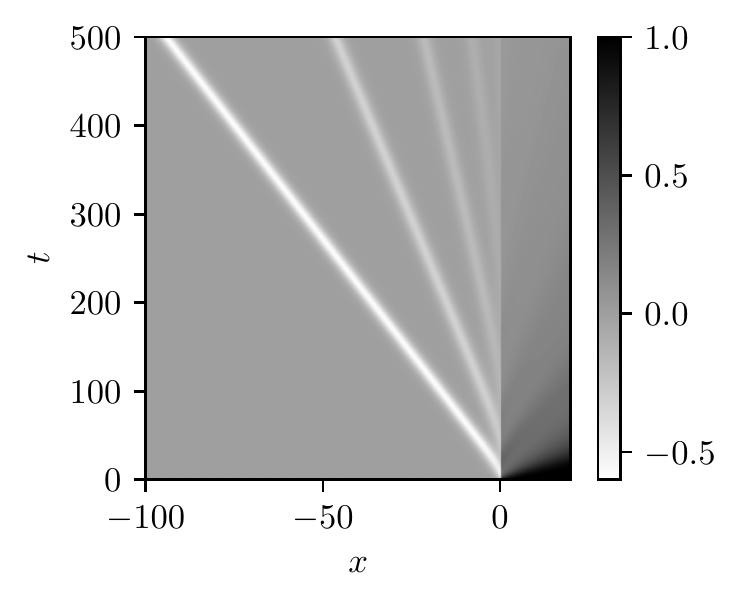}
  \caption{Spatio-temporal evolution of the solitary wavetrain
    generated for $\xi=0.1$ and $u_-=0$. The colors are inverted:
    light (dark) colors correspond to small (large) values of the
    solution $u(x,t)$. The smoothed step or dip in the vicinity of
    $x=0$ is radiating negative solitary waves with different speeds
    in the region $x<0$.}
\label{decay0b}
\end{SCfigure}

Figure \ref{decay} displays different evolution scenarios of smoothed
step initial data for $\xi=0.1$ depending on the value of $u_-$.  If
$u_-^{\rm (2)} <u_-$ (case a) the smoothed step emits a linear
wavepacket.  Additionally, solitary waves are generated if $0< u_-
\leq u_-^{\rm (2)} $ as depicted in case b.  If $u_-= 0$ (case c), the
smoothed step decays into a solitary wavetrain. The solitary wavetrain
behavior can be recognized by monitoring the region close to the
expansion shock in the long time regime, as in the contour plot of
Fig.~\ref{decay0b}.  The early time evolution changes for $u_-^{(1)} \leq u_-<0$ (case d) where the smoothed step also generates
solitary waves but fewer than the $u_- = 0$ case. For $u_- < 0$, the initial
smoothed step persists in the form of the approximate truncated
expansion shock solution~\eqref{trunc}.  In all of these cases, the
solution $u(x,t) $ converges to the rarefaction wave
solution~\eqref{RW_solution} as $t \to \infty,$ for each fixed~$x.$

\subsection{The threshold for generation of a negative solitary wave}
\label{threshold}

In this subsection, we derive the conditions for and properties of the
emitted solitary wave in the simplest case where only one solitary
wave is shedded, cf.~Fig.~\ref{soliton_amp}.  We also determine the
threshold $u_- = u_-^{\rm (2)} > 0$ for which solitary waves are
generated adjacent to the RW for any value of $\xi \ll 1$.  Up to now
the value of the parameter $u_-^{\rm (2)}$ has been obtained
numerically for the value $\xi=0.1$.

The derivation of the solitary wave's speed (or equivalently amplitude
using the speed-amplitude relation~\eqref{c}) is achieved using the
{\it mass} conservation law (the BBM equation itself)
\begin{equation}
\label{consN0}
u_t + \left(\tfrac12 u^2 - u_{xt} \right)_x = 0.
\end{equation}
and an additional {\it energy} conservation
law satisfied by smooth solutions \cite{olver_euler_1979}
\begin{equation}
\label{consE0}
\left(\tfrac12 u^2 + \tfrac12 u_x^2 \right)_t + \left(\tfrac13 u^3 -
u u_{xt} \right)_x = 0.
\end{equation}
Integration of~\eqref{consN0} and~\eqref{consE0} between the limits
$x=-L$ and $x=+L$ yields
\begin{align}
\label{consN}
&\frac{dN}{dt} + \left[ \tfrac12 u^2 -
u_{xt} \right]_{-L}^{+L} = 0,\\
\label{N}
&N(t) = \int \limits_{-L}^{+L}
u \, dx,\\
\label{consE}
&\frac{dE}{dt} + \left[ \tfrac13 u^3 -
u u_{xt} \right]_{-L}^{+L} = 0,\\
\label{E}
&E(t) = \int \limits_{-L}^{+L}
\frac{u^2+u_x^2}{2} dx.
\end{align}
Because of the term $u_x^2$, the initial energy $E(t=0)$, obtained by
substituting the initial smoothed step~\eqref{step} in~\eqref{E},
strongly depends on the transition width~$\xi,$ which plays a crucial
role in the appearance of negative solitary waves, as was demonstrated
numerically in Fig.~\ref{noRW}.

\begin{figure}[h]
\centering
\begin{subfigure}[t]{.49\textwidth}
\centering
\includegraphics{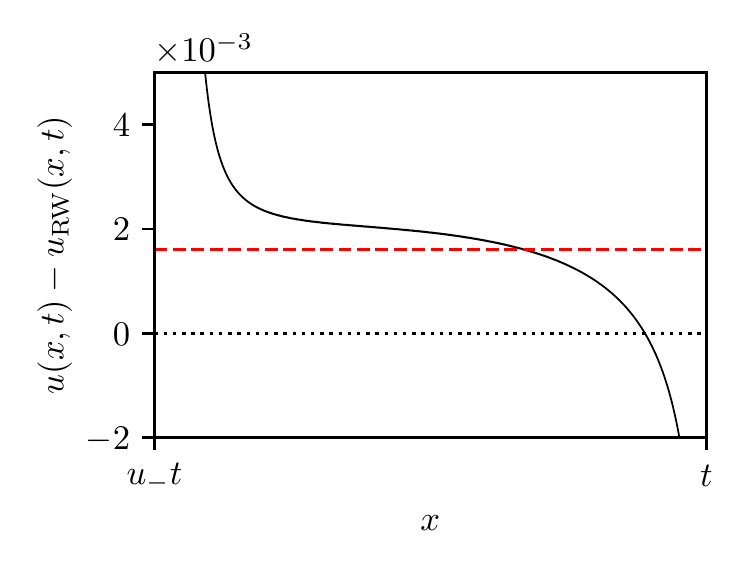}
\caption{RW solution $(\xi,u_-)=(0.1,0.18)$. The dashed red line
corresponds to the shift
$u_{\rm RW}(x-b,t) - u_{\rm RW}(x,t)= -b/t$; $b\approx -1.61$ is given by
the resolution of~\eqref{Esol1}.}
\label{mass}
\end{subfigure}\hfill
\begin{subfigure}[t]{.49\textwidth}
\centering
\includegraphics{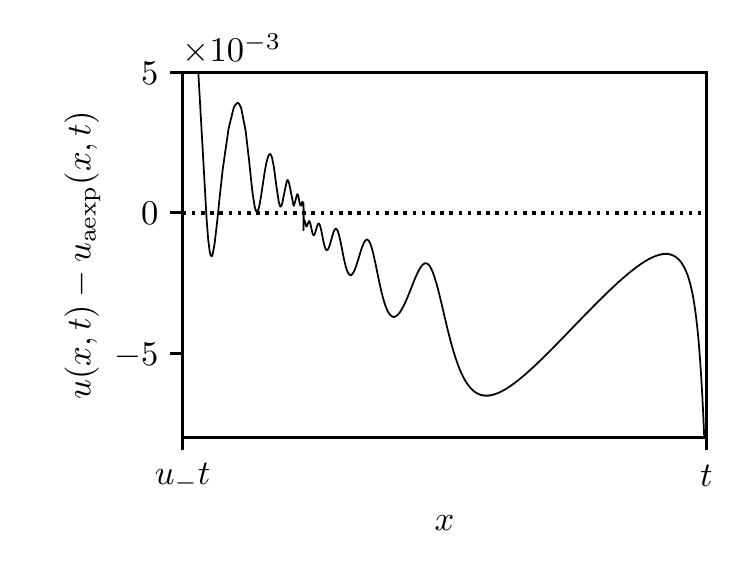}
\caption{Asymmetric expansion wave solution
$(\xi,u_-)=(0.1,-0.3)$.
}
\label{mass2}
\end{subfigure}%
\caption{Absolute error of the hydrodynamic state at $t=1000$ when a
solitary wave is generated.}
\end{figure}

Motivated by observations in Sec.~\ref{numsols}, we suppose that when
$u_-$ is very close to $u_-^{\rm (2)}$, the solution $u(x,t)$ can be
approximated for large $t$ by
\begin{equation}
\label{ansatz1}
u(x,t) =  u_{\rm RW}(x-b,t) + u_{\rm sol}(x,t;c,u_-) + \varphi(x,t),\qquad
c<s_{\rm min} = -u_-/8<0,\qquad t \gg 1,
\end{equation}
where $b$ and $c$ remain to be determined.  Expressions for $u_{\rm
  RW}$ and $u_{\rm sol}$ are given in \eqref{RW_solution} and
\eqref{sol_wave}, respectively, and $\varphi(x,t)$ is a correction
term.  We suppose here that the RW is not centered and denote $b$ the
value of the shift. The incorporation of the RW shift is crucial for
the precise determination of the threshold for solitary wave emission
and the accompanying solitary wave's speed $c$.  Figure \ref{mass}
shows the necessity of incorporating a vertical shift $-b/t$ of the
centered RW solution $u_{\rm RW}(x,t)$ in order to achieve good
agreement with the numerical simulation.  We also hypothesize that the
dominant contribution to the correction $\varphi(x,t)$ is a linear
wavetrain, analogous to that studied in Sec.~\ref{linear_wave}.  Note
that the solitary wave and the linear wavetrain correction propagate
on the background $\ubar = u_-$, fixed by the boundary
condition. Thus, the condition $c<s_{\rm min}$ ensures that the
solitary wave is negative and that it is well-separated from the RW
and the linear wavetrain for $t\gg 1$.  This assumption is consistent
with numerical computations.  We now show that the formula
\eqref{ansatz1} represents a solution of the dispersive Riemann
problem~\eqref{bbm},\eqref{step} for a certain value of the solitary
wave speed $c$ depending on $\xi \ll 1$ and $u_-$.

In a first approximation, we suppose that the correction
$\varphi(x,t)$ has a negligible mass and energy contribution:
\begin{equation}
\label{ansatz1b}
u(x,t) \sim u_{\rm RW}(x-b,t) + u_{\rm sol}(x,t;c,u_-) ,
\qquad t \gg 1.
\end{equation}
We evaluate the mass $N(t)$ and the energy $E(t)$ in two different
ways: i) by direct substitution of~\eqref{ansatz1b} in~\eqref{N}
and~\eqref{E} and ii) by solving~\eqref{consN} and ~\eqref{consE} with
the initial condition~\eqref{step} and the boundary conditions
$\lim_{x \to \pm \infty}u(x,t) = u_\pm$,
$\lim_{x \to \pm \infty}u_{xt}(x,t) = 0$. Equating the two expressions
for $N(t)$ and $E(t)$ then yields an approximation to the solitary
wave speed $c$ and the RW shift $b$.

(i) \
The direct substitution of the ansatz~\eqref{ansatz1b}
into~\eqref{N} yields
\begin{equation}
\label{N2}
N =N_{\rm RW} + N_{\rm sol},\quad
N_{\rm RW} = \int \limits_{-L}^{+L} u_{\rm RW} dx,\quad
N_{\rm sol} = \int \limits_{-L}^{+L} (u_{\rm sol}-u_-) dx,
\end{equation}
and the direct substitution of the ansatz~\eqref{ansatz1b}
into~\eqref{E} yields
\begin{equation}
\label{E2}
E =E_{\rm RW} + E_{\rm sol},\quad
E_{\rm RW} = \int \limits_{-L}^{+L} \frac{u_{\rm RW}^2+(\partial_x
u_{\rm RW})^2}{2} dx,\quad
E_{\rm sol} = \int \limits_{-L}^{+L} \frac{u_{\rm sol}^2+2 u_- u_{\rm
sol} + (\partial_x u_{\rm sol})^2}{2} dx.
\end{equation}
We choose $L$ and $t$ such that: $L \gg t \gg 1$. The contribution of
the RW to the energy then simplifies to
\begin{align}
&N_{\rm RW} = \frac{u_-^2-1}{2}
t  +  (u_-+1) L + (u_--1) b,\\
&E_{\rm RW} = \frac{u_-^3-1}{3}
t + \frac{1-u_-}{2t}  +  \frac{u_-^2+1}{2} L +  \frac{u_-^2-1}{2} b
\sim \frac{u_-^3-1}{3}
t + \frac{u_-^2+1}{2} L +  \frac{u_-^2-1}{2} b.
\end{align}
Since $L \gg 1$ and $u_{\rm sol}(x,t;c)$ decays exponentially to $0$
when $|x| \to \infty$, we can allow $L \to \infty$ in the integral for
$N_{\rm sol}$ and $E_{\rm sol}$
\begin{align}
N_{\rm sol}(c,u_-) &\sim \int \limits_{-\infty}^{+\infty} (u_{\rm
sol}-u_-) dx = 12 \sqrt{1-\frac {u_-} {c}} c\\
\label{solc}
E_{\rm sol}(c,u_-) &\sim \int \limits_{-\infty}^{+\infty}
\frac{u_{\rm sol}^2+2 u_- u_{\rm sol} + (\partial_x u_{\rm
sol})^2}{2} dx = \frac{12}{5} \sqrt{1 - \frac{u_-}{c}}
\left(\frac{6 c^2}{u_-^2} - \frac{2 c}{u_-} + 1 \right) u_-^2.
\end{align}

(ii) \ Now we consider the conservation laws~\eqref{consN}
and~\eqref{consE}. Since $L \gg t$, the RW and the solitary wave have
not reached the positions $x=\pm L$ at the time $t$ and the solution
$u(x,t)$ decays rapidly to $u_\pm$ for $|x| \sim L$.  In particular,
we have $u \sim u_\pm$ and $|u_{xt}| \ll u_\pm^2$ at $x=\pm L$. Thus
the conservation laws read
\begin{align}
\label{N1}
&\frac{dN}{dt}
\sim \frac{u_-^2- 1}{2},\quad \text{which yields}\quad
N \sim \frac{u_-^2-1}{2} t + N(0),\quad
N(0) = \int \limits_{-L}^{+L} u_0 \, dx,\\
\label{E1}
&\frac{dE}{dt}
\sim \frac{u_-^3- 1}{3},\quad \text{which yields}\quad
E \sim \frac{u_-^3-1}{3} t + E(0),\quad
E(0) = \int \limits_{-L}^{+L} \frac{u_0^2+(\partial_x
u_0)^2}{2} dx,
\end{align}
where $u_0$ is the smoothed step initial data~\eqref{step}.
We calculate
\begin{equation}
N(0) = (u_-+1) L,\quad
E(0) = \frac{(u_-^2+1)L}{2} +
\frac{(u_-+1)^2 \tanh(L/\xi)\left[3-3\xi^2-\tanh^2(L/\xi) \right]}
{12 \xi} .
\end{equation}
Since $L \gg \xi$, the initial energy $E(0)$ simplifies to
\begin{equation}
E(0) \sim \frac{(u_-^2+1)L}{2} + (u_--1)^2 \left(\frac{1}{6\xi}-
\frac{\xi}{4} \right) .
\end{equation}

\begin{figure}[h]
\centering
\begin{subfigure}[t]{.49\textwidth}
\centering
\includegraphics{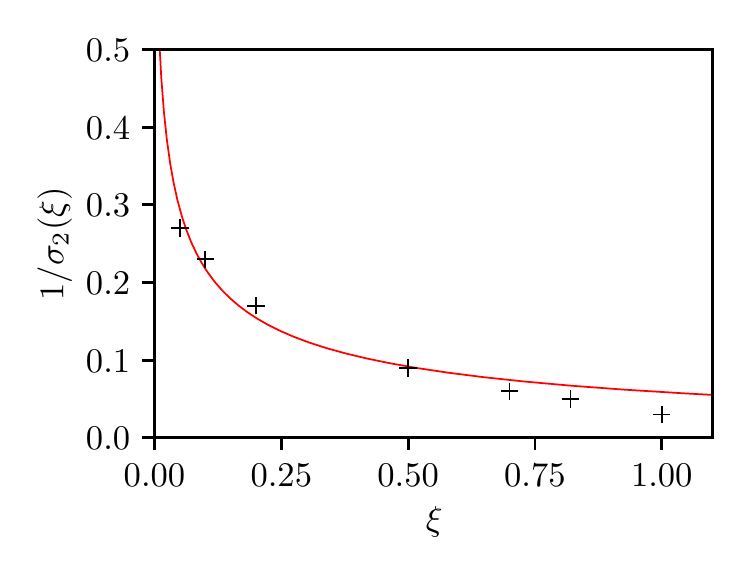}
\caption{Variation of the threshold for the generation of a
solitary wave when $u_- < 1/\sigma_2(\xi)$.  The red solid line
corresponds to~\eqref{umax} and the crosses to the thresholds determined
numerically. }
\label{fig:umax}
\end{subfigure}\hfill
\begin{subfigure}[t]{.49\textwidth}
\centering
\includegraphics{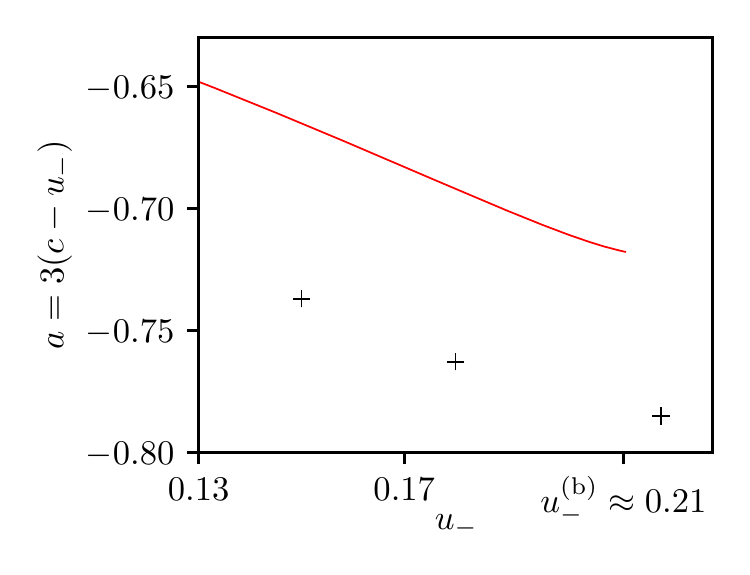}
\caption{Amplitude of the shedded solitary wave for $\xi=0.1$.  The
red solid line corresponds to the solution of~\eqref{Esol2}.
}
\label{soliton_amp2}
\end{subfigure}%
\caption{Solitary wave shedding with $u_->0$
in the regime $\xi \ll 1$}
\label{soliton_threshold}
\end{figure}

Equating~\eqref{N2} and~\eqref{N1}, and ~\eqref{E2}
and~\eqref{E1} yields
\begin{align}
\label{Esol1}
N_{\rm sol}(c,u_-) + (u_--1) b =0,\quad
E_{\rm sol}(c,u_-) + \frac{u_-^2-1}{2} b =(u_--1)^2
\left(\frac{1}{6\xi}- \frac{\xi}{4} \right) ,
\end{align}
which combine into a single equation for the solitary wave speed $c$
\begin{equation}
\label{Esol2}
E_{\rm sol}(c,u_-) - \frac{u_-+1}{2} N_{\rm sol}(c,u_-) =
(u_--1)^2 \left(\frac{1}{6\xi}-
\frac{\xi}{4} \right).
\end{equation}
A solitary wave emerges to the left of the linear wavetrain
if~\eqref{Esol2} has a solution~$c < -u_-/8$.  This is the case if and
only if $u_- < u_-^{\rm (2)}$ where $u_-^{\rm (2)}$ solves the
equation:
\begin{equation}
E_{\rm sol}\left(-\frac{u_-^{\rm (2)}}{8},u_-^{\rm (2)}\right) -
\frac{u_-^{\rm (2)}+1}{2} N_{\rm
sol}\left(-\frac{u_-^{\rm (2)}}{8},u_-^{\rm (2)}\right) =
\left(u_-^{\rm (2)}-1 \right)^2 \left(\frac{1}{6\xi}-
\frac{\xi}{4} \right) .
\end{equation}
Since $\xi \ll 1$, we have the approximation
\begin{equation}
\label{umax}
u_- \leq u_-^{\rm (2)} = 1/\sigma_2(\xi),\quad
\sigma_2(\xi) = 1+\frac{9}{2}
\sqrt{\frac{21}{5} \xi} +\frac{27}{4} \xi + {\cal O}(\xi^{3/2})
.
\end{equation}
The critical value $u_-^{\rm (2)} = 1/\sigma_2(\xi)$ constitutes the
threshold for the emission of one solitary wave in the dispersive
Riemann problem. This formula compares quantitatively with the
threshold obtained numerically for different values of $\xi$ as
displayed in Fig.~\ref{fig:umax}.

Suppose now $\sigma_2(\xi) u_-\leq 1 $. Solving~\eqref{Esol2} for
$c(u_-)$, we obtain the solitary wave's amplitude
$a(u_-) = 3[c(u_-)-u_-]$ in function of $u_-$.
Figure \ref{soliton_amp2} displays the variation of $c(u_-)$ when
$\xi=0.1$, which compares reasonably well with the amplitude of the
emitted solitary wave obtained numerically.

Discrepancy between the analytical results and the numerics in
Fig.~\ref{soliton_threshold} is due to neglecting the linear
corrective term $\varphi$ in the ansatz~\eqref{ansatz1}.  Because it
separates from the RW and solitary wave, the contribution of the
linear wave to the mass and the energy are
\begin{equation}
N_\varphi = \int_{s_{\rm min} t}^{u_- t} (u-u_-) dx,\quad E_\varphi =
\int_{s_{\rm min} t}^{u_- t} \frac{u^2+(\partial_x u)^2-u_-^2}{2} dx .
\end{equation}
For example, we obtain for $(u_-,\xi)=(0.18,0.1)$ and $t=1000$,
$(N_\varphi,E_\varphi)\approx (12,0.21)$ from the numerical simulation
that, when compared to the contribution of the solitary wave $(N_{\rm
  sol},E_{\rm sol}) = (36,0.41)$, is comparable.  Thus, neglecting
$\varphi$ in~\eqref{ansatz1} overestimates the solitary wave mass and
energy.  An improved estimate for the solitary wave amplitude is
possible if the variation of $\varphi(x,t)$ for $x \in (s_{\rm min} t,
u_- t)$ can be determined.  While we have obtained an asymptotic
description of linear wavetrains that are generated by smoothed step
dispersive Riemann problems (cf.~eq.~\eqref{sol_lin}), our analysis
did not include the generation of a solitary wave, which is a
significantly more complex problem.  In particular, the initial data
for the linear wavetrain~\eqref{phi0} would require modification due
to initial solitary wave-linear wavetrain interaction.  Therefore, we
have not found a simple analytical estimation of $N_\varphi$ and
$E_\varphi$.

We emphasize that both computations presented in this section predict
solitary wave emission for $u_-$ close to the threshold $u_-
\sigma_2(\xi) = 1$. Indeed, we observe in the previous section that
the numerical solution involves the generation of multiple solitary
waves for sufficiently small values of $|u_-|$, which is not taken
into account by the ansatz~\eqref{ansatz1} considered here.  For
example, multiple solitary waves are generated for $u_-<0.13$ when
$\xi=0.1$.

A similar argument could be used to explain the emission of solitary
waves for $u_-^{(1)} \leq u_-<0$, where the solution could be
approximated by
\begin{equation}
u(x,t) = u_{\rm aexp}(x-b,t) + u_{\rm sol}(x,t;c,u_-) +
\varphi(x,t),\qquad c<u_-<0,\qquad t \gg 1,
\end{equation}
with $u_{\rm aexp}$ given by~\eqref{trunc} and $\varphi$ a corrective
term. Contributions to the mass and the energy from the correction
term are of the same order as the contributions from the solitary
wave.  The determination of $\varphi$ is, as demonstrated in the
previous computation, necessary to compute the solitary wave's
amplitude. Typical variations of~$\varphi(x,t)$ within the interior
region of the asymmetric expansion shock $x \in (u_-t,t)$ are
displayed in Fig.~\ref{mass2}.  Contrary to the previous computation,
we have not found a simple ansatz that can describe the variation of
$\varphi(x,t)$.  The determination of the solitary wave's amplitude
for $u_-<0$ is left as a subject for future work.

\section{Dispersive Shock Waves}
\label{DSW}

DSWs are expanding modulated nonlinear wavetrains regularizing
wavebreaking singularities in dispersive hydrodynamics, a conservative
counterpart of regularizing classical shock waves in viscous fluid
dynamics. The shock structure of a DSW is more complex than the shock
structure of a viscous shock wave.  In particular, a DSW cannot be
described by a traveling wave solution of the nonlinear wave equation
\cite{el_dispersive_2017}.  We refer the reader to the recent review
\cite{el_dispersive_2016} for the principal ideas and applications of
DSW theory.  In this section, we analyze the DSWs generated in the BBM
dispersive Riemann problem for $|u_+|<u_-$; see regions $b$ and $e$ in
Figs.~\ref{fig:xi10} and \ref{fig:xi01}, respectively. We first
consider the region
\begin{equation}\label{R_data}
0<\mu u_- < u_+<u_-,\ \ \mbox{where} \ \mu = e^{-3/2}/4.
\end{equation}
This restriction on the initial data (the formula for $\mu$ is derived
later) is related to DSW admissibility conditions specific to the BBM
equation. As an aside, we note that for the KdV equation \eqref{kdv},
the only requirement on the Riemann data for a DSW solution is
$u_->u_+$.  Although the structure of the BBM DSW within the
admissibility region \eqref{R_data} is qualitatively similar to that
of the KdV DSW, the quantitative description of BBM DSWs requires a
modified approach due to non-integrability of the BBM equation. The
DSW fitting method introduced in \cite{el_resolution_2005} (see also
\cite{el_dispersive_2016}) enables the analytical determination of the
DSW edge speeds in non-integrable dispersive-hydrodynamic systems. A
recent extension of DSW fitting \cite{congy_nonlinear_2019} also
enables partial determination of the interior DSW structure. In
Section~\ref{sec:BBM_fitting}, we apply the extended DSW fitting
method to the BBM equation and compare the obtained analytical results
with numerical simulations of the smoothed dispersive Riemann problem
~\eqref{step}.

In the complementary region $-u_-<u_+<\mu u_-$ of Riemann data,
DSW fitting fails due to the development of nonlinear two-phase
oscillations at one of the DSW's edges. This phenomenon, termed DSW
implosion, is a nonlinear counterpart of the two-phase interference
pattern occurring in linear wavetrains near the zero dispersion point,
described earlier in Section~\ref{linear_wave}.  Numerical simulations
and the onset of DSW implosion were first studied in
\cite{lowman_dispersive_2013} for the viscous fluid conduit equation.
In Section~\ref{implosion}, we develop an analytical framework for the
structure of DSW implosion by deriving a coupled NLS equation for a
nonlinear superposition of two BBM Stokes waves that asymptotically
describe the nonlinear two-phase dynamics of the imploded region.

\subsection{Classical, convex DSWs:  general properties, fitting relations and  interior modulation}
\label{sec:dsw_gen}

It is convenient to describe properties of DSWs in the framework of a
fairly general scalar dispersive equation
\begin{equation}\label{dh}
u_t+f(u)_x=D[u]_x,
\end{equation}
where $f(u)$ is the hyperbolic flux and $D[u]$ is a dispersion
operator, giving rise to a real valued linear dispersion relation
$\omega=\omega_0(k, \ubar)$. For the BBM equation,
$f(u)=\tfrac12 u^2$, $D[u]=u_{xt}$ and $\omega_0(k, \ubar)$ is given
by \eqref{dispersion}.

\begin{figure}
\centering
\includegraphics[width=0.5\textwidth]{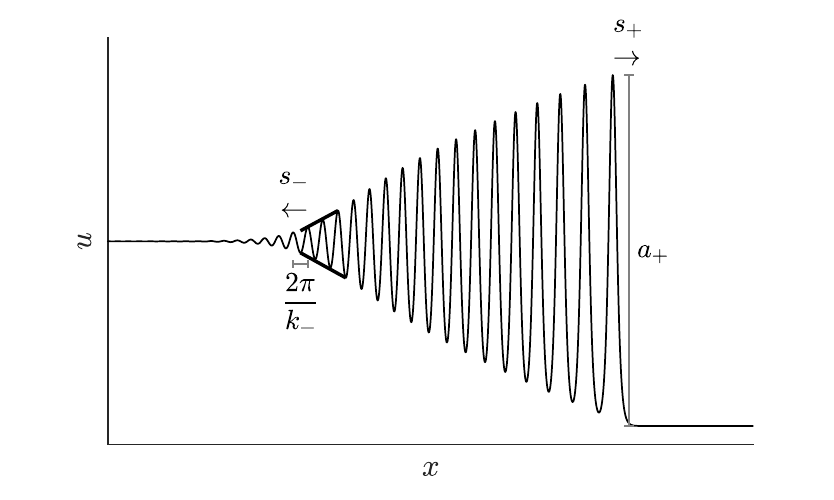}
\caption{KdV-type DSW in convex dispersive hydrodynamics with negative
  dispersion.  The DSW fitting relations determine the soliton edge
  velocity $s_+$ and amplitude $a_+$ as well as the harmonic edge
  velocity $s_-$ and wavenumber $k_-$.}
\label{fig:Fig1}
\end{figure}

The rapidly oscillating structure of DSWs---see, e.g.,
Fig.~\ref{fig:Fig1}---motivates the use of asymptotic, WKB-type,
methods for their analytical description. One such method, known as
Whitham modulation theory \cite{whitham_linear_1999} (see also
\cite{kamchatnov_nonlinear_2000}), is based on the averaging of
dispersive hydrodynamic conservation laws over nonlinear periodic
wavetrains, leading to a system of first order quasilinear PDEs, known
as the Whitham modulation system.  Classical DSW theory has been
developed for KdV-type equations.  More generally, it is useful to
define the dispersive hydrodynamic equations
\eqref{dh} to be  {\em classical or convex}  if  the associated Whitham modulation system 
has the properties of  strict hyperbolicity and genuine
nonlinearity;   see
\cite{el_dispersive_2016}.  For the scalar equation \eqref{dh},
convexity of the flux, i.e., $f''(u) \ne 0$, and of the linear
dispersion relation, i.e., $\partial_{kk}\omega_0(k, \ubar) \ne 0$ for
$k \ne 0$ and $\ubar \in \mathbb{R}$, are defining conditions for
convexity of the dispersive hydrodynamic equation
\cite{el_dispersive_2016}. This convexity property provides for the
existence of a class of solutions to the Whitham modulation system
that describe the DSW structure.

For the BBM equation, the hyperbolic flux is indeed convex, but the
linear dispersion relation is not, because
$\partial_{kk}\omega_0(k, \ubar)=0$ at the {\em zero dispersion
points} $k=\sqrt{3}$ and $\ubar = 0$ (additionally $k= 0$), see
\eqref{disp_curv}.  As a result, the classical, convex DSW regime is
realized \emph{only} for a restricted domain of Riemann data $u_\pm$ given by  \eqref{R_data},
and established below.

Whitham theory has proven particularly effective for the description
of DSWs in both integrable and non-integrable systems.  If the
dispersive hydrodynamics are described by an integrable equation such
as the KdV equation, the associated Whitham system can be represented
in a diagonal, Riemann invariant form \cite{whitham_non-linear_1965,
  flaschka_multiphase_1980,kamchatnov_nonlinear_2000}. This fact
enabled Gurevich and Pitaevskii (GP)
\cite{gurevich_nonstationary_1974} to construct an explicit modulation
solution for a DSW generated by a dispersive Riemann problem for the
KdV equation. The GP construction is based on a self-similar,
rarefaction wave solution of the KdV-Whitham system of three
modulation equations for three slowly varying parameters that locally
describe the oscillatory wave.  For our purposes, it is convenient to
choose the mean $\ubar$, the wavenumber $k$ and the amplitude $a$ as
the three parameters, all expressible in terms of the Riemann
invariants. The interior shock structure of a DSW is then described by
a self-similar, centered solution of the modulation equations whose
values, $(\ubar, k, a)(x,t)$ for $t \gg 1$, lie on a 2-wave
rarefaction curve connecting the edge points $(\ubar, k, a)=(u_-, k_-,
0)$ and $(\ubar, k, a)=(u_+, 0, a_+)$.  These edge points are
associated with a harmonic, small-amplitude wave ($a \to 0$) and a
solitary wave ($k \to 0$), respectively.  The two integrals associated with
the 2-wave curve generate a family of Riemann problems that are
parametrized by $(u_-,u_+)$, i.e.~$k_- = k_-(u_-,u_+)$ and $a_+ =
a_+(u_-,u_+)$.  We emphasize that the GP, self-similar solution
results from a genuine Riemann problem for the Whitham modulation
equations, i.e.~\textit{discontinuous} initial data, in contrast to an
initial smoothed step profile like \eqref{step} posited for the
dispersive Riemann problem.  For convex dispersive hydrodynamic
problems, this distinction is not so important because the long-time
evolution of a smoothed step profile is described by the GP solution.
But for nonconvex problems such as BBM, there are subtleties
associated with the smoothing length scale $\xi$---possibly beyond the
application of leading order Whitham modulation theory---that need to
be considered.

A DSW in a general convex, scalar (i.e., unidirectional) dispersive
hydrodynamic system such as \eqref{dh} has a multi-scale structure
that is similar to the KdV DSW.  It consists of an oscillatory
transition between two non-oscillatory---e.g., slowly varying or
constant---states: one edge is associated with a solitary wave that is
connected, via a slowly modulated periodic wavetrain, to a harmonic,
small-amplitude wave at the opposite edge. The relative position
(left/trailing or right/leading) of the soliton and harmonic edges
determines the DSW orientation $d \in \{\pm 1\},$ where $d = +1$ if
the solitary wave is on the right/leading edge.  DSW orientation is
determined by the dispersion sign as
$d=-\hbox{sgn}[\partial_{kk}\omega_0(k, \ubar)]$
\cite{el_dispersive_2016,el_dispersive_2017}. The DSW polarity $p \in
\{\pm 1\}$ is defined by the polarity of the solitary wave edge, where
$p = +1$ if the solitary wave is a wave of elevation relative to its
background.  DSW polarity is given by $p = - \sgn \ [ {\partial_{kk}
  \omega_0} f''(u)]$.  The DSW shown in Fig.~\ref{fig:Fig1} has
$d=p=+1$.

By direct evaluation of the BBM linear dispersion relation
\eqref{dispersion}, we observe that the BBM equation is expected to
support DSWs with positive orientation/polarity $d = p = +1$ and
negative orientation/polarity $d=p=-1$.  These two kinds of DSWs map
to each other through the BBM invariant transformation
$x \to -x, u \to -u$. Hence, it is sufficient to only consider DSWs
with $d=p=+1$.

\bigskip
{\it DSW fitting relations}

\medskip For non-integrable dispersive equations such as the BBM
equation, diagonalization of the associated Whitham modulation system
in terms of Riemann invariants is generally not possible. Thus, the
explicit determination of the Whitham system's simple wave solution
corresponding to a DSW is problematic, even though its existence
requires only strict hyperbolicity and genuine nonlinearity in the
2-wave characteristic family.  One can, however, explicitly determine
key observables associated to each DSW edge for integrable and
non-integrable equations provided the dispersive hydrodynamics are
convex.  These observables include the DSW edge speeds $s_{\pm}$ and
their associated wave parameters---the harmonic edge wavenumber $k_-$
and the soliton edge amplitude $a_+,$ as functions of the initial data
$u_\pm$.  The determination of these observables represents the
fitting of a DSW to the long-time dynamics of dispersive Riemann initial data.
The DSW fitting method proposed in \cite{el_resolution_2005} (see also
\cite{el_dispersive_2016}) is based on a fundamental, generic
property: the Whitham modulation equations admit exact reductions to a
set of common, much simpler, analytically tractable equations in the
limits of vanishing amplitude and vanishing wavenumber, which
correspond to the harmonic and soliton DSW edges, respectively.  DSW
fitting has also been successfully applied to the propagation of a
broad localized pulse into a constant state \cite{el_asymptotic_2008,kamchatnov_dispersive_2019,maiden_solitary_2020}.

We now formulate the set of DSW fitting relations for an initial step
in the mean $\ubar$, a dispersive hydrodynamic analogue of the
classical Rankine-Hugoniot jump conditions for viscously regularized
shocks. As outlined above, the DSW fitting relations specify the
speeds $s_-$ and $s_+$ of the DSW edges and the associated wave
parameters $k_-$ and $a_+$ in terms of the Riemann data $(u_-,u_+)$.
Since we focus on positive DSW polarity and orientation, the leftmost
trailing edge, propagating with constant speed $s_-$, is associated
with the vanishing amplitude harmonic wave, while the rightmost
leading edge, propagating with speed $s_+,$ is associated with the
solitary wave. Thus, $s_+>s_-$.  In what follows, the Riemann
step data $(u_-, u_+)$ of \eqref{step} are fixed in the regime of
interest \eqref{R_data}, where we assume the existence of the integral
2-wave curve of the modulation system connecting the harmonic and
soliton edges.  The DSW fitting relations are summarized in three
steps (see \cite{el_resolution_2005, el_dispersive_2016} for
additional details):

\medskip (i) \ {\em The DSW harmonic edge} is the modulation
characteristic $x=s_-t$ with speed $s_{-} = s_-(u_-, u_+)$ determined
by the linear group velocity for the edge mean $\ubar=u_-$ and
wavenumber $k=k_-$:
\begin{equation}
\label{eq:harm_def1}
s_{-} = \partial_k \omega_0(k_{-}, u_{-}).
\end{equation}
Given $u_-$ and $u_+$, the harmonic edge wavenumber $k_-$ is
determined
by solving the initial value problem
\begin{equation}
\label{eq:harm_ODE2}
\frac{\rmd k}{\rmd \ubar} =  \frac{\partial_{\ubar}
\omega_0(k,\ubar)}{f'(\ubar) - \partial_k \omega_0(k,\ubar)},
\quad  k(u_+)= 0,
\end{equation} 
for $K(\ubar, u_+) \equiv k(\ubar)$, and setting $k_-=K(u_-, u_+)$.

\medskip (ii) \ {\em The DSW soliton edge} is the characteristic
$x=s_+ t$ whose speed $s_+ = s_{+}(u_{-}, u_{+})$ coincides with the
solitary wave velocity for the edge mean $\ubar = u_+$ and amplitude
$a = a_+$
\begin{equation}
\label{soli_edge_speed1}
s_{+} = c(a_+, u_+),
\end{equation}
where $c(a,\ubar)$ is the velocity of a solitary wave with amplitude $a$ on
the background $\ubar$.  For the description of the soliton edge, it
is instructive to make the change of variables
$(\ubar, a) \to (\ubar,\tilde k)$, where the \emph{conjugate
wavenumber} $\tilde k$ is defined implicitly by
\begin{equation}
\label{ak}
c(a, \ubar)=\frac{\tilde \omega_0 (\tilde k, \ubar)}{\tilde k},
\end{equation}
and $\tilde \omega_0(\tilde k, \ubar)$ is the conjugate dispersion
relation defined in terms of the linear dispersion relation according
to
$\tilde \omega_0(\tilde k, \ubar) = - i \omega_0(i \tilde k, \ubar)$.
Given $u_-$, $u_+$, the value $\tilde k_{+}$ at the soliton edge of
the DSW is determined by solving the initial value problem
\begin{equation}
\label{eq:sol_ODE2}
\frac{\rmd \tilde k}{\rmd \ubar} =  \frac{\partial_{\ubar} \tilde
\omega_0(\tilde k,\ubar)}{f'({\ubar}) - \partial_{\tilde k} \tilde
\omega_0(\tilde k, \ubar)},
\quad \tilde  k(u_-)= 0 
\end{equation}  
for $\tilde K(u_-,\ubar) \equiv \tilde k(\ubar)$, and setting
$\tilde k_+=\tilde K(u_-, u_+)$.  The soliton edge speed $s_+$ and the
solitary wave amplitude $a_+$ are then found from \eqref{soli_edge_speed1},
\eqref{ak}, with $\tilde k=\tilde k_+$, $\ubar=u_+$.

Note that we have used a slightly modified presentation of DSW fitting
as compared to Refs.~\cite{el_resolution_2005, el_dispersive_2016} in
order to explicitly highlight the dependence of the edge velocities
$s_\pm$ and edge wave parameters $k_-$, $a_+$ on the initial Riemann
data $u_-$, $u_+$.

\medskip (iii) \ {\em Admissibility conditions:} \ The DSW fitting
relations are subject to the existence of a 2-wave curve of the
Whitham modulation equations continuously connecting
$(\ubar,k,a) = (u_-,k_-,0)$ to $(\ubar,k,a) = (u_+,0,a_+)$.  The
following admissibility conditions are necessary for the existence of
a 2-wave solution \cite{el_resolution_2005,hoefer_shock_2014}
\begin{align}
\label{admis1}
\textrm{Causality:}  \quad &s_-< f'(u_-), \ \ s_+ > f'(u_+), \ \ s_+ >
s_-. \\[3mm]
\label{admis2}
\textrm{Convexity:}  \quad &\frac{\partial s_{-}}{\partial u_{-}}
\ne 
0, \quad \frac{\partial s_{-}}{\partial
u_{+}} \ne 
0, \quad   \frac{\partial
s_{+}}{\partial u_{+}} \ne 0, \quad
\frac{\partial s_{+}}{\partial u_{-}}
\ne 0.
\end{align}

\bigskip
{\it DSW interior modulation}

\medskip

It was shown in \cite{congy_nonlinear_2019} that the DSW fitting
procedure can be extended into the DSW interior $x>s_-t$ adjacent to
the harmonic edge, where the wave amplitude is small. This is done by
exploiting the overlap domain for the asymptotic applicability of
weakly nonlinear modulated (Stokes) waves in the Whitham modulation
equations (describing slow modulations of arbitrary amplitude waves)
and the nonlinear Schr\"odinger equation (describing slow modulations
of finite but small amplitude waves).  A brief outline of the results
of \cite{congy_nonlinear_2019} relevant to this paper is presented
below.

The weakly nonlinear regime of the DSW $|u(x,t)-u_-|\ll u_-$ is well
described by a slowly modulated Stokes wave
\begin{equation}
\label{stokes}
u(x,t) \sim u_- + \left[ A(x,t) e^{i(k_- x -\omega_0(k_-,u_-) t)} + \cc
\right] + B(x,t),
\end{equation}
where $k_- =K(u_-, u_+)$ is determined by DSW fitting, $A(x,t) \in
\mathbb{C}$ and $B(x,t) \in \mathbb{R}$ are slowly varying fields in
comparison to the fast oscillations of the carrier wave. Note that
$B(x,t)$ describes the induced mean flow.  The crest to trough
amplitude $a(x,t)$ and the wavenumber $k(x,t) = k_- + v(x,t)$ of the
Stokes wave are related to the complex envelope $A(x,t)$ by
\begin{equation}
\label{Amad}
A(x,t) = \frac{a(x,t)}{4} \exp \left[i \int v(x,t) dx \right].
\end{equation}
The dynamics of $A(x,t)$ are governed by the nonlinear Schr\"odinger
(NLS) equation
\begin{equation}
\label{NLS}
i A_t + i s_- A_x + \beta_- A_{xx}
+\gamma_- |A|^2 A=0,\quad
B = b(k_-,u_-) |A|^2,
\end{equation}
where $\beta_-=\beta(k_-, u_-)=\tfrac12 \partial_{kk}\omega_0(k_-,
u_-)$ and $\gamma_- = \gamma(k_-, u_-)$ are determined as part of the
standard multiple scales derivation of the NLS equation (see for
instance \cite{benney_propagation_1967}).  Note that the slow
variation of the background $\ubar(x,t)$ is solely due to the induced
mean flow and is constrained to the variation of the amplitude
according to
\begin{equation}
\label{ubar}
\ubar = u_- + b_- a^2/16,
\end{equation}
where $b_- = b(k_-, \ubar_-)$ is also determined as part of the NLS
derivation.  The NLS equation describes a stable modulation iff
$\beta_-\gamma_-<0$ (defocusing regime). Also, the derivation requires
that $\beta_-$ does not vanish, which is the case in the convex
modulation regime considered here.

For the description of a DSW's structure, the NLS equation \eqref{NLS}
is considered with boundary conditions specified at the harmonic edge
\begin{equation}\label{bc_NLS}
x = s_- t: \qquad a(x,t) = 0,\ \   k(x,t) = k_- .\; 
\end{equation}
As a result, the universal DSW modulation near the trailing edge is
given by the following special self-similar rarefaction wave solution
of~\eqref{NLS}
\begin{equation}
  \label{DSW_sol}
  a(x,t) \sim \frac{4}{3 \sqrt{-2 \beta_-\gamma_-}} \left[\frac{x}{t} -
    s_- \right],\; \quad
  k(x,t) \sim k_- + \frac{1}{3\beta_-} \left[\frac{x}{t} -
    s_- \right].
\end{equation}
The formulae \eqref{DSW_sol} represent a first-order approximation of
the DSW's structure near its harmonic edge. A more accurate
description of the DSW modulation, which asymptotically extends deeper
into the DSW's interior, is achieved by including higher-order terms
in the derivation of the NLS equation.  The resulting HNLS equation
(see Appendix) subject to the boundary conditions \eqref{bc_NLS}
yields a rarefaction wave solution for $a$ and $k$ that includes terms
$\mathcal{O}(\frac{x}{t} -s_- )^2$ in addition to the linear terms of
eq.~\eqref{DSW_sol}.

\subsection{BBM DSW fitting}
\label{sec:BBM_fitting}

We now apply the DSW fitting relations
\eqref{eq:harm_def1}-\eqref{eq:sol_ODE2} and the universal modulation
formulae \eqref{DSW_sol} (and its HNLS improvement) to the description
of BBM DSWs generated within the region of Riemann data $u_-, u_+$
defined by the admissibility conditions \eqref{admis1},
\eqref{admis2}. The BBM equation satisfies the prerequisites for DSW
fitting \cite{el_resolution_2005, el_dispersive_2016}: periodic
travelling waves parameterised by three independent integrals of
motion exist (with harmonic wave and solitary wave limits) and BBM has
at least two conservation laws~\eqref{bbm} and~\eqref{consE0}. The
speed-amplitude relation for solitary waves is given by \eqref{c}.  We
initially consider the Riemann data \eqref{step2} with $u_- > u_+ > 0$
(other configurations will be considered later).

Using the BBM linear dispersion relation \eqref{dispersion}, the
characteristic equation \eqref{eq:harm_ODE2} for the harmonic edge is
\begin{equation}
\frac{\rmd  k}{\rmd  \ubar} = \frac{1+k^2}{\ubar k(3+k^2)},
\quad k(u_{+})=0.
\end{equation}
Integration yields
\begin{equation}
\label{harm_int}
\ubar=u_{+}(1+k^2)e^{k^2/2}.
\end{equation}
The $k$-locus $k_{-}=K(u_{-}, u_+)$ at the DSW harmonic edge is then
found from \eqref{harm_int} by setting $\ubar=u_{-}$, $k=k_-$, and is
given by the implicit equation
\begin{equation}
\label{bbm_k_locus}
(1+k_-^2)e^{k_-^2/2} =  \frac{u_{-}}{u_{+}} \, .
\end{equation}
The harmonic edge speed is then found from~\eqref{vg} as
\begin{equation}
\label{bbm_s_harm}
s_{-} = \partial_k \omega_0(k_{-},u_{-})= u_{-}
\frac{1-k_{-}^2}{(1+ k_{-}^2)^2}.
\end{equation}

To describe the solitary wave edge, we introduce the BBM conjugate
dispersion relation
$\tilde \omega_0( \tilde k,\ubar)= - i \omega_0(i \tilde k,\ubar)$ to
obtain
\begin{equation}
\tilde \omega_0( \tilde k,\ubar) =\ubar\frac{\tilde k}{1- \tilde  k^2}.
\end{equation}
The characteristic equation \eqref{eq:sol_ODE2} for the solitary wave
edge then becomes
\begin{equation}\label{sol_int}
\frac{\rmd  \tilde k}{\rmd  \ubar}=\frac{1- \tilde k^2}{\ubar \tilde
k(\tilde k^2-3)}, \quad \tilde k(u_{-})=0.
\end{equation}
Its solution has the form
\begin{equation}\label{kofu1}
\ubar=u_{-} (1- \tilde k^2) e^{-\tilde k^2/2}
\end{equation}
The $\tilde k$-locus $\tilde k_{+} =\tilde K(u_{-}, u_{+})$ of the DSW
soliton edge is obtained by setting $\ubar =u_+$,
$\tilde k = \tilde k_+$ in \eqref{kofu1}, and is given by the implicit
equation (cf.~\eqref{bbm_k_locus})
\begin{equation}
\label{tilde_locus}
(1-\tilde k_{+}^2) e^{-\tilde k_{+}^2/2} = \frac{u_{+}}{u_{-}}. 
\end{equation}
The solitary wave edge speed $s_{+}$ is found from
\eqref{soli_edge_speed1}, \eqref{ak} by setting
$\tilde k =\tilde k_{+}$, $\ubar = u_{+}$:
\begin{equation}
\label{bbm_s_soli}
s_{\rm +}= \frac {\tilde \omega_0 (\tilde k_{+}, u_{+})}{\tilde k_+} =
\frac{u_{+}}{1- \tilde k_{+}^2}\, .
\end{equation}
The DSW amplitude $a_{+}$ at the soliton edge is then determined via
$s_{\rm +}= c(a_{+},u_{+})$ with $c(a,\ubar) $ given by \eqref{c},
which yields
\begin{equation}\label{a+}
a_{+} = 3u_{+}\frac{\tilde k_{+}^2}{1- \tilde k_{+}^2}.
\end{equation}
Positivity of the solitary wave amplitude requires $\tilde k_{+}<1$.

We now verify the DSW fitting admissibility conditions \eqref{admis1},
\eqref{admis2}. The first two causality conditions \eqref{admis1} for
BBM assume the form $s_-< u_-, \ \ s_+ > u_+$ and are readily verified
provided $\tilde k_+ < 1$, which is true so long as $u_- > u_+ >
0$.  Next, a calculation yields
\begin{equation}
\label{eq:1}
\frac{\partial s_-}{\partial u_-} = -
\frac{3+k_-^4}{(1+k_-^2)^2(3+k_-^2)}, \quad \frac{\partial s_-}{\partial
u_+} = \frac{2 (3-k_-^2) e^{k_-^2/2}}{3+4k_-^2+k_-^4} .
\end{equation}
While the first convexity condition in \eqref{admis2} is always
satisfied, the second condition requires $k_- \ne \sqrt{3}$, which,
after substitution in \eqref{bbm_k_locus}, gives
$({u_-}/{u_+}) \ne 4e^{3/2}$.  The critical line
\begin{equation}
\label{crossover}
u_{+}= \mu \,u_{-}, \quad \mu = e^{-3/2}/4 \approx 0.056
\end{equation}
corresponds to the zero dispersion point $k_-=\sqrt{3}$ and delimits
the applicable region for the DSW fitting method
\begin{equation}\label{restr_reg1}
0 < \mu u_- <  u_+ .
\end{equation}
Another calculation yields
\begin{equation}
\label{eq:2}
\frac{\partial s_+}{\partial u_+} = \frac{1}{3-\tilde k_+^2}, \quad
\frac{\partial s_+}{\partial u_-} =  \frac{2 e^{-\tilde
k_+^2/2}}{3-\tilde k_+^2} .
\end{equation}
Equation \eqref{tilde_locus} implies that $0 < \tilde k_+ < 1$ within
the region \eqref{restr_reg1}.  Then the third and fourth convexity
conditions in \eqref{admis2} are readily verified to hold true for all
$u_-, u_+$ in \eqref{restr_reg1}.

It follows that within the admissibility region \eqref{restr_reg1},
$k_-<\sqrt{3}$ so the dispersion sign is
\begin{equation}
\sgn \, \beta(k_-, u_-) = \sgn \left ( u_- \frac{2 k_-
(k_-^2-3)}{(1+ k_-^2)^3} \right )<0 .
\end{equation}    
Consequently, the DSW orientation and polarity are $d=p=1$, consistent
with our original assumption $s_+>s_-$, which ensures the third
causality condition \eqref{admis1} within the region
\eqref{restr_reg1} for $\mu$ defined in \eqref{crossover}.

\begin{figure}[h]
\begin{subfigure}{0.32\linewidth}
\centering
\includegraphics{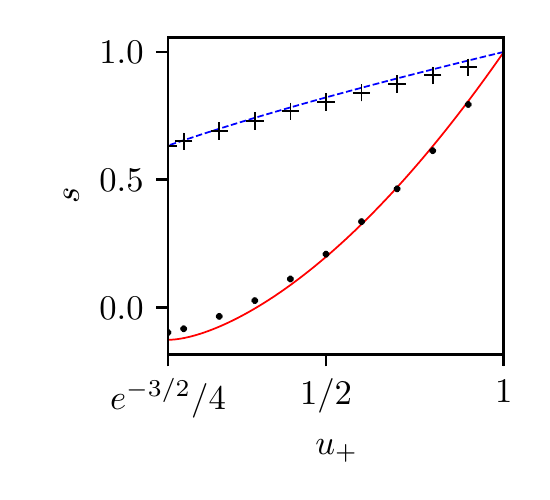}
\caption{Edge velocities~\eqref{bbm_s_harm} and~\eqref{bbm_s_soli}.}
\label{fig:bbm_edge}
\end{subfigure}
\begin{subfigure}{0.32\linewidth}
\centering
\includegraphics{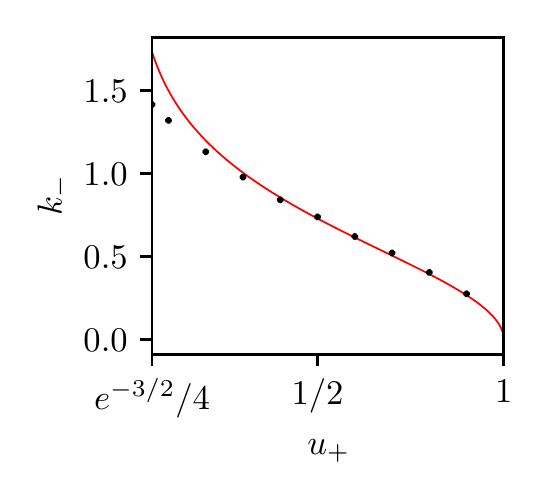}
\caption{Harmonic edge wavenumber~\eqref{bbm_k_locus}.}
\label{fig:bbm_wavenumber}
\end{subfigure}\hfill
\begin{subfigure}{0.32\linewidth}
\centering
\includegraphics{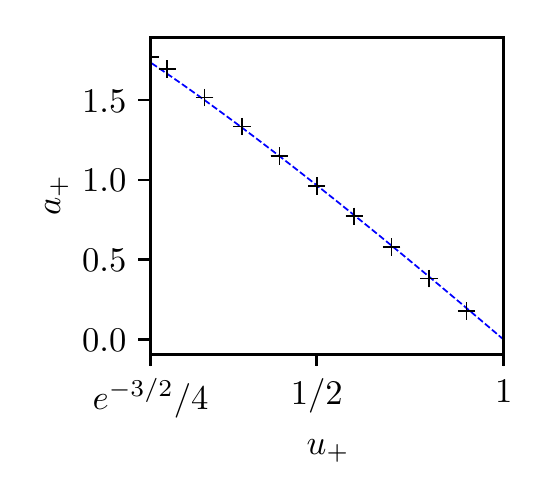}
\caption{Solitary wave edge amplitude~\eqref{a+}.}
\label{fig:bbm_amplitude}
\end{subfigure}\hfill
\centering
\caption{Edge parameters as functions of $u_+$ for fixed
  $u_-=1$. Solid (resp.~dashed) lines represent DSW fitting formulas
  for the harmonic (resp.~solitary wave) edge. Markers represent the
  parameter values extracted from direct numerical simulation of the
  BBM smoothed dispersive Riemann problem at $t=1000$ with initial
  data \eqref{step} where $\xi=10$. }
\label{fig:bbm_riemann1}
\end{figure}

Figure \ref{fig:bbm_edge} shows excellent agreement between the
analytical expressions \eqref{bbm_s_harm} and \eqref{bbm_s_soli} and
the velocities of the DSW edges extracted from numerical
simulations. Additionally, Fig.~\ref{fig:bbm_wavenumber} displays the
comparison between~\eqref{bbm_k_locus} and the harmonic edge
wavenumber extracted numerically, and Fig.~\ref{fig:bbm_amplitude}
displays the comparison between~\eqref{a+} and the solitary wave edge
amplitude.

The interior BBM DSW modulation in the vicinity of the harmonic edge
is given by \eqref{DSW_sol}. It is determined by the coefficients
$\beta(k, \ubar)$ and $\gamma(k, \ubar)$ of the NLS equation
\eqref{NLS} for slowly varying modulations of the weakly nonlinear
periodic traveling wave of the BBM equation with $k=k_-$,
$\ubar=u_-$.  The derivation of the NLS equation is standard and we
only present here the expressions for $\beta$ and $\gamma$
\begin{equation}
\label{coeff_NLS}
\beta(k,\ubar) =  \ubar\frac{ k
(k^2-3)}{(1+ k^2)^3} ,\; \quad
\gamma(k,\ubar) = \frac{3+5k^2}{6\ubar k(3+k^2)} .
\end{equation}
Within the admissible region \eqref{restr_reg1}, $k_-< \sqrt{3}$ and
so $\beta(k_-, u_-) \gamma(k_-, u_-) <0$, thus ensuring modulational
stability of the DSW's harmonic edge.  Additionally, the function
$b(k, \ubar)$ determining the variation of the induced mean flow
\eqref{ubar} near $x=s_-t$ is given by
\begin{equation}\label{b_BBM}
b(k ,\ubar) = -  \frac{(1+k^2)^2}{\ubar k^2(3+k^2)}.
\end{equation}

A comparison between the modulation solution~\eqref{DSW_sol},
\eqref{ubar}, and the DSW structure extracted from numerical
simulations is displayed in Fig.~\ref{fig:BBM}.  The figure also
displays the modulation solution obtained using the HNLS
approximation, see Appendix~\ref{HNLSb} and
\cite{congy_nonlinear_2019} for details.  The HNLS modulation solution
provides the best agreement with the numerical DSW structure near the
harmonic edge and deeper into the DSW interior.  Near the soliton
edge, there is a departure from the HNLS description, which is to be
expected because the (H)NLS descriptions are based upon modulations of
a weakly nonlinear Stokes wave, which does not describe the DSW in the
strongly nonlinear, or solitary wavetrain regime.
\begin{figure}[h]
\begin{subfigure}[t]{.33\textwidth}
\centering
\includegraphics{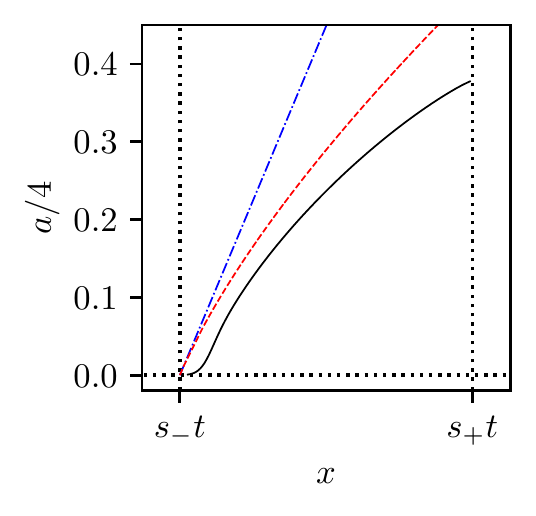}
\caption{amplitude}
\label{fig:BBM_amplitude_structure}
\end{subfigure}%
\hfill
\begin{subfigure}[t]{.33\textwidth}
\centering
\includegraphics{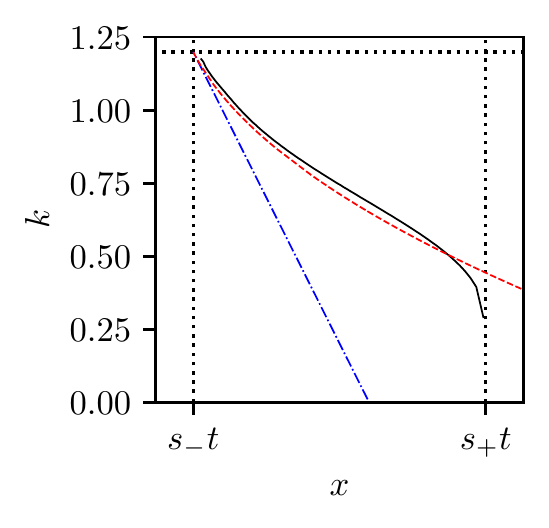}
\caption{wavenumber}
\label{fig:BBM_wavenumber_structure}
\end{subfigure}%
\hfill
\begin{subfigure}[t]{.33\textwidth}
\centering
\includegraphics{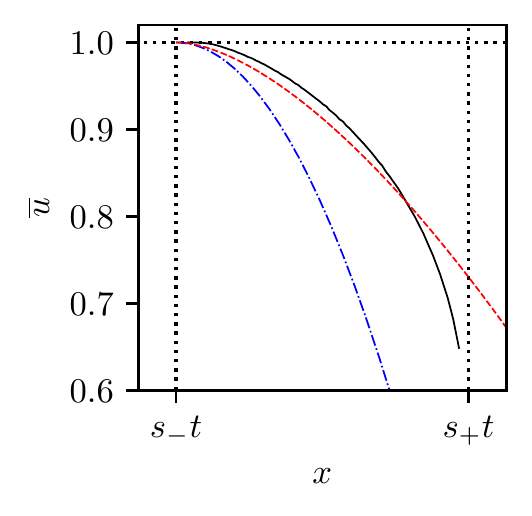}
\caption{mean value}
\label{fig:BBM_mean_structure}
\end{subfigure}
\caption{Comparison between the asymptotic prediction for the DSW
  structure and the numerical solution of the smoothed dispersive
  Riemann problem~\eqref{bbm}, \eqref{step} at $t=1000$ with
  $(u_-,u_+,\xi)=(1,0.2,10)$. The black solid line is the numerical
  solution, the dash-dotted blue line is the NLS prediction and the
  dashed red line is the HNLS prediction.}
\label{fig:BBM}
\end{figure}

The wavenumber and mean value predictions in Figures
\ref{fig:BBM_wavenumber_structure}, \ref{fig:BBM_mean_structure}
visually exhibit better agreement with the numerically extracted DSW
parameters than the DSW amplitude in
\ref{fig:BBM_amplitude_structure}.  There is some deviation in the
numerically extracted oscillation envelope amplitude depicted in
Fig.~\ref{fig:BBM_amplitude_structure} near the trailing edge from
predicted linear behavior $a \propto \frac{x}{t} - s_-$ in
\eqref{DSW_sol}. This is not surprising given that our analysis in
Section \ref{linear_wave} demonstrates that the transition width for
smoothed step initial data significantly alters the DSW's oscillatory
structure in the neighborhood of the DSW harmonic edge.  In fact, a
fundamental discrepancy between the DSW modulation solution and the
numerical oscillation behavior is known and even occurs for the KdV
equation \cite{grava_numerical_2008}.  In contrast to the well-defined
DSW soliton edge, the DSW harmonic edge discrepancy from leading order
modulation theory leads to ad hoc approaches for comparing the
harmonic edge velocity and wavenumber with numerical simulations.
Nevertheless, modulation theory provides a reasonable prediction of
the DSW's amplitude, wavenumber, and mean structure near the DSW
harmonic edge.

\subsection{DSW implosion}
\label{implosion}

\begin{figure}[h]
  \centering
  \begin{subfigure}[t]{.5\textwidth}
    \centering
    \includegraphics{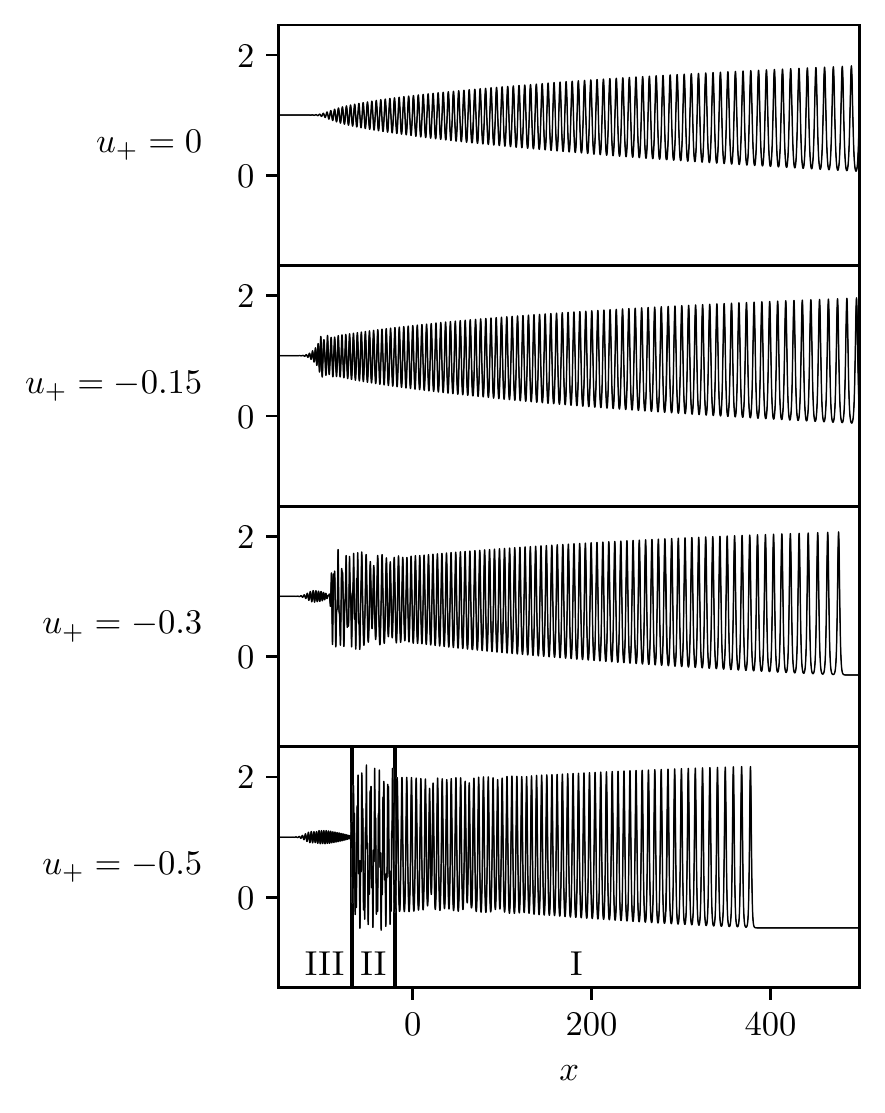}
    \caption{Variation of the whole solution.}
  \end{subfigure}%
  \begin{subfigure}[t]{.5\textwidth}
    \centering
    \includegraphics{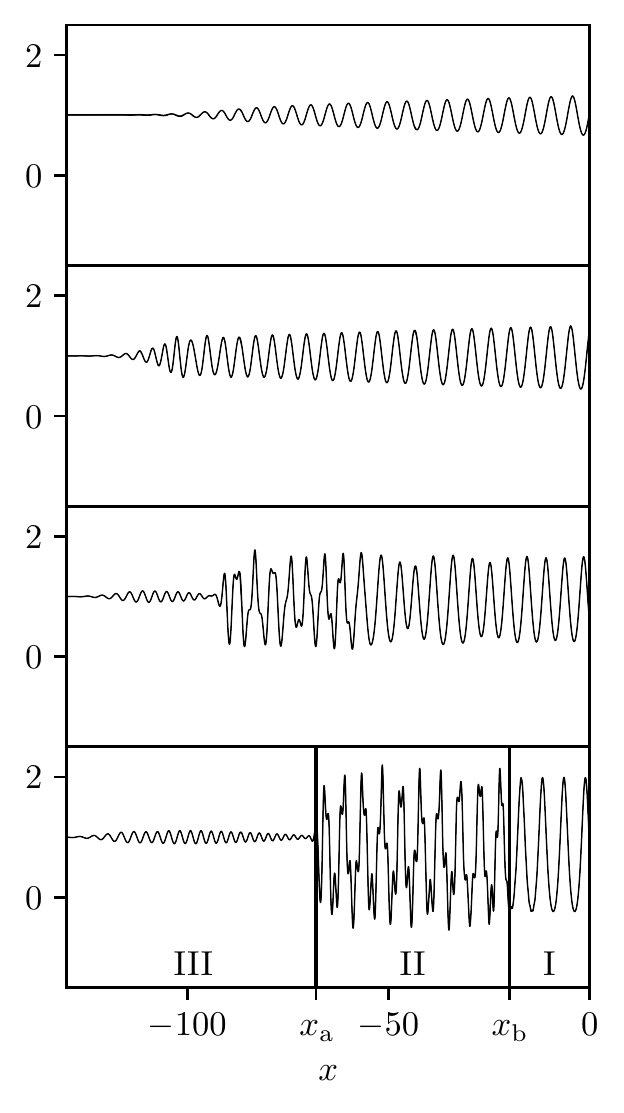}
    \caption{Close up of the region: $x<0$.}
  \end{subfigure}
  \caption{Numerical solution of the dispersive Riemann problem with
    $u_+\leq 0<u_-=1$ at $t=1000$. When $u_+ \ll \mu u_-$, the
    solution divides into three distinct regions identified in the
    example $u_+=-0.5$. Vertical solid lines correspond to the left
    and right boundaries of region II: $x=x_{\rm a}(t)$ and $x=x_{\rm
      b}(t)$.}
  \label{fig:implosion}
\end{figure}

Figure \ref{fig:implosion} shows that the solution of the dispersive
Riemann problem is no longer a single-phase DSW when $u_+<\mu u_-$. A
beating pattern develops close to the DSW small-amplitude edge that is
clearly depicted in the examples for which $u_+\leq -0.3$. This
phenomenon has been previously identified
\cite{lowman_dispersive_2013} as DSW implosion in which the group
velocity of linear waves at the harmonic edge experiences a minimum,
i.e.~at a zero dispersion point.  In this section, we investigate the
solution for $u_+<\mu u_-$ in greater detail. In the numerical
examples presented in this section, $\xi=10$ unless otherwise stated.
The variation of a typical solution when $u_+=-0.5$ in
Fig.~\ref{fig:implosion} shows that 3 distinct modulation regions
exist: a finite amplitude, stable, single-phase wave in region~I, a
beating pattern in region II and a small amplitude wave in region III;
the boundaries between regions I and II denoted $x=x_{\rm b}$, and
regions II and III denoted $x=x_{\rm a}$, separate the spatial regions
where the wave is single-phase and two-phase.

\begin{figure}[h]
  \centering
  \includegraphics{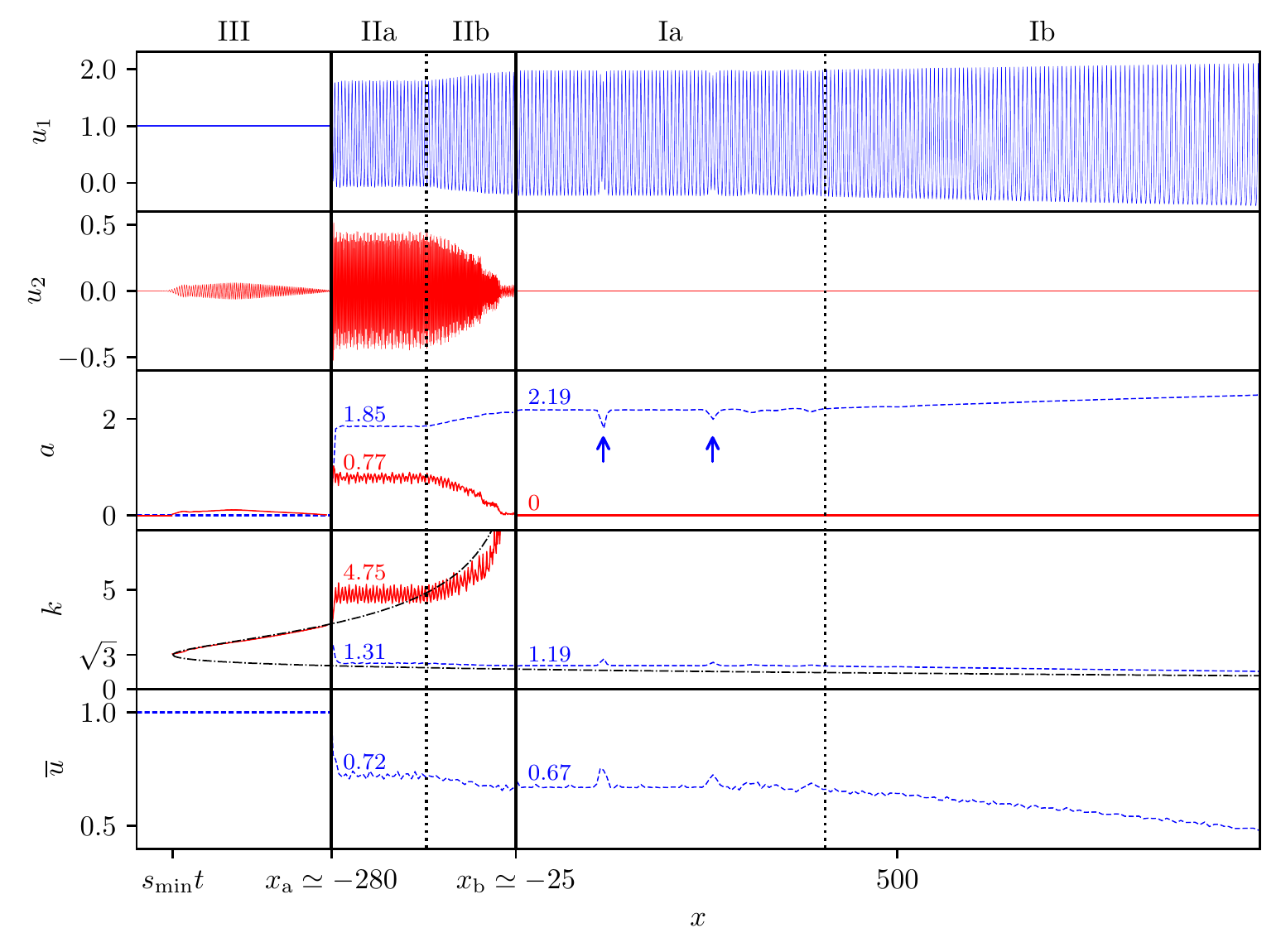}
  \caption{Variation of the slow and fast oscillatory fields $u_1$ and
    $u_2$, respectively, extracted from the numerical solution
    $u(x,t)$ at $t=4000$ when $(\xi,u_-,u_+)=(10,1,-0.5)$. Variation
    of the fields $u_1$, $u_2$ are displayed in the first, second rows
    with dashed, solid curves, respectively. Their extracted amplitude
    $a_1,a_2$ and wavenumber $k_1,k_2$ modulations are shown in the
    third and fourth rows, respectively with dashed (resp.~solid)
    curves corresponding to parameters for the field $u_1$
    (resp.~$u_2$).  The dash-dotted curve in the fourth row is the
    linear, stationary phase solution~\eqref{k1},\eqref{k2}.  The
    fifth row depicts the mean of the solution $\bar{u}$.  Solid
    vertical lines represent the boundaries $x_{\rm a}$ and $x_{\rm
      b}$ and dash-dotted vertical lines identify internal
    subdivisions of regions I (Sec.~\ref{regionI}) and II
    (Sec.~\ref{regII}). Vertical arrows indicate the position of dark
    solitary waves in region Ia.}
\label{fourier5b}
\end{figure}

\subsubsection{Region I: stable single-phase wave}
\label{regionI}

We first consider the modulation of the wave in region I. The
variations of the crest to trough amplitude $a(x,t)$, the wavenumber
$k(x,t)$, and the mean $\overline{u}(x,t)$ obtained numerically are
displayed in Fig.~\ref{fourier5b} (where $u_1=u$ in region~I, as
detailed in Sec.~\ref{regII}).  Region I can be subdivided into two
subregions: a region denoted Ia where $a(x,t)$, $k(x,t)$, and
$\bar{u}(x,t)$ are approximately constant, apart from localized,
small-width modulations, and a region denoted Ib where $a(x,t)$,
$k(x,t)$, and $\overline{u}(x,t)$ vary monotonically. The modulation
in region Ib is similar to the DSW modulation investigated in
Secs.~\ref{sec:dsw_gen} and~\ref{sec:BBM_fitting} where the wavenumber
$k(x,t)$ monotonically decreases until it reaches the solitary wave
limit $k=0$, while the amplitude monotonically increases toward the
soliton edge. However, the DSW fitting of Sec.~\ref{sec:dsw_gen} does
not apply to this region since the admissibility
condition~\eqref{restr_reg1} is not fulfilled, and the parameters of
the soliton edge cannot be obtained using this method.  In fact, the
wave structure in region Ib can be identified as a \textit{partial
  DSW}, where the nonlinear modulated wave does not reach the
small-amplitude, harmonic limit.  Instead, it terminates at some
nonzero amplitude $a_0$, mean $\bar{u}_0$, and wavenumber $\kappa_0$
that are matched to the right edge of region Ia.  These types of
structures occur in initial-boundary value problems for nonlinear
dispersive PDEs (see, e.g.~\cite{marchant_initial-boundary_1991}) and,
in particular, are realized in problems involving resonant or
transcritical shallow-water flows past localized bathymetry
\cite{grimshaw_resonant_1986,smyth_modulation_1987,el_transcritical_2009}. We
do not present a quantitative description of partial BBM DSWs here as
this would require the development of a full nonlinear modulation
theory for the BBM equation, a significant undertaking that deserves a
dedicated study.

The specific modulation of the wave in region Ia is a new feature that
only develops when $\sigma_3 u_- < u_+<\mu u_-$. The localized
depletions of the amplitude $a(x,t)$ and the corresponding
augmentations of the wavenumber $k(x,t)$ propagate on the
``background'' $(a,k)=(a_0,\kappa_0)$ at constant velocity; we have,
for instance in the example of Fig.~\ref{fourier5b} where $u_+=-0.5$,
$(a_0,\kappa_0) \approx (2.19,1.19)$. In the following, we investigate
the modulation of the field $u(x,t)$ using a weakly nonlinear NLS
description that was introduced in the previous sections.  The
nonlinear wave can be approximated in the weakly nonlinear regime by
the Stokes wave (cf.~for instance \eqref{stokes})
\begin{equation}
%\label{stokes2}
u(x,t) \sim \ubar(x,t) + \left[ A(x,t) e^{i(\kappa_0 x
-\omega_0(\kappa_0,u_-) t)} + \cc \right],
\end{equation}
where $\ubar(x,t) \sim u_-$ is a slowly varying field. The crest to
trough amplitude $a$ and the wavenumber $k$ are given by
\begin{align}
  a(x,t) = 4|A(x,t)|,\;
  k(x,t) = \kappa_0 + \partial_x \arg A.
\end{align}
Since $\kappa_0 < \sqrt 3$, the complex envelope of the Stokes wave
$A(x,t)$ solves the defocusing NLS equation
($\beta(\kappa_0,u_-)\gamma(\kappa_0,u_-) < 0$):
\begin{equation}
  \label{NLS2}
  i A_t + i \partial_k\omega_0(\kappa_0,u_-) A_x + \beta(\kappa_0,u_-)
  A_{xx} +\gamma(\kappa_0,u_-) |A|^2 A=0.
\end{equation}

\begin{figure}[h]
\centering
\begin{subfigure}[t]{.5\textwidth}
\centering
\includegraphics{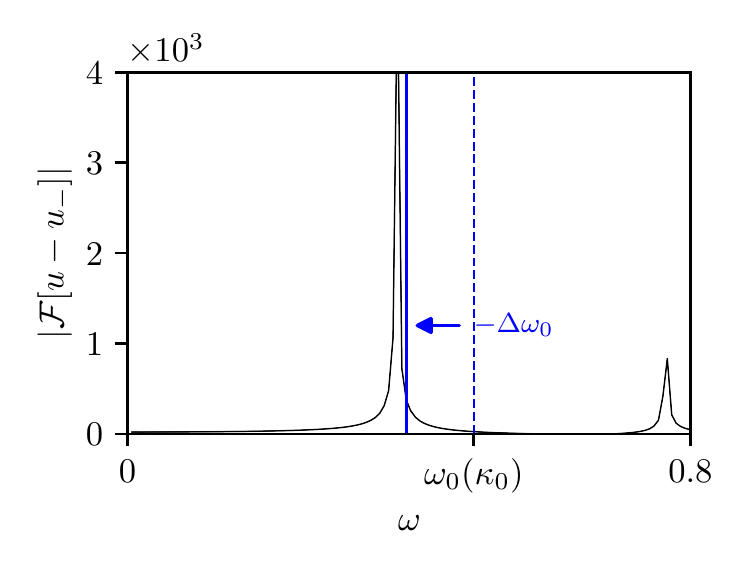}
\caption{$x=0$ (region Ia)}
\label{omega2}
\end{subfigure}%
\begin{subfigure}[t]{.5\textwidth}
\centering
\includegraphics{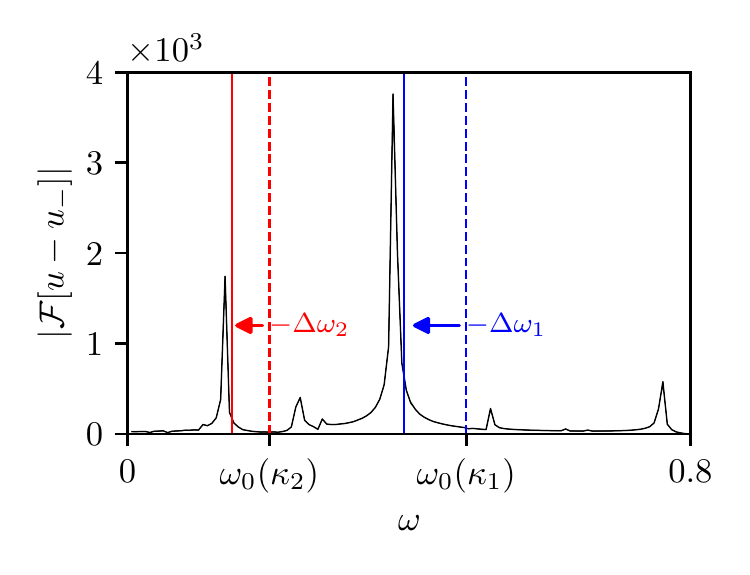}
\caption{$x=-300$ (region IIa)}
\label{omega}
\end{subfigure}
\caption{Temporal Fourier transform of the numerical solution
  $u(x,t)-u_-$ [$(u_-,u_+,\xi)=(1,-0.5,10)$] at a fixed position $x$
  and the time-sampling $t \in (5000, 6000)$. Peaks of the Fourier
  transform indicate the dominant frequency of the weakly nonlinear
  wave. Dashed vertical lines indicate the linear frequencies
  $\omega_0(\kappa_i)$ and solid vertical lines the shifted
  frequencies $\omega_0(\kappa_i)-\Delta \omega_i$.}
\end{figure}

We first consider the nonlinear plane wave solution
\begin{equation}
  \label{acst2}
  a = a_0={\rm const},\quad k = \kappa_0 =  {\rm const}.
\end{equation}
Substituting~\eqref{acst2} into the NLS equation~\eqref{NLS2}, we
obtain the modulation of the complex envelope
\begin{equation}
\label{deltaw2}
A(x,t) = \frac{a_0}{4} \exp\left(i \Delta \omega_0 t
\right),\quad \Delta \omega_0 = \gamma(\kappa_0,u_-) \frac{a_0^2}{16}.
\end{equation}
As a result, the weakly nonlinear wave in region Ia is approximately
\begin{equation}
  \label{shift}
  u(x,t) = \ubar + \frac{a_0}{4} \cos\left[ \kappa_0 x -
  (\omega_0(\kappa_0,u_-)-\Delta \omega_0) t\right],\quad
  \ubar = u_- + b(\kappa_0,u_-) \frac{a_0^2}{16}
\end{equation}
showing that weakly nonlinear interaction shifts the linear frequency
$\omega_0(\kappa_0,u_-)$ by $-\Delta \omega_0$.  Figure \ref{omega2}
displays the frequency shift computed numerically for the example
$u_+=-0.5$, and shows reasonable agreement with the NLS equation
description~\eqref{NLS2}. Equation \eqref{shift} also yields $\ubar =
0.77$ whereas the average extracted numerically is $\ubar \approx
0.67$ as shown in Fig.~\ref{fourier5b}. A better analytical
approximation of $\Delta \omega_0$ and $\ubar$ necessitates higher
order terms in the NLS description.

Localized perturbations of the parameters $a(x,t)$ and $k(x,t)$ can be
identified with dark solitary wave solution of the NLS
equation~\eqref{NLS2}
\begin{equation}
\label{dark_sol}
A_{\rm DS}(x,t) = \frac{a_0}{4} \left( i \,c\, \sqrt{\frac{8}{-\beta_0
\gamma_0 a_0^2}} + \sqrt{1+\frac{8c^2}{\beta_0 \gamma_0 a_0^2} }\tanh
\left[ \sqrt{-\frac{\gamma_0 a_0^2}{32 \beta_0}-\frac{c^2}{4\beta_0^2}
} (x-ct-x_0) \right] \right) {\rm e}^{i a_0^2 t/16},
\end{equation}
where $\beta_0 = \beta(\kappa_0,u_-)$ and $\gamma_0 =
\gamma(\kappa_0,u_-)$ (cf.~\eqref{coeff_NLS}).  The dark solitary wave is
parameterized by the speed of the solitary wave $c$ and $x_0$, its position
at $t = 0$.  The phase of the dark solitary wave experiences a shift when
the localized amplitude depression is traversed
\begin{equation}
  \label{delta_DS}
  \lim_{x\to +\infty} \arg A_{\rm DS}(x,t) -
  \lim_{x\to -\infty}\arg A_{\rm DS}(x,t) = -2 {\rm sgn}(\beta_0 c)
  \cos^{-1} \left( \frac{4
      c}{\sqrt{-2\beta_0 \gamma_0 a_0^2}}\right).
\end{equation}
This phase-shift can be observed in the numerical solution but differs
from the analytical formula~\eqref{delta_DS}. This indicates that the
wave in region I is governed by the NLS equation only to a first
approximation, and the description of traveling waves of the envelope
$A(x,t)$ necessitates a higher order description such as the HNLS
equation~\eqref{HNLS}. We leave this comparison as a subject for
future work.

The emergence of envelope solitary waves is reminiscent of the
shedding of solitary waves investigated in Sec.~\ref{shedding} because
solitary waves are only generated for a sufficiently large jump
$|u_--u_+|$.

\subsubsection{Region II: stable two-phase wave}
\label{regII}

Figure \ref{fig:implosion} shows that the implosion of the DSW occurs
in regions II and III, which broadly overlap with the two-phase region
$x \in (s_{\rm min} t,0)$ identified in the linear analysis of
Sec.~\ref{linear_wave}.  In this section, we focus on region II. We
show that $u(x,t)$ displays a beating pattern that can be approximated
by a superposition of two single-phase, weakly nonlinear waves
\begin{equation}
  %\label{u1u2}
  u(x,t) \sim u_1(x,t)+u_2(x,t),
\end{equation}
were $u_1$ and $u_2$ correspond, respectively, to slow and fast
oscillating waves.

\begin{SCfigure}[][h]
  \centering
  \includegraphics{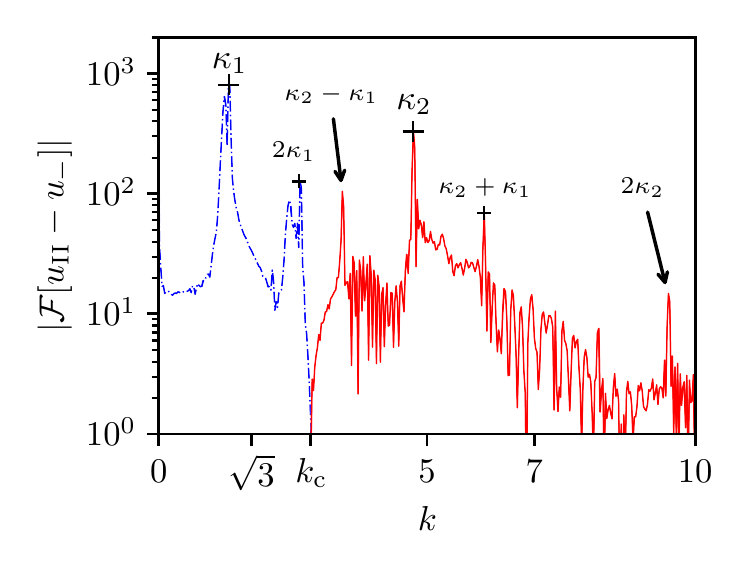}
  \caption{Fourier transform of $u_{\rm II}(x,t)-u_-$ for
    $(\xi,u_-,u_+) = (10,1,-0.5)$ at $t=4000$.  $k_{\rm c}
    \sim 2.83$ divides the spectrum in a small-wavenumber region
    (dashed blue line) and large-wavenumber region (solid red
    line). The crosses $+$ identified the dominant wavenumbers of the
    component $u_1$ and $u_2$ as well as the corresponding higher
    order harmonics.}
\label{fourier1}
\end{SCfigure}

In order to determine $u_1$ and $u_2$ from our numerical simulations,
we compute the spatial Fourier transform of $u_{\rm II}(x,t)$, which
is the restriction of $u(x,t)$ to region II.  The example $u_+=-0.5$
is shown in Fig.~\ref{fourier1} depicting multiple maxima of the
Fourier transform. The two largest spectral contributions correspond
to the dominant wavenumber $\kappa_1$ of the slowly oscillating wave
$u_1$ and the dominant wavenumber $\kappa_2>\kappa_1$ of the fast
oscillating wave $u_2$. Other maxima correspond to harmonics that
result from weakly nonlinear interactions such as second harmonic
generation at $2\kappa_1$ and $2\kappa_2$, and two wave interaction
$\kappa_2 \pm \kappa_1$.  We then identify the small-wavenumber
content of the Fourier spectrum ($k<k_{\rm c}$) as the Fourier
transform of $u_1(x,t)$ and the large-wavenumber content as the
Fourier transform of $u_2(x,t)$, where $k_{\rm c}$ is the minimum
identified for $\kappa_1 < k < \kappa_2$ ($k_{\rm c} \approx 2.8$ in
the example depicted in Fig.~\ref{fourier1}). The variation of the
fields $u_1$ and $u_2$ is obtained by computing the inverse Fourier
transforms of the band limited functions
\begin{equation}
  %\label{u1u2}
  u_1(x,t) = {\cal F}^{-1} \left[
    \begin{cases}
      {\cal F}[u_{\rm II}(x,t)], &k<k_{\rm c}\\
      0, & k \geq k_{\rm c}
    \end{cases}
  \right],\quad
  u_2(x,t) = {\cal F}^{-1} \left[
    \begin{cases}
      0, &k<k_{\rm c}\\
      {\cal F}[u_{\rm II}(x,t)], & k \geq k_{\rm c}
    \end{cases}
  \right].
\end{equation}
The division into small and large wavenumber bands of the Fourier
spectrum was examined in Sec.~\ref{linear_wave} for linear waves.
There, $k_{\rm c}=\sqrt 3$ separated the small and large wavenumber
bands.  In the weakly nonlinear regime, $k_{\rm c} \neq \sqrt 3$
because the second harmonic $2\kappa_1$ can be larger than $\sqrt 3$
($2\kappa_1 \simeq 2.6 $ in the example presented in
Fig.~\ref{fourier1}). The variation of $u_1(x,t)$ and $u_2(x,t)$
extracted from the numerical solution for $u_-=-0.5$ and their
modulation parameters in region II are depicted in
Fig.~\ref{fourier5b}. Note that the numerical method presented here to
extract the two single-phase waves only yields an approximation to
$u_1$ and $u_2$.  For instance, the higher harmonic
$\kappa_2-\kappa_1$ contributes to the Fourier spectrum of $u_2$ in
the example considered in Fig.~\ref{fourier1}, even though it
corresponds to the nonlinear interaction between $u_1$ and $u_2$. In
practice, the quasi-monochromatic aspect of the extracted waves $u_1$
and $u_2$ (cf.~Fig.~\ref{fourier5b}) indicates that the numerical
method presented above yields a good approximation.

We now make the ansatz that the two modulated wavetrains $u_1$ and
$u_2$ can be modeled by two Stokes waves
\begin{equation}
  \label{stokes2}
  \begin{split}
    u_1(x,t) = \left[ A_1(x,t) e^{i(\kappa_1 x -\omega_0(\kappa_1,u_-)
        t)} + \cc \right] + \ubar(x,t),\; u_2(x,t) = A_2(x,t)
    e^{i(\kappa_2 x -\omega_0(\kappa_2,u_-) t)} + \cc
    ,\\
  \end{split}
\end{equation}
where $A_1(x,t),A_2(x,t)$ are complex valued and $\ubar(x,t)$ is real
valued, all of which are slowly varying.  The dominant wavenumbers of
the Stoke waves satisfying $\kappa_1<\sqrt 3$ and $\kappa_2 > \sqrt 3$
are extracted numerically using the procedure already described.  Note
that by the definition of $u_1$ and $u_2$, the slow modulation of the
background $\ubar(x,t)$ is necessarily incorporated in the variation
of $u_1(x,t)$ so that $u_2(x,t)$ has zero mean. The crest to trough
amplitudes $a_j$ and wavenumbers $k_j$ are
\begin{align}
  a_j(x,t) = 4|A_j(x,t)|, \quad k_j(x,t) = \kappa_j + \partial_x \arg
  A_j, \quad j = 1,2.
\end{align}
The modulation equations for the complex envelopes $A_1$ and $A_2$ are
two coupled NLS equations
\begin{equation}
  \label{coupled_NLS}
  \begin{split}
    &i A_{1t} + i \partial_k \omega_0(\kappa_1,u_{-}) A_{1x} +
    \beta(\kappa_1,u_-) A_{1xx} + \left[\gamma(\kappa_1,u_-) |A_1|^2 +
      \nu(\kappa_1,\kappa_2,u_-) |A_2|^2 \right] A_1=0,\\
    &i A_{2t} + i \partial_k \omega_0(\kappa_2,u_{-}) A_{2x} +
    \beta(\kappa_2,u_-) A_{2xx} + \left[\gamma(\kappa_2,u_-) |A_2|^2 +
      \nu(\kappa_2,\kappa_1,u_-) |A_1|^2 \right] A_2=0,\\
    &\ubar(x,t) = u_-+ b(\kappa_1,u_-) |A_1|^2 + b(\kappa_2,u_-)
    |A_2|^2,
  \end{split}
\end{equation}
obtained with a standard multiple-scale computation. The coefficients
$\beta(\kappa,u_-)$ and $\gamma(\kappa,u_-)$ are given
in~\eqref{coeff_NLS} and $b(\kappa,u_-)$ in~\eqref{b_BBM}. The
nonlinear interaction between the two components $A_1$ and $A_2$ is
characterized by the coefficient
\begin{equation}
  \nu(\kappa,\lambda,u_-) = \frac{\kappa (1 + \lambda^2) (9 + 6 \kappa^2 +
    \kappa^4 + 21 \lambda^2 + 13 \kappa^2 \lambda^2 + \kappa^4
    \lambda^2 + 9 \lambda^4 + 3 \kappa^2 \lambda^4 + \lambda^6)}{(1
    + \kappa^2) \lambda^2 (3 + \lambda^2) (3 + \kappa^2 - \kappa
    \lambda + \lambda^2) (3 + \kappa^2 + \kappa \lambda +
    \lambda^2) u_-} > 0.
\end{equation}
The modulation description~\eqref{stokes2}, \eqref{coupled_NLS} is the
weakly nonlinear extension of the modulated two-phase linear wave
solution~\eqref{sol3}, which applies when the two wavenumbers $k_1$
and $k_2$ are distinct.

As indicated in Fig.~\ref{fourier5b}, Region II can be divided into
two subregions: region IIa where the wave parameters are approximately
constant, and region IIb where the variations of $a_j(x,t)$, $j = 1,2$
are approximately monotone.

We first focus on the modulation in region IIb where the amplitudes
$a_1(x,t)$ and $a_2(x,t)$ are, respectively increasing and decreasing
functions of $x$. Ultimately $a_2(x,t)=0$ at the boundary with region
I ($x=x_{\rm b}$) and the amplitude of the slow varying wave
$a_1(x,t)$ matches with $a(x,t) = a_0$ in region Ia. We thus extend the
definition of $u_1$ and $u_2$ to region I by setting:
\begin{equation}
u_1(x,t) = u(x,t),\quad u_2(x,t) = 0,\quad \forall x \geq x_{\rm b}.
\end{equation}
The modulation close to $x=x_{\rm b}$ can thus be uniformly described,
across the boundary between the region IIb and Ia, by the simplified
coupled equations
\begin{equation}
\label{coupled_NLS2}
\begin{split}
&i A_{1t} + i \partial_k \omega_0(\kappa_1,u_{-}) A_{1x} +
\beta(\kappa_1,u_-) A_{1xx} + \gamma(\kappa_1,u_-) |A_1|^2 A_1=0,\\
&i A_{2t} + i \partial_k \omega_0(\kappa_2,u_{-}) A_{2x} +
\beta(\kappa_2,u_-) A_{2xx} + \nu(\kappa_2,\kappa_1,u_-) |A_1|^2 A_2=0.
\end{split}
\end{equation}
The approximation of \eqref{coupled_NLS} invoked here is $|A_2| \ll
|A_1|$, as suggested by Fig.~\ref{fourier5b}.  The modulation of $A_1$
is independent of $A_2$ and is governed by the, modulationally stable,
defocusing NLS equation as described in Sec.~\ref{regionI}. The
modulation of $A_2$ is a linear Schr\"odinger equation with potential
$-\nu(\kappa_2,\kappa_1,u_-)|A_1(x,t)|^2$. Consequently, although the
wavenumber $\kappa_2$ is larger than $\sqrt 3$, the modulation
dynamics of $u_2$ remain stable in the vicinity of $x=x_{\rm b}$.
This highlights the stabilizing effect of the slowly-oscillating
component $u_1(x,t)$ on the fast-oscillating component $u_2(x,t)$ in
region IIb.  We do not carry out a detailed analysis of the
modulations $A_j$ in region IIb, leaving that for a separate,
dedicated study.

We now consider region IIa for which the modulation variables are
approximately constant
\begin{equation}
  \label{acst}
  a_1  = {\rm const},\quad a_2 = {\rm const},\quad
  k_1 = \kappa_1,\quad k_2  =\kappa_2,
\end{equation}
where we identify the modulation wavenumber $k_j$ with the constant,
dominant wavenumber $\kappa_j$ extracted from the numerical simulation
so that $\partial_x \arg A_j=0$. The values of the parameters $a_j$
are also extracted from the numerical solution.  In
Fig.~\ref{fourier5b}, we have $(a_1,k_1) \approx (1.85,1.31)$ and
$(a_2,k_2) \approx (0.77,4.75)$. Substituting~\eqref{acst} into the
coupled NLS equations~\eqref{coupled_NLS}, we obtain the modulation of
the complex envelopes in the form of two plane waves
\begin{equation}
  \label{deltaw}
  \begin{split}
    &A_1(x,t) = \frac{a_1}{4} \exp\left(i \Delta \omega_1 t
    \right),\quad \Delta \omega_1 =
    \frac{1}{16}\left(\gamma(\kappa_1,u_-) a_1^2 + 
      \nu(\kappa_1,\kappa_2,u_-) a_2^2 \right),\\
    &A_2(x,t) = \frac{a_2}{4} \exp\left( i \Delta \omega_2 t \right),\quad
    \Delta \omega_2 = \frac{1}{16} \left(\gamma(\kappa_2,u_-) a_2^2 +
      \nu(\kappa_2,\kappa_1,u_-) a_1^2 \right).
  \end{split}
\end{equation}
As a result, the two-phase weakly nonlinear wave in region IIa for the
BBM equation is approximately
\begin{equation}
\label{shift2}
\begin{split}
&u(x,t) = \ubar + \frac{a_1}{4} \cos\left[ \kappa_1 x -
    (\omega_0(\kappa_1,u_-)-\Delta \omega_1) t\right] + \frac{a_2}{4}
  \cos\left[ \kappa_2 x -
  (\omega_0(\kappa_2,u_-)-\Delta \omega_2) t\right],\\
  & \ubar = u_- + b(\kappa_1,u_-) \frac{a_1^2}{16}+ b(\kappa_2,u_-)
  \frac{a_2^2}{16} 
\end{split}
\end{equation}
showing that weakly nonlinear interactions between the two wave modes
shift the linear frequencies $\omega_0(\kappa_1,u_-)$ and
$\omega_0(\kappa_2,u_-)$ by $-\Delta \omega_1$ and $-\Delta\omega_2$,
respectively.  Figure \ref{omega} displays good agreement between the
weakly nonlinear frequency predictions \eqref{shift} and the temporal
Fourier spectrum of the BBM numerical simulation at $x = -300$ for the
case $u_+ = -0.5$.  Equation \eqref{shift2} also yields $\ubar = 0.86$
whereas the average extracted numerically in Fig.~\ref{fourier5b} is
$\ubar \approx 0.72$.  A better approximation of $\Delta \omega_1$,
$\Delta \omega_2$ and $\ubar$ could be achieved with higher order
terms in the NLS description.

It is important to point out that, in the absence of two-phase
coupling, the modulation $A_2$ satisfies the scalar NLS equation
\eqref{NLS2} with $\kappa_0 \to \kappa_2$.  Since $\kappa_2 >
\sqrt{3}$, $\beta(\kappa_2,u_-)\gamma(\kappa_2,u_-) > 0$ so that the
equation is of focusing type and the plane wave solution \eqref{acst2}
(with $a_0 \to a_2$) for $A_2$ is linearly unstable to infinitesimal,
long wavelength perturbations, also known as modulationally unstable.
However, the nonlinear coupling between the modulations in the coupled
NLS equations \eqref{coupled_NLS} can stabilize the plane wave
solution \eqref{deltaw}.  For the specific example shown in region IIa
of Fig.~\ref{fourier5b}, we have verified that this is indeed the case
with a calculation described in Appendix \ref{sec:stab-nonl-two}.

\subsubsection{Region III: linear wave}
\label{regIII}

Since the carrier wavenumber of the field $u(x,t)$ is larger than
$\sqrt 3$ in region III---cf.~Fig.~\ref{fourier5b} where
$(u_1,u_2)=(u_-,u-u_-)$ in region III---$u$ represents a fast
oscillating wave according to the criterion introduced in
Sec.~\ref{regII}, and we set
\begin{equation}
  u_1(x,t) = u_-,\quad u_2(x,t) = u(x,t),\quad \forall x < x_{\rm a}.
\end{equation}
Contrary to the neighborhood of the boundary $x=x_{\rm b}$, the fields
$u_1$ and $u_2$ are rapidly changing at the boundary $x=x_{\rm a}$.

The amplitude of $u(x,t)$ in region III is small and its modulation
can be described by the linear theory introduced in
Sec.~\ref{linear_wave}. The bottom row of Fig.~\ref{fourier5b} shows
that the modulation of the wavenumber $k(x,t)$ obtained numerically
lies on the stationary point $k_2(x/t)$ from linear theory given
by~\eqref{k2}.

\subsubsection{Summary of DSW implosion}

Here we summarize our description of DSW implosion for dispersive
Riemann problems with $\sigma_3u_- < u_+ < \mu u_-$ in regions c and f
of Figs.~\ref{fig:xi10} and \ref{fig:xi01}, respectively ($\sigma_3$
is defined in the next subsection).  The entire coherent wave
structure arises due to compressive initial data that cannot be
resolved by an admissible DSW alone.  Instead, both short and long
waves are generated that represent a nonlinear generalization of the
two-phase, linear wave modulation investigated in
Sec.~\ref{linear_wave}.  From right to left, the solution is described
by modulated long waves in a partial DSW (region I) that transitions
to region II where long and short waves coexist, significantly
influencing one another.  This two-phase modulation abruptly
terminates at a short wavelength, linear wavepacket (region III)
propagating on the background $u_-$, hence the entire structure
transitions from modulated, long nonlinear waves to modulated, short
linear waves.

Similar to region I identified in linear theory, the nonlinear wave in
region I is a modulated one-phase wave with a small wavenumber ($k<
\sqrt 3$).  The one-phase wave in region Ib is a partial DSW, a
carryover from dispersive Riemann problems with $\mu u_- < u_+ < u_-$
in which DSWs are admissible.  Its wave parameters $a$, $k$, and
$\overline{u}$ are strictly monotone functions of $x$ in this region.
Adjacent to the partial DSW is a stable nonlinear plane wave with
multiple localized, depression modulations of the wave envelope (dark
envelope solitary waves) that stably propagate in region Ia.

The wavetrain in region Ia matches smoothly to a nonlinear beating
pattern in region II at $x=x_{\rm b}$.  In a first approximation,
$u(x,t)$ is the superposition of two modulated, weakly nonlinear,
monochromatic (Stokes) waves $u_1(x,t)$ and $u_2(x,t)$, each with
distinct dominant wavenumbers $\kappa_1<\sqrt 3$ and $\kappa_2>\sqrt
3$.  This is a generalization of the superposition of two modulated,
linear wavetrains with distinct dominant wavenumbers (see region II
described in Sec.~\ref{linear_wave}). The weakly nonlinear components
$u_1$ and $u_2$ are approximately governed by coupled NLS equations,
as demonstrated by their nonlinear frequency shifts.  Although a
weakly nonlinear wave of dominant wavenumber larger than $\sqrt 3$ is
modulationally unstable, the component $u_2(x,t)$ is shown to be
stable when nonlinearly coupled to $u_1(x,t)$.

Region III displays a principal difference from linear theory.  In the
linear theory, the two wavenumbers $k_1$ and $k_2$ coalesce at $\sqrt
3$ when $x=-u_-/8 t$, and the corresponding linear wave $u(x,t)$
continuously changes from a beating two-wave structure to an Airy
wave. In the dispersive Riemann problem considered here, the
modulation changes very rapidly, in fact, discontinuously on the
modulation length scale at $x=x_{\rm a}$.  The adjacent wave in region
III is a linear, one-phase wave with wavenumber $k_2>\sqrt 3$. The
discontinuity at $x=x_{\rm a}$ displayed in Fig.~\ref{fourier5b}
resembles a so-called Whitham shock wave structure that was recently
identified and investigated for several nonlinear dispersive PDEs in
\cite{sprenger_discontinuous_2020}.  Higher order dispersion was shown
to be essential for the generation of Whitham shocks.  The coupled NLS
equations \eqref{coupled_NLS} we have used to effectively describe
region II do not incorporate higher order dispersion.

We have focused in this subsection on DSW implosion resulting from
broad initial data $\xi = 10$.  Therefore, this nonclassical wave
pattern emerges during the course of nonlinear evolution and does not
require the imposition of a small scale feature in the initial data.
In fact, DSW implosion for narrow data $\xi \ll 1$ results in
essentially the same nonlinear wave coherent structure.  See
Fig.~\ref{fig:xi01}f.

\begin{figure}
\centering
\begin{subfigure}[t]{.5\textwidth}
\centering
\includegraphics{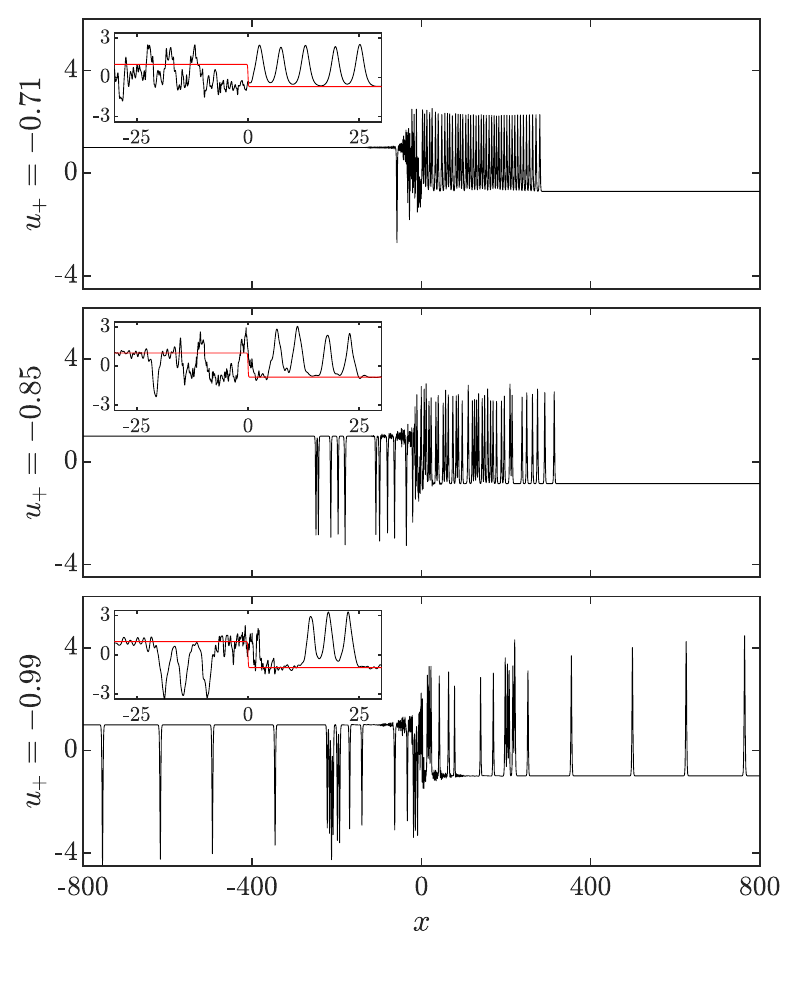}
\caption{$\xi = 0.1$}
\label{fig:ex_DSWa}
\end{subfigure}%
\begin{subfigure}[t]{.5\textwidth}
\centering
\includegraphics[clip=true,trim=13 0 0 0]{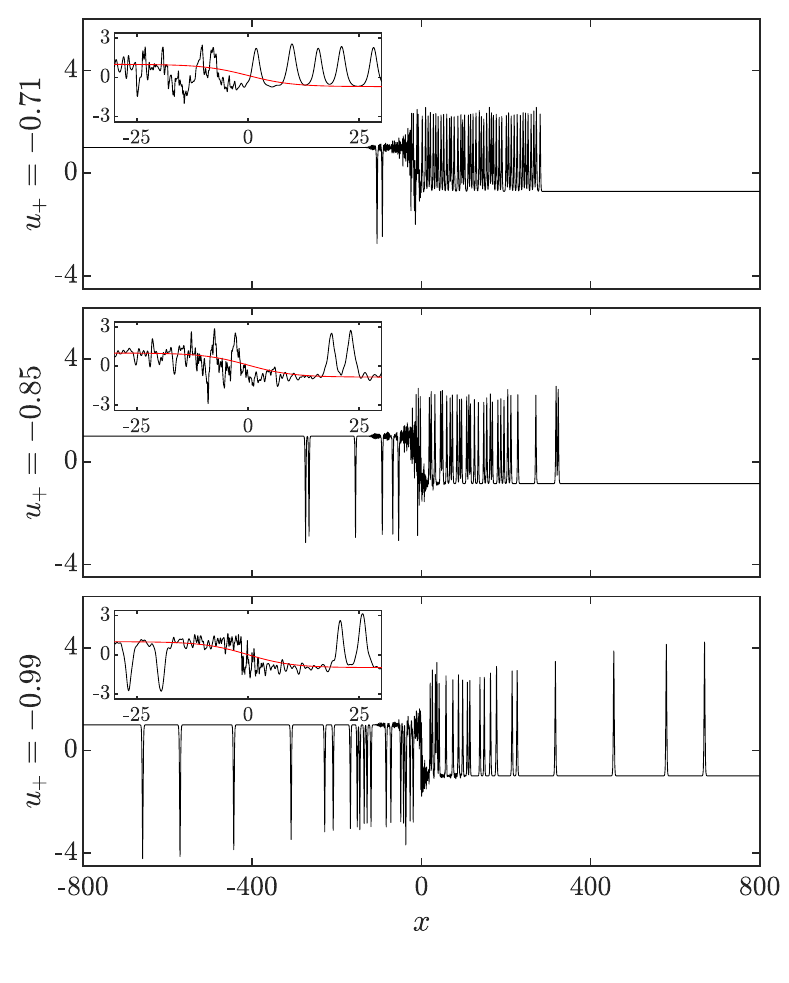}
\caption{$\xi = 10$}
\label{fig:ex_DSWb}
\end{subfigure}
\caption{Numerical solution at $t=1000$ for narrow (a) and broad (b)
  smoothed step initial data in the incoherent regime: $u_+\lessapprox
  -0.7$.  The insets are zoomed-in and include the initial condition
  (red).}
\label{ex_DSW}
\end{figure}

\subsection{Incoherent solitary wavetrain}
\label{sec:incoh-solit-wave}

As the initial smoothed step height is increased through the DSW
implosion regime that is described in the previous subsection, a new
feature emerges when $-u_- < u_+ < \sigma_3(\xi) u_-$
(cf.~Fig.~\ref{fig:xi10}d): the emission of solitary waves.  This
effect is thoroughly described in section \ref{shedding} for expansive
initial data.  We now examine solitary wave emission accompanying
compressive data.  This subsection is somewhat speculative and brief
because we base our description of the dynamics solely upon numerical
simulation.

Figure \ref{ex_DSW} depicts the numerical solution for narrow
\ref{fig:ex_DSWa} and broad \ref{fig:ex_DSWb}, smoothed step initial
data.  In the vicinity of $u_+ = -0.7$, we observe a depression
solitary wave emitted to the left edge of the two-phase oscillatory
region II (recall Fig.~\ref{fourier1}) by $t = 1000$.  Other
simulations, not shown, exhibit solitary wave emission at later times
for $u_+ \approx -0.65$.  As the right edge $u_+$ is reduced below
$-0.7$, more solitary waves, including elevation solitary waves
propagating to the right, are emitted.  These simulations depict a
gradual transition as $u_+$ is decreased from DSW implosion to the
emission of solitary waves of an increasingly incoherent character.
As $u_+$ approaches $-u_-$, the solitary waves exhibit different
amplitudes and are emitted intermittently as shown in the simulation
animations that are included as supplementary material to this
manuscript.  While the simulations for different smoothed step width
$\xi$ shown in Figs.~\ref{fig:ex_DSWa} and \ref{fig:ex_DSWb} exhibit
differences, their qualitative character is the same.  This stands in
contrast to the case of solitary wave shedding for expansive data
where we showed in section \ref{shedding} that the initial smoothed
step width must be sufficiently narrow, i.e.~the initial data must
include a sufficiently small scale feature.  We remark that
qualitatively similar dynamics were observed numerically for broad,
localized initial and periodic data in \cite{francius_wave_2001}.

These pseudospectral simulations were highly resolved.  The numerical
method is different from that utilized in other sections of the paper
(see Appendix \ref{sec:scheme-bbm-equation}) and is described in
\cite{el_dispersive_2017}.  The grid spacing $\Delta x = 0.00343$ and
timestep $\Delta t = 0.002$ ensured that the conserved energy $E(t)$
in eq.~\eqref{E} remains numerically within $10^{-10}$ of its
predicted value $E(t) = \tfrac{1}{3}(u_+^3 - u_-^3)t + E(0)$
(eq.~\eqref{consE}) for the simulated times $t \in [0,1000]$ and
domain $x \in [-900,900]$.  In all cases, the relative magnitude of
the Fourier coefficients $|\hat{u}(k,t)/\hat{u}(0,t)|$ decays to
$10^{-15}$ for sufficiently large $k$.  Furthermore, the large spatial
domain ensures that the boundary deviations from $u_\pm$ remain below
$10^{-15}$ in magnitude for the simulated times.

One important finding from the simulations in Figure \ref{ex_DSW} is
that even a slight asymmetry in the initial smoothed step ($u_+ =
-0.99 u_-$) leads to a significant departure from the symmetric
dispersive Lax shock that is generated when $u_+ = -u_-$ (recall
section \ref{sec:stat-lax-shocks}).  The animations for this case that
are included as supplementary material are particularly revealing in
this regard.  Thus, the dispersive Lax shock lacks robustness to small
changes in the boundary conditions.

\section{Conclusions and Outlook}
\label{sec:conclusions-outlook}

In this paper, we have developed a classification of solutions to the
dispersive Riemann problem for the BBM equation.  The BBM equation
\eqref{BBM_original} as a mathematical model for unidirectional,
weakly nonlinear dispersive shallow water wave propagation is
asymptotically equivalent to the celebrated KdV equation while
providing more satisfactory short-wave/high-frequency behavior in the
sense that the linear dispersion relation is bounded for the BBM
equation, but unbounded for the KdV equation.  However, the bounded
dispersion relation is nonconvex, a property that gives rise to a
number of intriguing features that are markedly different from those
found in the KdV equation.  Some of these features exemplify phenomena
previously observed in other nonlinear dispersive equations, but some
are new, providing the motivation for the study of the reduced BBM
equation \eqref{bbm} as a distinct dispersive regularization of the
Hopf equation.

The main feature of the BBM equation \eqref{bbm} that distinguishes it
from the KdV equation \eqref{kdv} is the non-evolutionary dispersive
term $-u_{xxt}$.  This term introduces an intrinsic nonlocality length
scale $\ell$ and results in a qualitative and quantitative dependence
of solutions on the comparison between $\ell$ and the typical spatial
scale $\xi$ associated with the initial condition.  The nonlocality
length $\ell$ coincides with the BBM coherence length (the unit
amplitude solitary wave width) defined by the balance of nonlinearity
and dispersion. In the normalization of \eqref{bbm}, $\ell=1$.  The
nonlocal nature of the BBM equation \eqref{bbm} imparts a bounded
linear dispersion relation $\omega=\omega_0(k,\overline{u})$ that is
nonconvex, exhibiting zero-dispersion points
$\partial_{kk}\omega_0(k,\overline{u})=0$ for particular values of the
wavenumber and background.  Generally, we find that the complex
interplay between nonlocality, nonlinearity and nonconvex dispersion
in the BBM equation gives rise to remarkably rich dynamics for the
smoothed dispersive Riemann problem with initial data \eqref{step} in
the form of a monotone transition between two constant states $u_-$ at
$x \to -\infty$ and $u_+$ at $x \to +\infty$.  We identify regions in
the plane of boundary states $(u_-,u_+)$ where BBM solutions exhibit
qualitatively different behaviors. Additionally, the evolution of data
within these regions often depends significantly upon the smoothing of
the initial data characterized by its transition width $\xi$, thus
introducing a further distinction between various regions.  To capture
the smoothing-dependent features, we construct two separate dispersive
Riemann problem classifications: for $\xi \gg 1$ and for $\xi \ll 1$
shown in Figures~\ref{fig:xi10} and \ref{fig:xi01}, respectively. A
detailed guide to the wave patterns presented in these figures is
contained in Section~\ref{bbm_riemann}, which summarizes our findings.

Emergent wave phenomena for the BBM dispersive Riemann problem can be
roughly split into two categories: classical and nonclassical.
Classical phenomena include dispersive shock waves (DSWs) and
rarefaction waves, which are also observed in convex, KdV-type
dispersive hydrodynamics, hence termed classical.  The nonclassical
features are due to nonconvex dispersion and include the generation of
two-phase linear wavetrains, expansion shocks, solitary wave shedding,
dispersive Lax shocks (in a conservative dispersive equation!), DSW
implosion and the generation of incoherent solitary wavetrains. Some
of these striking nonclassical features are present only in the
evolution of narrow initial steps, $\xi \ll 1$. We stress that all the
observed nonclassical phenomena involve short wave features that are
formally not applicable in the traditional asymptotic regime of the
original BBM equation \eqref{BBM_original} in which $u_{xxt} \sim
-u_{xxx}$.

The description of a diverse range of linear and nonlinear wave
patterns emerging from the BBM dispersive Riemann problem has required
an equally diverse set of approaches---analytical and numerical---to
their analysis and description. We analyze linear wavetrains generated
near the zero dispersion point using the stationary phase method and
matched asymptotics, while solitary wave shedding is tackled by a
combination of short time analysis and energy arguments. DSWs in the
convex propagation regime, defined by strict hyperbolicity and genuine
nonlinearity of the associated Whitham modulation system, are analyzed
using the extended DSW fitting method while DSW implosion dynamics are
shown to be approximately described by the coupled NLS equation
derived from BBM via multiple-scale expansions for weakly nonlinear,
two-phase waves. All analytical results are accompanied by detailed
numerical simulations. The numerical methods and approaches are
presented in the Appendix.

Overall, our analysis shows that the BBM equation describes
nonclassical dispersive hydrodynamics that are distinct from
previously studied systems with nonconvex flux (modified KdV and
Gardner equations \cite{el_dispersive_2017,kamchatnov_undular_2012},
Miyata-Choi-Camassa system \cite{esler_dispersive_2011}, derivative
NLS equation \cite{ivanov_riemann_2017}) and higher order/nonconvex
dispersion (Kawahara equation
\cite{sprenger_shock_2017,sprenger_discontinuous_2020}, nonlocal NLS
equation \cite{el_radiating_2016,baqer_modulation_2020}).

Another interpretation of these results incorporates the distinction
between long and short waves, evident in the aforementioned difference
between the linear dispersion relations for KdV \eqref{kdv} and BBM
\eqref{bbm} in which the phase and group velocities, although
asymptotically equivalent for long waves, are bounded for BBM and
unbounded for KdV in the short wave regime.  While linear theory
predicts that short waves in BBM are feeble, in the words of the
equation's namesake \cite{benjamin_model_1972}, nonlinear wave
propagation does not suffer from this restriction
\cite{manna_asymptotic_1998}.  The prevalence of nonlinear, short-long
wave interaction in the dispersive Riemann problem for BBM is
striking.  When a short scale is encoded in the initial condition, new
features arise in the solution, including expansion shocks, the
emission of solitary waves, and 2-phase linear wavepackets.  Short
wave effects also arise during the course of nonlinear evolution,
independent of the spatial scale of the initial data.  Dispersive
shock wave implosion, the dispersive Lax shock, and incoherent
solitary wavetrains are notable examples.  In fact, the self-similar
dispersive Lax shock $u(x,t) = g(xt)$ exhibits short waves that are
unbounded in wavenumber as $t \to \infty$.  While a slight asymmetry
in the boundary conditions arrests the unbounded wavenumber production
in the dispersive Lax shock, the resulting intermittent solitary wave
emission suggests a conservative counterpart to the chaos observed in
numerical simulations of a damped-driven BBM equation
\cite{rempel_origin_2007}.  None of these short wave effects occur in
the KdV equation.

Some of the results obtained in this paper carry over to other
dispersive hydrodynamic systems.  The observed prominent dependence of
the long-time dynamics on the initial transition width points to a
leading order effect that has mostly been ignored in previous DSW
studies.  The analysis of two-phase linear wavetrains near a zero
dispersion point is general, and the coupled NLS description of DSW
implosion can be applied to the conduit and magma equations where this
effect is known to occur. Expansion shocks and solitary wave shedding have
been observed in the regularized Boussinesq shallow water system
\cite{el_stationary_2018} albeit in a nonphysical, short-wave regime.
Still, it would be fair to say that many of the striking nonconvex
wave regimes generated in the BBM dispersive Riemann problem await
their realization in other systems.

Our work poses a number of interesting questions that will hopefully
be addressed in future research.  These include a better understanding
of solitary wave shedding and the intermittency associated with the
incoherent solitary wave regime, stability analysis of dispersive Lax
shocks, and a full analytical description of the nonlinear two-phase
structures and envelope solitary waves associated with DSW implosion.

\bibliographystyle{plain}
%\bibliography{biblio}

\appendix

\section{Numerical method}

\subsection{Scheme for the BBM equation}
\label{sec:scheme-bbm-equation}

The numerical method described here has been used to solves
numerically the Riemann problem~\eqref{bbm},\eqref{step}. It
incorporates a standard fourth-order Runge--Kutta timestepper and a
pseudo-spectral Fourier spatial discretization. A different method is
used in Sec.~\ref{sec:incoh-solit-wave}.

The initial condition~\eqref{step} is implemented with the $L$-periodic
function:
\begin{equation}
u_0^{\rm num.}(x) = \frac{u_+-u_-}{2} \left( \tanh \left[
\frac{x}{\xi} \right ] + \tanh \left[
\frac{x-L/2}{\xi} \right ] \right) + u_-.
\end{equation}
$L$ is sufficiently large such that the hydrodynamic states generated
by the steps at $x=0$ and $x=L/2$ do not overlap at time $t$. Note
that $L \gg \xi$ and $|u_0^{\rm num.}(x)-u_\pm| < 10^{-15} $ far from
the step locations. The grid spacing is $\Delta x \approx 0.0114$.

The spatial Fourier transform of the BBM equation reads:
\begin{equation}
\label{bbm_num}
\frac{\rmd \hat u}{\rmd t} = \frac{-i k}{2(1+k^2)} \widehat{u^2},
\end{equation}
where $\hat u(k,t)$ is the spatial Fourier transform of $u(x,t)$.  The
truncation of the Fourier domain of $\hat u$ turns~\eqref{bbm_num}
into a nonlinear ordinary differential equation which is temporally
evolved according to the fourth-order Runge--Kutta method with a
timestep $\Delta t = 0.1$.  The computation of the Fourier transform
of $\widehat{u^2}$ is implemented using the fast Fourier transform.

\subsection{Determination of the linear wave parameters}
\label{wave_param}

We detail in this section the procedure to extract from the numerical
solution the wave parameters of the, beating, two-phase wave observed
in the region $x<0$ in Sec.~\ref{linear_wave}.  In the following the
index $i=1$ ($i=2$) denotes the wave with a modulated wavenumber
smaller (larger) than~$\sqrt 3$.

\begin{figure}[h]
\includegraphics{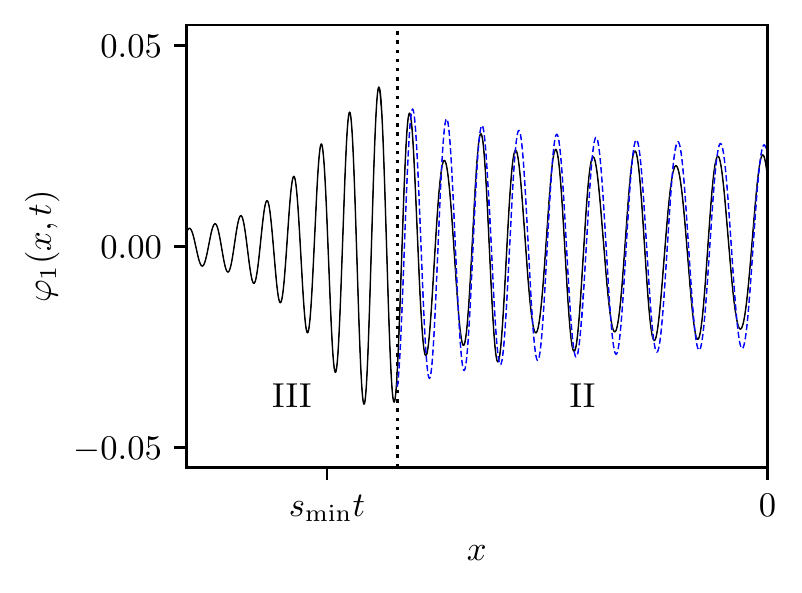}
\includegraphics{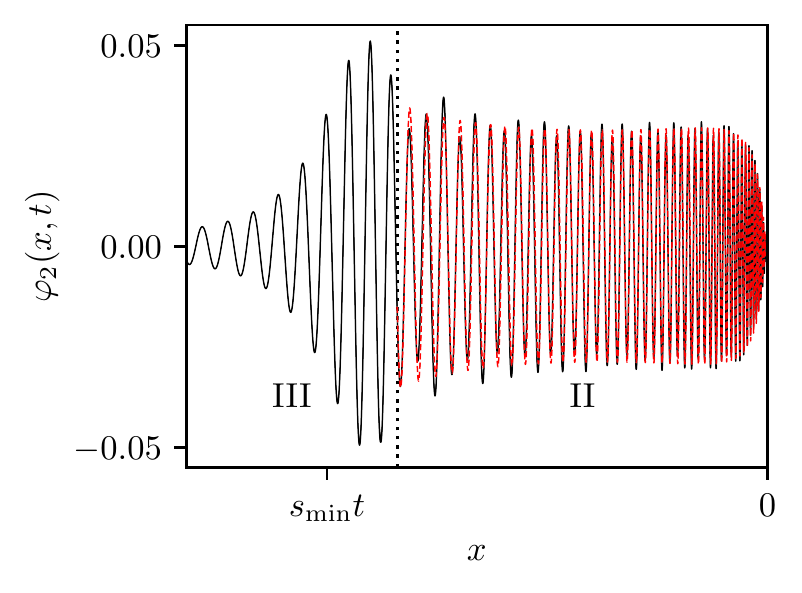}
\caption{Variation of the linear waves $\varphi_1(x,t)$ and
  $\varphi_2(x,t)$ (black solid lines) for the dispersive Riemann
  problem $(\xi,u_-,u_+) = (0.1,1,1.5)$ at $t=500$. For
  comparison we plot the corresponding stationary phase solution in
  dashed line: the left (right) plot displays the first (second) term
  of~\eqref{sol30}. Note that the discrepancy between the analytical
  solution and the numerical solution and is due to the violation of
  the small amplitude, cf. discussion in Sec.\ref{linear_wave}. }
\label{spm_k_filter}
\end{figure}

We first compute the Fourier transform $\hat \varphi(k,t)$ of the
numerical solution $\varphi(x,t) = u(x,t) - u_-$. $\varphi(x,t)$ is
given by:
\begin{equation}
\label{sol30}
\varphi(x,t) = \frac{1}{2\pi} \int_{\mathbb{R}} \hat \varphi(k,t)
e^{ikx} dk = \varphi_1(x,t) + \varphi_2(x,t),
\end{equation}
where
$\varphi_1(x,t) = 1/2\pi \int_{|k| < \sqrt 3} \hat \varphi(k,t)
e^{ikx} dk$ and
$\varphi_2(x,t) = 1/2\pi \int_{|k| > \sqrt 3} \hat \varphi(k,t)
e^{ikx} dk$ are, with a good approximation, two distinct one-phase
linear waves, cf. Fig.~\ref{spm_k_filter}. We then extract from the
variations of $\varphi_1$ and $\varphi_2$ the modulated wavenumbers
$k_1(x,t)$ and $k_2(x,t)$.
By definition the distinction between $\varphi_1$ and $\varphi_2$
should only hold in the $(x,t)$-region where
$k_1(x,t) \ll \sqrt 3 \ll k_2(x,t)$. In practice we see in
Fig.~\ref{spm_k} that the two linear waves are still distinct
even if $k_1$ and $k_2$ are close to $\sqrt 3$.

Note that we also obtain numerically
$k_1 \sim k_2 \sim \sqrt 3$ for $x \ll s_{\rm min} t$ even if no
linear waves can propagate in the region. This is a numerical
artifact and one can check that $\varphi_1$ and $\varphi_2$ are out of
phase in this region such that $\varphi = \varphi_1+\varphi_2 = 0$ as
expected.

\section{HNLS description}
\label{HNLSb}

\subsection{HNLS equation}

The NLS equation displays both a dispersive effect term
$A_{xx} \propto \varepsilon^3$ and a weakly nonlinear term
$|A|^2A \propto \varepsilon^3$, where $\varepsilon$ is the small
parameter describing the slow spatio-temporal dependence and small
amplitude of the envelope $A(x,t)$,
cf. \cite{benney_propagation_1967}. The next order of the envelope
dynamics of the weakly nonlinear wave:
\begin{equation}
u(x,t) \sim u_0 + \left[ A(x,t) e^{i(\kappa_0 x
-\omega_0(\kappa_0,u_0) t)} + \cc \right] + B(x,t),
\end{equation}
is commonly called higher order NLS (HNLS) equation, includes the
terms: $A_{xxx}\propto \varepsilon^4$,
$|A|^2 A_x\propto \varepsilon^4$ and $A^2 A^*_x \propto \varepsilon^4$
(cf. {our paper} and references therein). The HNLS equation for the
BBM equation reads:
\begin{equation}\label{HNLS}
\begin{split}
&i A_t +  i \partial_k\omega_0 A_x + \beta
A_{xx}  + \gamma | A|^2
A  + i \delta
A_{xxx} + i \lambda  |A|^2 A_x
+ i \nu  A^2  A^*_x = 0,\\
&B = b |A|^2 + \frac i
2 b_2 (A A^*_x - A^* A_x) = b \frac{a^2}{16} + b_2 \frac{a^2 v}{16}.,
\end{split}
\end{equation}
where the coefficients $\beta$, $\gamma$ and $b$ are given
in~\eqref{coeff_NLS}, and $a$ and $v$ are defined in~\eqref{Amad}.
$\delta$, $\lambda$, $\nu$ and $b_2$ are
function of the parameters $\kappa_0,u_0$ given by:
\begin{equation}
\begin{split}
&\delta = \frac{u_0(1-6\kappa_0^2+\kappa_0^4)}{(1+\kappa_0^2)^4},\;
\lambda = \frac{-9-9\kappa_0^2+9\kappa_0^4+5\kappa_0^6}
{3u_0(3+\kappa_0^2)^2 (\kappa_0^2+\kappa_0^4)},\;\\
&\nu = \frac{-27-21\kappa_0^2+19\kappa_0^4+5\kappa_0^6}
{6u_0(3+\kappa_0^2)^2 (\kappa_0^2+\kappa_0^4)},\;
b_2 = \frac{-2(\kappa_0^2-3)(1+\kappa_0^2)}{u_0
\kappa_0^3(3+\kappa_0^2)^2}.
\end{split}
\end{equation}

\subsection{DSW modulation}

The universal DSW modulation near the trailing edge is given by the
following special self-similar rarefaction wave solution
of~\eqref{HNLS}, cf.~\cite{congy_nonlinear_2019}:
\begin{equation}
\begin{split}
&\lambda_- \frac{a^2}{16} + 2\beta_- v - 3\delta_- v^2 - {\rm
sgn}(\beta_-) \frac{a^2}{16} \sqrt{D(a,v) } = \frac{x}{t} - s_-,\\
&D(a,v) = \nu_-^2-\frac{32(\beta_--3\delta_-
v)(\gamma_--(\lambda_--\nu_-)v)}{a^2},
\end{split}
\end{equation}
where $a$ and $v$ are related by the differential equation:
\begin{equation}
\frac{\rmd v}{\rmd a} + \frac{\nu_- + {\rm sgn}(\beta_-) a
\sqrt{D(a,v)} }{16 (\beta_--3\delta_- v)},\quad v(0) = 0.
\end{equation}
The parameters $\beta_-$, $\gamma_-$, $\delta_-$, $\lambda_-$ and
$\nu_-$ are the coefficients $\beta$, $\gamma$, $\delta$, $\lambda$ and
$\nu$ evaluated at $(\kappa_0,u_0)=(k_-,u_-)$ where $k_-$ is given
by~\eqref{bbm_k_locus}.

\section{Stability of nonlinear two-phase solution}
\label{sec:stab-nonl-two}

In \cite{griffiths_modulational_2006}, it was shown that the stability
of the plane wave solution \eqref{deltaw} to the coupled NLS equations
\eqref{coupled_NLS} boils down to determining the roots $v$ of the
quartic polynomial
\begin{equation}
  \label{eq:16}
  \begin{split}
    &\left ( (v - v_{g1})^2 + Q_1(q) \right ) \left ( (v - v_{g2})^2 +
      Q_2(q) \right ) - R = 0, \\
    &Q_j(q) = - \beta_j^2q^2+\frac{1}{8} a_j^2 \beta_j \gamma_j, \quad R =
    \frac{1}{64} \beta_1 \beta_2 \nu_{12} \nu_{21} a_1^2 a_2^2 ,
  \end{split}
\end{equation}
where we have introduced the notation $v_{gj} = \partial_k
\omega_0(\kappa_j,u_-)$, $\beta_j \equiv \beta(\kappa_j,u_-)$,
$\gamma_j \equiv \gamma(\kappa_j,u_-)$, and $\nu_{ij} \equiv
\nu(\kappa_i,\kappa_j,u_-)$ for brevity.  A root $v$ of the quartic
\eqref{eq:16} is the phase velocity and $q$ is the wavenumber, both of
the infinitesimal plane wave perturbation $\propto \exp(iq(x-vt))$ of
the nonlinear solution \eqref{deltaw}.  Consequently, if all four
roots of \eqref{eq:16} are real for every $q \in \mathbb{R}$, then the
weakly nonlinear two-phase wavetrain \eqref{deltaw} is modulationally
stable.  

\begin{figure}
  \centering
  \includegraphics[scale=0.333333]{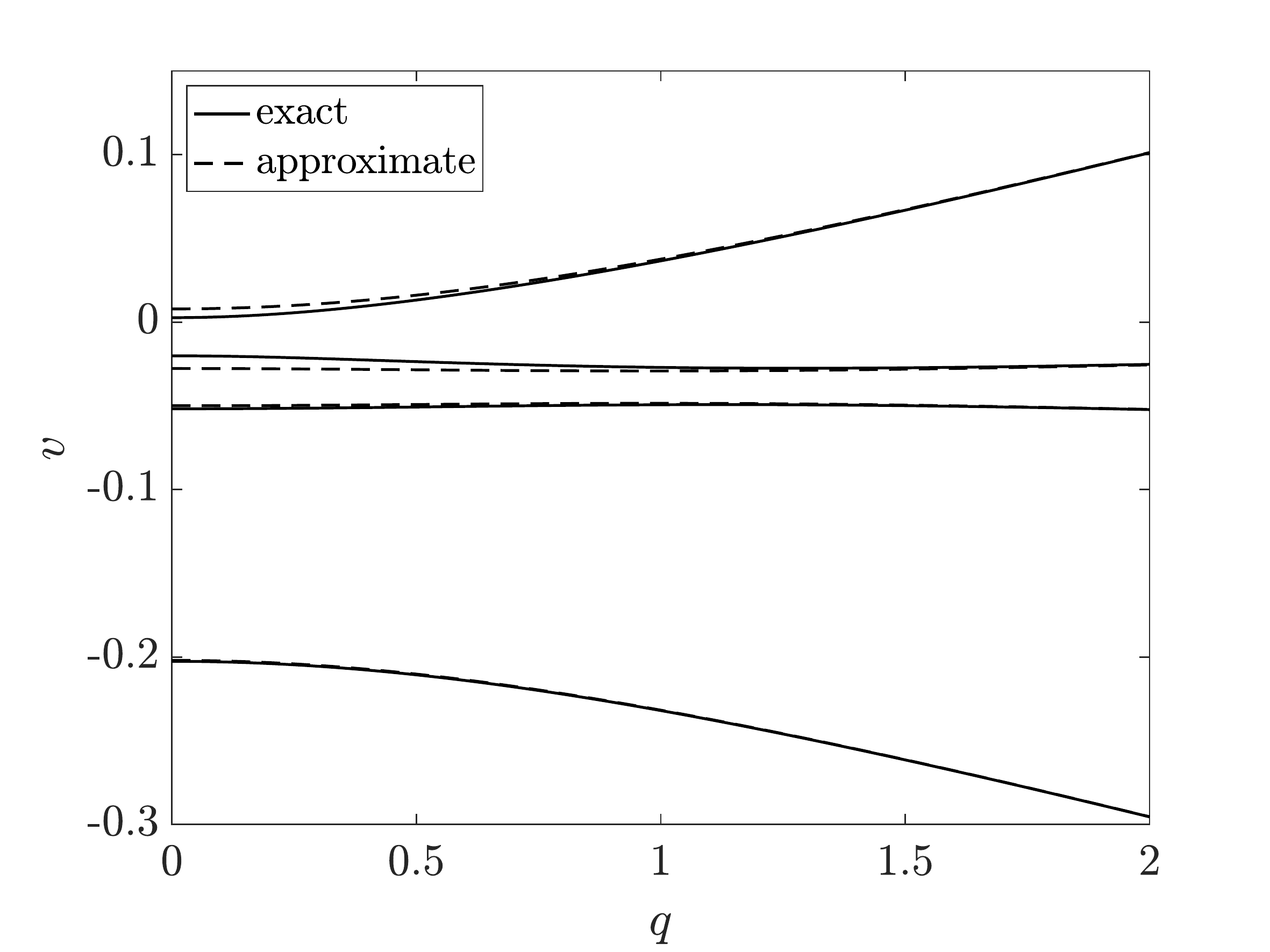}
  \caption{Linear dispersion relation $v(q)$ for perturbations to the
    plane wave solution \eqref{deltaw} corresponding to
    Fig.~\ref{fourier5b}.  The exact roots of \eqref{eq:16} (solid)
    and their approximation \eqref{eq:18} (dashed) are shown.  All are
    purely real, implying modulational stability of region IIa.}
  \label{fig:stability_coupled_nls}
\end{figure}

In general, there are no simple, explicit expressions for the roots of
\eqref{eq:16}.  However, a simplification of the general calculations
in \cite{griffiths_modulational_2006} can be made when the two group
velocities $v_{g1}$, $v_{g2}$ are close so that \eqref{eq:16} can be
approximated by taking $(v-v_{gj})^2 \to (v-\bar{v}_g)^2$ where
$\bar{v}_g = \frac{1}{2}(v_{g1}+v_{g2})$ is the average of the group
velocities.  Then $(v-\bar{v}_g)^2$ in the approximation of
\eqref{eq:16} satisfies a quadratic equation.  Introducing the
difference of the group velocities $\Delta v_g =
\frac{1}{2}(v_{g2}-v_{g1})$ and assuming $|\Delta v_g/\bar{v}_g| \ll
1$, the four roots of \eqref{eq:16} are approximately
\begin{equation}
  \label{eq:18}
  2(v - \bar{v}_{g} \pm \Delta v_g)^2 = -Q_1(q)-Q_2(q) \pm \sqrt{(Q_2(q) -
    Q_1(q))^2 + 4 R } .
\end{equation}
Then the necessary and sufficient condition for stability is the real,
non-negativity of $-Q_1(q)-Q_2(q) \pm \sqrt{(Q_2(q) - Q_1(q))^2 + 4 R
}$.  Since $Q_1(q) < 0$ and $Q_2'(q) < 0$, it suffices to consider
only the long wavelength case $q = 0$.  For the simulation shown in
Fig.~\ref{fourier5b}, we calculate
\begin{equation}
  \label{eq:19}
  -Q_1(0)-Q_2(0) \pm \sqrt{(Q_2(0) - Q_1(0))^2 + 4 R} \approx 0.0111
  \pm 0.0109 > 0 ,
\end{equation}
hence the two-phase plane wave solution \eqref{deltaw} is stable.  A
plot of the linear dispersion relation $v = v(q)$ for perturbations to
\eqref{deltaw} as roots of \eqref{eq:16} and their approximation
\eqref{eq:18} for the extracted simulation parameters in
Fig.~\ref{fourier5b} is shown in Fig.~\ref{fig:stability_coupled_nls}.

\end{document}